\documentclass[12pt]{iopart}
\usepackage{iopams}
\usepackage{graphicx,float}
\usepackage{color}
\usepackage{psfrag}
\newcommand{\be}{\begin{equation}}
\newcommand{\ee}{\end{equation}}
\newcommand{\bea}{\begin{eqnarray}}
\newcommand{\eea}{\end{eqnarray}}

\eqnobysec
\begin{document}
\bibliographystyle{jpa}

\title{Dynamical structure factor at small $q$ for the XXZ spin-1/2 chain}

\author{R G Pereira$^{1\dagger}$, J Sirker$^2$, J-S Caux$^3$, R Hagemans$^3$, J M Maillet$^4$, S R White$^5$ and I Affleck$^1$}

\address{$^1$ Department of Physics and Astronomy, University of British Columbia,
Vancouver, British Columbia, Canada V6T 1Z1}

\address{$^2$ Max-Planck-Institute for Solid State Research, Heisenbergstr. 1,
  70569 Stuttgart, Germany}

\address{$^3$ Institute for Theoretical Physics, University of Amsterdam, 1018
XE Amsterdam, The Netherlands}

\address{$^4$ Laboratoire de Physique, \'{E}cole Normale Sup\'{e}rieure de Lyon et CNRS,
69364 Lyon C\'{e}dex 07, France}

\address{$^5$ Department of Physics and Astronomy, University of California, Irvine
CA 92697, USA}

\ead{$^{\dagger}$rpereira@phas.ubc.ca}

\date{\today{}}

\begin{abstract}
We combine Bethe Ansatz and field theory methods to study the longitudinal
  dynamical structure factor $S^{zz}\left(q,\omega\right)$ for the anisotropic
  spin-1/2 chain in the gapless regime. Using bosonization, we derive a low
  energy effective model, including the leading irrelevant operators (band
  curvature terms) which account for boson decay processes. The coupling
  constants of the effective model for finite anisotropy and finite magnetic
  field are determined exactly by comparison with corrections to thermodynamic
  quantities calculated by Bethe Ansatz. We show that a good approximation for
  the shape of the on-shell peak of $S^{zz}\left(q,\omega\right)$ in the
  interacting case is obtained by rescaling the result for free fermions by
  certain coefficients extracted from the effective Hamiltonian. In
  particular, the width of the on-shell peak is argued to scale like
  $\delta\omega_{q}\sim q^{2}$ and this prediction is shown to agree with the
  width of the two-particle continuum at finite fields calculated from the
  Bethe Ansatz equations. An exception to the $q^2$ scaling  is found at finite field and large anisotropy parameter (near the isotropic point). We also present the calculation of the
  high-frequency tail of $S^{zz}\left(q,\omega\right)$ in the region
  $\delta\omega_{q}\ll\omega-vq\ll J$ using finite-order perturbation theory
  in the band curvature terms. Both the width of the on-shell peak and the
  high-frequency tail are compared with 
  $S^{zz}\left(q,\omega\right)$ calculated by Bethe Ansatz for finite chains
  using determinant expressions for the form factors and excellent
  agreement is obtained. Finally, the accuracy of the form factors is
  checked against the exact first moment sum rule and the static structure
  factor calculated by Density Matrix Renormalization Group (DMRG).

\end{abstract}

\maketitle

\section{Introduction}

The problem of a spin-1/2 chain with anisotropic antiferromagnetic
exchange interaction has been extensively studied \cite{IanLesHouches} and constitutes one of best known examples of strongly correlated one-dimensional systems \cite{giamarchi}. The XXZ model is integrable and exactly solvable by Bethe Ansatz \cite{BetheZP71,OrbachPR112}, which makes it possible to calculate exact ground state properties as well as thermodynamic quantities. At the same time, it exhibits a critical regime as a function of the anisotropy parameter, in which the system falls into the universality class of the Luttinger liquids. The long distance asymptotics of correlation functions can then be calculated by applying field theory methods. The combination of field theory and Bethe Ansatz has proved quite successful in explaining low energy properties of spin chain compounds such as Sr$_2$CuO$_3$ and KCuF$_3$  \cite{motoyama}.

Recently, most of the interest in the XXZ model has turned to the study of dynamical correlation functions. The relevant quantities for spin chains are the dynamical structure factors $S^{\mu\mu}(q,\omega)$, $\mu=x,y,z$, defined as the Fourier transform of the spin-spin correlation functions \cite{muller}. These are directly probed by inelastic neutron scattering experiments \cite{nagler,stone}. They are also probed indirectly by nuclear magnetic resonance \cite{thurber}, since the spin lattice relaxation rate is proportional to the integral of the transverse structure factor over momentum \cite{moryia,sirker}. 

Even though one can use the Bethe Ansatz to construct the exact eigenstates, the evaluation of matrix elements, which still need to be summed up in order to obtain the correlation functions, turns out to be very complicated in general. In the last ten years significant progress has been made with the help of quantum group methods \cite{JimboBOOK}. It is now possible to write down analytical expressions for the form factors for the class of two-spinon excitations for the Heisenberg chain (the isotropic point) at zero field \cite{BougourziPRB54,bougourzi,BougourziPRB57}, as well as for four-spinon ones \cite{BougourziMPLB10,AbadaNPB497,CauxJSTATP12013}. No such expressions are available for general anisotropy in the gapless regime or for finite magnetic field, but in those cases the form factors can be expressed in terms of determinant formulas \cite{KitanineNPB554,KitanineNPB567,CauxJSTATP09003} which can then be evaluated numerically for finite chains for two-particle states \cite{BiegelEL59,BiegelJPA36,SatoJPSJ73} or for the general multiparticle contributions throughout the Brillouin zone \cite{CauxPRL95,CauxJSTATP09003}.

From a field theory standpoint, dynamical correlations can be calculated fairly easily using bosonization \cite{schulz}. However, this approach is only asymptotically exact in the limit of very low energies and relies on the approximation of linear dispersion for the elementary excitations. In some cases, the main features of a dynamical response depend on more detailed information about the excitation spectrum of the system at finite energies -- namely the breaking of Lorenz invariance by band curvature effects. That poses a problem to the standard bosonization approach, in which nonlinear dispersion and interaction effects cannot be accommodated simultaneously. For that reason, a lot of effort has been put into understanding 1D physics beyond the Luttinger model \cite{pustilnik1,pustilnik2,pustilnik3,wiegmann,kopietz,pereira,rozhkov,teber,Aristov,khodas}.

In particular, using the bosonization prescription one can relate the longitudinal dynamical structure factor $S^{zz}(q,\omega)$ at small momentum $q$ to the spectral function of the bosonic modes of the Luttinger model. In the linear dispersion approximation, the conventional answer is that $S^{zz}(q,\omega)$ is a delta function peak at the energy carried by the noninteracting bosons \cite{giamarchi}. As in the higher-dimensional counterparts, the broadening of the peak is a signature of a finite lifetime. The problem of calculating the actual lineshape of $S^{zz}(q,\omega)$ at small $q$ is thus related to the fundamental question of the decay of elementary excitations in 1D. 

In the bosonization approach, interactions are included exactly, but band curvature effects must be treated perturbatively. All the difficulties stem from the fact that band curvature operators introduce interactions between the bosons and ruin the exact solvability of the Luttinger model. To make things worse, perturbation theory in those operators breaks down near the mass shell of the bosonic excitations \cite{samokhin} and no proper resummation scheme is known to date. The best alternative seems to be guided by the fermionic approach, which treats band curvature exactly but applies perturbation theory in the interaction \cite{pustilnik2}. 

In this paper we address this question using both  bosonization and Bethe Ansatz. Our goal is to make predictions about $S^{zz}(q,\omega)$ that are nonperturbative in the interaction ({\it i.e.}, anisotropy) parameter and are therefore valid in the entire gapless regime of the XXZ model (including the Heisenberg point). We focus on the finite field case, which in the bosonization approach is described by a simpler class of irrelevant operators. To go beyond the weakly interacting regime we can resort to the Bethe Ansatz equations in the thermodynamic limit to calculate the exact coupling constants of the low energy effective model. Our analysis is supported by another type of Bethe Ansatz based method, which calculates the exact form factors for finite chains. This provides a nontrivial consistency check of our results.

The outline of the paper is as follows. In section \ref{sec:XXZmodel}, we introduce the longitudinal dynamical structure factor for the XXZ model in a finite magnetic field and review the exact solution for the XX model. In section \ref{sec:bosonmodel} we describe the effective bosonic model and explain how to fix the coupling constants of the irrelevant operators. Section \ref{sec:BAsolution} provides a short description of the Bethe Ansatz framework which is relevant for our analysis. In section \ref{sec:Broadening-of-the}, we show how to obtain the broadening of $S^{zz}(q,\omega)$ in a finite magnetic field both from field theory and Bethe Ansatz and compare our formula with the exact form factors for finite chains. In section \ref{sec:High-frequency-tail} we present a more detailed derivation of the high-frequency tail of $S^{zz}(q,\omega)$ reported in \cite{pereira}. The zero field case is briefly addressed in section \ref{sec:zerofield}. Finally, we check the sum rules and discuss the finite size scaling of the form factors in section \ref{sec:sumrules}.

\section{XXZ model}\label{sec:XXZmodel}

We consider the XXZ spin-1/2 chain in a magnetic field\begin{equation}
H=J\sum_{j=1}^{N}\left[S_{j}^{x}S_{j+1}^{x}+S_{j}^{y}S_{j+1}^{y}+\Delta S_{j}^{z}S_{j+1}^{z}-hS_{j}^{z}\right].\label{eq:XXZ}\end{equation}
 Here, $J$ is the exchange coupling, $\Delta$ is the anisotropy
parameter, $h$ is the magnetic field in units of $J$ and $N$ is the number of sites in the chain with periodic boundary conditions. We focus
on the critical regime (given by $-1<\Delta\leq1$ for $h=0$). We
are interested in the longitudinal dynamical structure factor at zero
temperature\begin{equation}
S^{zz}\left(q,\omega\right)=\frac{1}{N}\sum_{j,j^{\prime}=1}^{N}e^{-iq\left(j-j^{\prime}\right)}\int_{-\infty}^{+\infty}dt\, e^{i\omega t}\left\langle S_{j}^{z}\left(t\right)S_{j^{\prime}}^{z}\left(0\right)\right\rangle ,\label{eq:dynamical}\end{equation}
where $q$ takes the discrete values $q=2\pi n/N$, $n\, \mathbb{\in Z}$. It is instructive to write down the Lehmann representation for $S^{zz}\left(q,\omega\right)$\begin{equation}
S^{zz}\left(q,\omega\right)=\frac{2\pi}{N}\sum_{\alpha}\left|\left\langle 0\left|S_{q}^{z}\right|\alpha\right\rangle \right|^{2}\delta\left(\omega-E_{\alpha}+E_{GS}\right),\label{eq:lehmann}\end{equation}
 where $S_{q}^{z}=\sum_{j}S_{j}^{z}e^{-iqj}$, $\left|\alpha\right\rangle $
is an eigenstate with energy $E_{\alpha}$ and $E_{GS}$ is the ground state
energy. The matrix elements $\left\langle 0\left|S_{q}^{z}\right|\alpha\right\rangle$
are called form factors. We denote by
 $F^2\equiv\left|\left\langle 0\left|S_{q}^{z}\right|\alpha\right\rangle \right|^2$ the transition probabilities that appear in (\ref{eq:lehmann}). For a finite system, $S^{zz}\left(q,\omega\right)$
is a sum of delta function peaks at the energies of the eigenstates with fixed
momentum $q$. In this sense, $S^{zz}\left(q,\omega\right)$ provides
direct information about the excitation spectrum of the spin chain.
In the thermodynamic limit $N\rightarrow\infty$, the spectrum is
continuous and $S^{zz}\left(q,\omega\right)$ becomes a smooth function
of $q$ and $\omega$. Equation (\ref{eq:lehmann}) also implies that $S^{zz}\left(q,\omega\right)$
is real and positive and can be expressed as a spectral function\begin{equation}
S^{zz}\left(q,\omega\right)=-2\,\textrm{Im}\chi^{ret}\left(q,\omega\right),\label{eq:spectralfunction}\end{equation}
 for $\omega>0$. $\chi^{ret}\left(q,\omega\right)$ is the retarded
spin-spin correlation function and can be obtained from the Matsubara
correlation function \begin{equation}
\chi\left(q,i\omega_{n}\right)=-\frac{1}{N}\sum_{j,j^{\prime}=1}^{N}e^{-iq\left(j-j^{\prime}\right)}\int_{0}^{\beta}d\tau\, e^{i\omega_{n}\tau}\left\langle S_{j}^{z}\left(\tau\right)S_{j^{\prime}}^{z}\left(0\right)\right\rangle ,\label{eq:densitycorr}\end{equation}
 where $\beta$ is the inverse temperature, by the analytical continuation
$i\omega_{n}\rightarrow\omega+i\varepsilon$. 

It is well known that the one-dimensional XXZ model is equivalent to interacting spinless
fermions on the lattice. The mapping is realized by the Wigner-Jordan
transformation\begin{eqnarray}
S_{j}^{z} & \rightarrow & n_{j}-\frac{1}{2},\nonumber \\
S_{j}^{+} & \rightarrow & \left(-1\right)^{j}\, c_{j}^{\dagger}e^{i\pi\phi_{j}},\label{eq:WignerJordan}\\
S_{j}^{-} & \rightarrow & \left(-1\right)^{j}\, c_{j}^{\phantom{\dagger}}e^{-i\pi\phi_{j}},\nonumber \end{eqnarray}
 where $c_{j}^{\phantom{\dagger}}$ is the annihilation operator for
fermions at site $j$, $n_{j}=c_{j}^{\dagger}c_{j}^{\phantom{\dagger}}$
and $\phi_{j}=\sum_{\ell=1}^{j-1}n_{\ell}$. In terms of fermionic
operators, the Hamiltonian (\ref{eq:XXZ}) is written as \bea
H&=&J\sum_{j=1}^{N}\left[-\frac{1}{2}\left(c_{j}^{\dagger}c_{j+1}^{\phantom{\dagger}}+h.c.\right)-h\left(c_{j}^{\dagger}c_{j}^{\phantom{\dagger}}-\frac{1}{2}\right)\right.\nonumber\\& &\left.+\Delta\left(n_{j}-\frac{1}{2}\right)\left(n_{j+1}-\frac{1}{2}\right)\right].\label{eq:Hfermion}\eea

\subsection{Exact solution for the XX model\label{sub:Exact-solution-for}}

One case of special interest is the XX point $\Delta=0$, at which
(\ref{eq:Hfermion}) reduces to a free fermion model \cite{Lieb}. As the free
fermion point will serve as a guide for the resummation of the bosonic
theory, we reproduce the solution in detail here. For $\Delta=0$
the Hamiltonian (\ref{eq:Hfermion}) can be easily diagonalized by
introducing the operators in momentum space\begin{equation}
c_{p}=\frac{1}{\sqrt{N}}\sum_{j=1}^{N}e^{-ipj}c_{j}.\end{equation}
 with $p=2\pi n/N$, $n\,\mathbb{\in Z}$, for periodic boundary conditions.
The free fermion Hamiltonian is then\begin{equation}
H_{0}=\sum_{p}\epsilon_{p}c_{p}^{\dagger}c_{p}^{\phantom{\dagger}},\label{eq:freeH0}\end{equation}
 where $\epsilon_{p}=-J\left(\cos p+h\right)$ is the fermion dispersion.
In the fermionic language, the dynamical structure factor reads \begin{eqnarray}
S^{zz}\left(q,\omega\right) & = & \frac{1}{N}\int_{-\infty}^{+\infty}dt\, e^{i\omega t}\left\langle n_{q}\left(t\right)n_{-q}\left(0\right)\right\rangle \nonumber \\
 & = & \frac{2\pi}{N}\sum_{\alpha}\left|\left\langle 0\left|n_{q}\right|\alpha\right\rangle \right|^{2}\delta\left(\omega-E_{\alpha}+E_{GS}\right),\label{eq:fermionS(q,w)}\end{eqnarray}
where $n_{q}=\sum_{j}e^{-iqj}n_{j}=\sum_{p}c_{p}^{\dagger}c_{p+q}^{\phantom{\dagger}}$.

\begin{figure}
\begin{center} \includegraphics[%
scale=0.4]{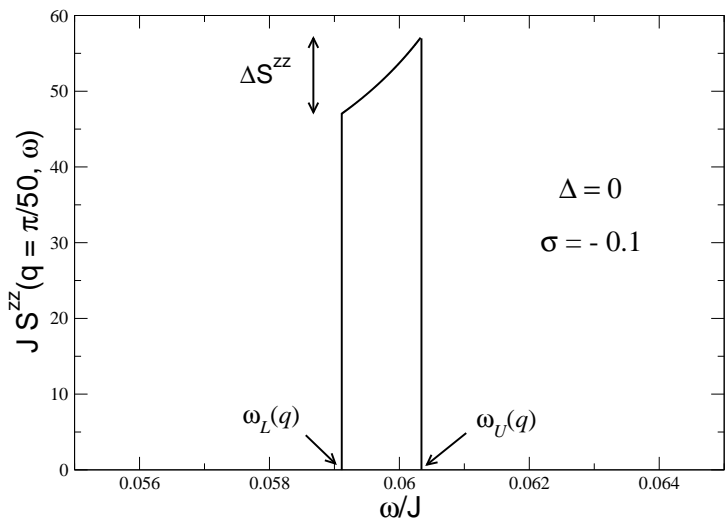}\end{center}

\caption{Exact dynamical structure factor $S^{zz}(q,\omega)$ for the free fermion point $\Delta=0$. For this graph we set $\sigma=-0.1$ ($k_F=2\pi/5$) and $q=\pi/50$.\label{freefermion}}
\end{figure}

We construct the ground state $\left|0\right\rangle $ by filling
all the single-particle states up to the Fermi momentum $k_{F}$. The
latter is determined by the condition $\epsilon_{k_{F}}=0$, which
gives\begin{equation}
k_{F}=\arccos\left(-h\right)=\pi\left(\frac{1}{2}+\sigma\right),\label{eq:kF}\end{equation}
where $\sigma\equiv\left\langle S_{j}^{z}\right\rangle =\left\langle n_{j}\right\rangle -\frac{1}{2}$
is the magnetization per site. We can also describe the excited states
in terms of particle-hole excitations created on the Fermi sea. The
only nonvanishing form factors appearing in $S^{zz}\left(q,\omega\right)$
are those for excited states with only one particle-hole pair carrying
total momentum $q$: $\left|\alpha\right\rangle =c_{p+q}^{\dagger}c_{p}^{\phantom{\dagger}}\left|0\right\rangle $.
The form factors are simply\begin{equation}
\left\langle 0\left|S_{q}^{z}\right|\alpha\right\rangle =\theta\left(k_{F}-\left|p\right|\right)\theta\left(\left|p+q\right|-k_{F}\right).\end{equation}
 For a finite system there are $qN/2\pi$ states with form factor 1, corresponding to different choices for the hole momentum $p$ below the Fermi surface. In the limit $N\rightarrow\infty$, (\ref{eq:fermionS(q,w)})
reduces to the integral\begin{eqnarray}
S^{zz}\left(q,\omega\right) & = & \int_{-\pi}^{\pi}dp\,\theta\left(k_{F}-\left|p\right|\right)\theta\left(\left|p+q\right|-k_{F}\right)\delta\left(\omega-\epsilon_{p+q}+\epsilon_{p}\right)\nonumber \\
 & = & \frac{\theta\left(\omega-\omega_{L}\left(q\right)\right)\theta\left(\omega_{U}\left(q\right)-\omega\right)}{\left.\left(d\omega_{pq}/dp\right)\right|_{\omega_{pq}=\omega}},\label{eq:ffresult}\end{eqnarray}
 where $\omega_{pq}=\epsilon_{p+q}-\epsilon_{p}$ is the energy of
the particle-hole pair and $\omega_{L}\left(q\right)$ and $\omega_{U}\left(q\right)$
are the lower and upper thresholds of the two-particle spectrum, respectively.
For the cosine dispersion, we have\begin{equation}
\omega_{pq}=2J\sin\left(p+\frac{q}{2}\right)\sin\frac{q}{2}.\end{equation}
 The expressions for the lower and upper thresholds depend on the
proximity to half-filling (zero magnetic field). Here we shall restrict
ourselves to finite field and small momentum $\left|q\right|\ll k_{F}$.
More precisely, we impose the condition\begin{equation}
|q|<\left|2k_{F}-\pi\right|=2\pi|\sigma|.\label{restrictFT}\end{equation}
 For $k_{F}<\pi/2$ ($\sigma<0$), we have \begin{eqnarray}
\omega_{L}\left(q\right) & = & 2J\sin\frac{\left|q\right|}{2}\sin\left(k_{F}-\frac{\left|q\right|}{2}\right),\label{XX_bounds}\\
\omega_{U}\left(q\right) & = & 2J\sin\frac{\left|q\right|}{2}\sin\left(k_{F}+\frac{\left|q\right|}{2}\right).\label{eq:upperbound}\end{eqnarray}
 If $k_{F}>\pi/2$, the above expressions for $\omega_{L}\left(q\right)$
and $\omega_{U}\left(q\right)$ are exchanged. Hereafter we take $k_F<\pi/2$ and $q>0$.
It follows from (\ref{XX_bounds}) and (\ref{eq:upperbound}) that
$S^{zz}\left(q,\omega\right)$ for fixed $q$ is finite within an
energy interval of width\begin{equation}
\delta\omega_{q}=\omega_U(q)-\omega_L(q)=4J\cos k_{F}\sin^{2}\left(\frac{q}{2}\right)\approx\left(J\cos k_{F}\right)q^{2}\label{eq:width}\end{equation}
 for small $q$. In fact, we can calculate $S^{zz}\left(q,\omega\right)$
explicitly using (\ref{eq:ffresult}). The result is \begin{equation}
S^{zz}\left(q,\omega\right)=\frac{\theta\left(\omega-\omega_{L}\left(q\right)\right)\theta\left(\omega_{U}\left(q\right)-\omega\right)}{\sqrt{\left(2J\sin\frac{q}{2}\right)^{2}-\omega^{2}}},\label{eq:szzcosine}\end{equation} which is illustrated in figure \ref{freefermion}. Note that, although the form factors are constant, $S^{zz}(q,\omega)$ is peaked at the upper threshold because of the larger density of states.
 The values of $S^{zz}\left(q,\omega\right)$ at the lower and upper
thresholds are both finite\begin{equation}
S^{zz}\left(q,\omega\rightarrow\omega_{L,U}\left(q\right)\right)=\left[2J\sin\frac{q}{2}\cos\left(k_{F}\mp\frac{q}{2}\right)\right]^{-1}.\end{equation}

In the small-$q$ limit, only excitations created around
the Fermi surface contribute to $S^{zz}\left(q,\omega\right)$. For
this reason, a simplifying approach would be to expand the fermion
dispersion around the Fermi points \begin{equation}
\epsilon_{k}^{R,L}\approx\pm v_{F}k+\frac{k^{2}}{2m}\mp\frac{\gamma k^3}{6}+\dots,\label{definevmgamma}\end{equation}
 where $k\equiv p\mp k_{F}$ for right ($R$) or left ($L$) movers,
$v_{F}=J\sin k_{F}$ is the Fermi velocity, $m=\left(J\cos k_{F}\right)^{-1}$
is the effective mass at the Fermi level and $\gamma=J\sin k_F$. The free fermion Hamiltonian
is then approximated by \begin{equation}
H_{0}=\sum_{k=-\infty}^{\infty}\left[\epsilon_{k}^{R}:c_{kR}^{\dagger}c_{kR}^{\phantom{\dagger}}:+\,\epsilon_{k}^{L}:c_{kL}^{\dagger}c_{kL}^{\phantom{\dagger}}:\right],\label{eq:kinetic}\end{equation}
 where $c_{kR,L}$ are the annhilation operators for fermions with momentum around $\pm k_F$, respectively, and : : denotes normal ordering with respect to the ground state.
If we retain only the linear term in the expansion, $\omega_{kq}$
turns out to be independent of $k$. This means that all particle-hole
excitations are degenerate, and $S^{zz}\left(q,\omega\right)$ is
given by a single delta function peak at the corresponding energy $\omega=v_{F}q$\begin{equation}
S^{zz}\left(q,\omega\right)=q\,\delta\left(\omega-v_{F}q\right).\label{eq:deltapeak}\end{equation}
 This is a direct consequence of the Lorentz invariance of the model
with linear dispersion. In order to get the broadening of $S^{zz}\left(q,\omega\right)$,
we must account for the nonlinearity of the dispersion, {\it i.e.}, band
curvature at the Fermi level. If we keep the next (quadratic) term
in $\epsilon_{k}^{R,L}$, we find\begin{equation}
S^{zz}\left(q,\omega\right)=\frac{m}{q}\,\theta\left(\frac{q^{2}}{2m}-\left|\omega-v_{F}q\right|\right).\label{eq:Szzquadratic}\end{equation}
 We note that this flat distribution of spectral weight is a good
approximation to the result in (\ref{eq:width}) and (\ref{eq:szzcosine})
in the limit $q\ll\cot k_F$, in the sense that the difference
between the values of $S^{zz}\left(q,\omega\right)$ at the lower
and upper thresholds is small compared to the average height of the peak (see figure \ref{freefermion}). This difference stems from the energy dependence of the density of states factor $1/(d\omega_{pq}/dp)|_{\omega_{pq}=\omega}$, which is recovered if we keep the $k^3$ term in the dispersion. It is easy to verify that for $q\ll \cot k_F$ ($\gamma m q\ll 1$)\begin{equation}\Delta S^{zz}\equiv S^{zz}(q,\omega_U(q))-S^{zz}(q,\omega_L(q))\approx \gamma m^2.\end{equation} $\Delta S^{zz}$ is $q$-independent, therefore $\Delta S^{zz}/(m/q)\sim q$ vanishes as $q\rightarrow 0$. This means that if we compare $S^{zz}(q,\omega)$ for different values of $q$ -- taking into account that $\delta\omega_q\sim q^2$ and $S^{zz}(q,\omega)\sim 1/q$ inside the peak and rescaling the functions accordingly -- the \emph{rescaled} function becomes flatter as $q\rightarrow 0$. On the other hand, the slope $\partial S^{zz}/\partial \omega$ near the center of the peak  diverges as $q\to 0$.

The thresholds for the two-particle
continuum, \begin{equation}
\omega_{U,L}\left(q\right)\approx v_Fq\pm\frac{q^{2}}{2m},\end{equation}
are easy to interpret. For $k_F<\pi/2$, the lower threshold corresponds to creating
a hole at the state with momentum $q$ below $k_{F}$ (a {}``deep
hole'') and placing the particle right above the Fermi surface, whereas
the upper one corresponds to the excitation composed of a {}``high-energy
particle'' at $k_{F}+q$ and a hole right at the Fermi surface \cite{pustilnik2}.

Alternatively, we could have calculated the density-density correlation
function, which for $\Delta=0$ is given by the fermionic bubble\begin{equation}
\chi\left(q,i\omega\right)=\int\frac{dk}{2\pi}\,\frac{\theta\left(-k\right)\theta\left(k+q\right)}{i\omega-\epsilon_{k+q}+\epsilon_{k}}-\left(\omega\rightarrow-\omega\right).\end{equation}
 Using the quadratic dispersion $\epsilon_{k}\approx v_{F}k+k^{2}/2m$,
we find \begin{equation}
\chi\left(q,i\omega\right)=\frac{m}{2\pi q}\,\log\left(\frac{i\omega-v_{F}q+q^{2}/2m}{i\omega-v_{F}q-q^{2}/2m}\right)-\left(\omega\rightarrow-\omega\right).\label{eq:fermionbubble}\end{equation}
The result (\ref{eq:Szzquadratic}) is then obtained by taking the
imaginary part of $\chi^{ret}\left(q,\omega\right)$ according to
(\ref{eq:spectralfunction}).

\section{Low energy effective Hamiltonian}\label{sec:bosonmodel}

\subsection{The free boson Hamiltonian\label{sub:The-free-boson}}

For a general anisotropy $\Delta\neq0$, the Hamiltonian (\ref{eq:Hfermion})
describes interacting spinless fermions. The standard approach to
study the low-energy (long-wavelength) limit of correlation functions
of interacting one-dimensional systems is to use bosonization to map
the problem to a free boson model -- the Luttinger model \cite{IanLesHouches}. This approach
has the advantage of treating interactions exactly. As a first step,
one introduces the fermionic field operators $\psi_{R,L}\left(x\right)$\begin{equation}
c_{j}\rightarrow\psi\left(x\right)=e^{ik_{F}x}\psi_{R}\left(x\right)+e^{-ik_{F}x}\psi_{L}\left(x\right),\end{equation}
\begin{eqnarray}
\psi_{R,L}\left(x\right) & = & \frac{1}{\sqrt{L}}\sum_{k=-\Lambda}^{+\Lambda}c_{kR,L}e^{\pm ikx},\end{eqnarray}
where $L=N$ is the system size (we set the lattice spacing to $1$) and $\Lambda<\pi$ is a momentum cutoff.
In the continuum limit, the kinetic energy part of the Hamiltonian
in (\ref{eq:kinetic}) can be written as \begin{eqnarray}
H_{0} & = & \int_{0}^{L}dx\,\left\{ :\psi_{R}^{\dagger}\left[v_{F}\left(-i\partial_{x}\right)+\frac{\left(-i\partial_{x}\right)^{2}}{2m}+\dots\right]\psi_{R}^{\phantom{\dagger}}:\,\right.\nonumber \\
 &  & \left.+\,:\psi_{L}^{\dagger}\left[v_{F}\left(-i\partial_{x}\right)+\frac{\left(-i\partial_{x}\right)^{2}}{2m}+\dots\right]\psi_{L}^{\phantom{\dagger}}:\right\} .\label{eq:kincontinuum}\end{eqnarray}
 The $1/m$ term is usually dropped using the argument that it has
a higher dimension and is irrelevant in the sense of the renormalization
group. However, it introduces corrections to the Luttinger liquid
fixed point which are associated with band curvature effects. Similarly,
if we write the interaction term in (\ref{eq:Hfermion}) in the continuum
limit, we get (following \cite{giamarchi})\begin{eqnarray}
H_{int} & = & \Delta J\int_{0}^{L}dx\,:\psi^{\dagger}\left(x\right)\psi^{\phantom{\dagger}}\left(x\right):\,:\psi^{\dagger}\left(x+1\right)\psi^{\phantom{\dagger}}\left(x+1\right):\nonumber \\
 & = & \Delta J\int_{0}^{L}dx\,\left\{\rho_{R}\left(x\right)\rho_{R}\left(x+1\right)+\rho_{L}\left(x\right)\rho_{L}\left(x+1\right)\right.\nonumber \\
 &  & +\rho_{R}\left(x\right)\rho_{L}\left(x+1\right)+\rho_{L}\left(x\right)\rho_{R}\left(x+1\right)\nonumber \\
 &  & +\left[ e^{i2k_{F}}\psi_{R}^{\dagger}\left(x\right)\psi_{L}^{\phantom{\dagger}}\left(x\right)\psi_{L}^{\dagger}\left(x+1\right)\psi_{R}^{\phantom{\dagger}}\left(x+1\right)+h.c.\right]\nonumber \\
 &  &  +\left. \left[e^{-i2k_{F}(2x+1)}\psi_{R}^{\dagger}\left(x\right)\psi_{L}^{\phantom{\dagger}}\left(x\right)\psi_{R}^{\dagger}\left(x+1\right)\psi_{L}^{\phantom{\dagger}}\left(x+1\right)+h.c. \right] \right\} \nonumber \\\label{eq:interaction}\end{eqnarray}
where $\rho_{R,L}\equiv\, :\psi_{R,L}^{\dagger}\psi_{R,L}^{\phantom{\dagger}}:$.
The last term corresponds to Umklapp scattering and is oscillating
except at half-filling (where $4k_{F}=2\pi$). We will neglect that term for the finite field case, but will restore it in section \ref{sec:zerofield} when we discuss the zero field case.

We now use Abelian bosonization and write the fermion fields as\begin{equation}
\psi_{R,L}\left(x\right)\sim\frac{1}{\sqrt{2\pi\alpha}}\, e^{-i\sqrt{2\pi}\phi_{R,L}\left(x\right)},\label{eq:mandelstam}\end{equation}
 where $\alpha\sim k_{F}^{-1}$ is a short-distance cutoff and $\phi_{R,L}$
are the right and left components of a bosonic field $\tilde{\phi}$
and its dual field $\tilde{\theta}$\begin{eqnarray}
\tilde{\phi} & = & \frac{\phi_{L}-\phi_{R}}{\sqrt{2}},\label{eq:phitilde}\\
\tilde{\theta} & = & \frac{\phi_{L}+\phi_{R}}{\sqrt{2}},\label{eq:thetatilde}\end{eqnarray}
 which satisfy $[\tilde{\phi}\left(x\right),\partial_{x^{\prime}}\tilde{\theta}\left(x^{\prime}\right)]=i\delta\left(x-x^{\prime}\right)$.
The density of right- and left-moving fermions can be shown to be
related to the derivative of the bosonic fields\begin{equation}
\rho_{R,L}\sim\mp\frac{1}{\sqrt{2\pi}}\partial_{x}\phi_{R,L},\label{eq:Jright/left}\end{equation}
 so that\begin{equation}
n\left(x\right)\sim\frac{1}{2}+\sigma+\frac{1}{\sqrt{\pi}}\partial_{x}\tilde{\phi}+\frac{1}{2\pi\alpha}\cos\left(\sqrt{4\pi}\tilde{\phi}-2k_{F}x\right).\label{eq:densityop}\end{equation}
 Here we are interested in the uniform (small $q$) part of the fluctuation
of $S_{j}^{z}\sim n\left(x\right)$, which is proportional to the
derivative of the bosonic field $\tilde{\phi}$. Bosonizing the linear
term in the kinetic energy (\ref{eq:kincontinuum}), we find\begin{eqnarray}
H_{0}^{lin} & = & \int_{0}^{L}dx\, iv_{F}\left(:\psi_{R}^{\dagger}\partial_{x}\psi_{R}^{\phantom{\dagger}}:\,-\,:\psi_{L}^{\dagger}\partial_{x}\psi_{L}^{\phantom{\dagger}}:\right)\nonumber \\
 & = & \frac{v_{F}}{2}\int_{0}^{L}dx\,\left[\left(\partial_{x}\phi_{R}\right)^{2}+\left(\partial_{x}\phi_{L}\right)^{2}\right].\label{eq:boskin}\end{eqnarray}
The terms that appear in the interaction part are\begin{eqnarray*}
\rho_{R,L}\left(x\right)\rho_{R,L}\left(x+1\right) & = & \frac{1}{2\pi}\left(\partial_{x}\phi_{R,L}\right)^{2},\\
\rho_{R}\left(x\right)\rho_{L}\left(x+1\right) & = & -\frac{1}{2\pi}\partial_{x}\phi_{R}\partial_{x}\phi_{L},\end{eqnarray*}
\begin{eqnarray}
 & \psi_{R}^{\dagger}\left(x\right)\psi_{L}^{\phantom{\dagger}}\left(x\right)\psi_{L}^{\dagger}\left(x+1\right)\psi_{R}^{\phantom{\dagger}}\left(x+1\right)=\nonumber \\
 & -\frac{\cos\left(2k_{F}\right)}{2\pi}\left(\partial_{x}\phi_{R}-\partial_{x}\phi_{L}\right)^{2}+\frac{\sin\left(2k_{F}\right)}{3\sqrt{2\pi}}\left(\partial_{x}\phi_{R}-\partial_{x}\phi_{L}\right)^{3}+\dots,\label{eq:expandinterac}\end{eqnarray}
where we have set $\alpha=1$ (equal to the level spacing; see \cite{giamarchi}).
If we keep only the marginal operators (quadratic in $\partial_{x}\phi_{R,L}$),
we get an exactly solvable model\bea
H_{LL}&=&\frac{v_{F}}{2}\int dx\left\{ \left(1+\frac{g_{4}}{2\pi v_{F}}\right)\left[\left(\partial_{x}\phi_{R}\right)^{2}+\left(\partial_{x}\phi_{L}\right)^{2}\right]\right.\nonumber\\& &\left. -\frac{g_{2}}{\pi v_{F}}\,\partial_{x}\phi_{L}\partial_{x}\phi_{R}\right\} ,\label{eq:quadratic}\eea
 where $g_{2}=g_{4}=2J\Delta[1-\cos(2k_{F})]=4J\Delta\sin^{2}k_{F}$.
The Hamiltonian (\ref{eq:quadratic}) can be rewritten in the form\begin{equation}
H_{LL}=\frac{1}{2}\int dx\,\left[vK\left(\partial_{x}\tilde{\theta}\right)^{2}+\frac{v}{K}\left(\partial_{x}\tilde{\phi}\right)^{2}\right],\label{eq:luttingermodel}\end{equation}
 where $v$ (the renormalized velocity) and $K$ (the Luttinger parameter)
are given by\begin{eqnarray}
v & = & v_{F}\sqrt{\left(1+\frac{g_{4}}{2\pi v_{F}}\right)^{2}-\left(\frac{g_{2}}{2\pi v_{F}}\right)^{2}}\approx v_{F}\left(1+\frac{2\Delta}{\pi}\sin k_{F}\right),\label{eq:velocity}\\
K & = & \sqrt{\frac{1+\frac{g_{4}}{2\pi v_{F}}-\frac{g_{2}}{2\pi v_{F}}}{1+\frac{g_{4}}{2\pi v_{F}}+\frac{g_{2}}{2\pi v_{F}}}}\approx1-\frac{2\Delta}{\pi}\sin k_{F}.\label{eq:luttingerK}\end{eqnarray}
 Expressions (\ref{eq:velocity}) and (\ref{eq:luttingerK}) are approximations
valid in the limit $\Delta\ll1$. The Luttinger model describes free
bosons that propagate with velocity $v$ and is the correct low energy
fixed point for the XXZ chain for any value of $\Delta$ and $h$
in the gapless regime. However, the correct values of $v$ and $K$
for finite $\Delta$ must be obtained by comparison with the exact
Bethe Ansatz (BA) solution. In the case $h=0$, the BA equations can
be solved analytically and yield\begin{eqnarray}
v\left(\Delta,h=0\right) & = & \frac{J\pi}{2}\frac{\sqrt{1-\Delta^{2}}}{\arccos\Delta},\label{eq:vBAzerofield}\\
K\left(\Delta,h=0\right) & = & \frac{\pi}{2\left(\pi-\arccos\Delta\right)}.\label{eq:KBAzerofield}\end{eqnarray}
 There are also analytical expressions for $h\approx0$ and $h$ close
to the critical field \cite{KorepinBOOK}. For arbitrary fields, one has
to solve the BA equations numerically in order to get the exact $v$
and $K$.

The Luttinger parameter in the Hamiltonian (\ref{eq:luttingermodel})
can be absorbed by performing a canonical transformation that rescales
the fields in the form $\tilde{\phi}\rightarrow\sqrt{K}\phi$ and
$\tilde{\theta}\rightarrow\theta/\sqrt{K}$. $H_{LL}$ then reads\begin{equation}
H_{LL}=\frac{v}{2}\int dx\,\left[\left(\partial_{x}\theta\right)^{2}+\left(\partial_{x}\phi\right)^{2}\right].\label{eq:Hluttingerrescaled}\end{equation}
 We can also define the right and left components of these rescaled
bosonic fields by\begin{equation}
\varphi_{R,L}=\frac{\theta\mp\phi}{\sqrt{2}}.\label{eq:phi_LR}\end{equation}
These are related to $\phi_{R,L}$ by a Bogoliubov transformation.
An explicit mode expansion (neglecting zero mode operators) is \begin{equation}
\varphi_{R,L}\left(x,\tau\right)=\sum_{q>0}\frac{1}{\sqrt{qL}}\, \left[a_{q}^{R,L}e^{-q\left(v\tau\mp ix\right)}+a_{q}^{R,L \dagger}e^{q\left(v\tau\mp ix\right)}\right],\label{eq:modeexpan}\end{equation}
 where $a_{q}^{R,L}$ are bosonic operators obeying $[a_{q}^{R,L\phantom{\dagger}},a_{q^{\prime}}^{R,L\dagger}]=\delta_{qq^{\prime}}$
and $q=2\pi n/L$, $n=1,2,\dots$, for periodic boundary conditions.
The Hamiltonian (\ref{eq:Hluttingerrescaled}) is then diagonal in
the boson operators \begin{equation}
H_{LL}=\sum_{q>0}vq\left[a_{q}^{R\dagger}a_{q}^{R\phantom{\dagger}}+a_{q}^{L\dagger}a_{q}^{L\phantom{\dagger}}\right].\end{equation}

We can calculate the propagators for the free fields $\partial_{x}\varphi_{R,L}$
from the mode expansion in (\ref{eq:modeexpan}). In real space, for
$L\rightarrow\infty$ and zero temperature ($\beta\rightarrow\infty$),
the propagators read\begin{equation}
D_{R,L}^{\left(0\right)}\left(x,\tau\right)=\left\langle \partial_{x}\varphi_{R,L}\left(x,\tau\right)\partial_{x}\varphi_{R,L}\left(0,0\right)\right\rangle _{0}=\frac{1}{2\pi}\frac{1}{\left(v\tau\mp ix\right)^{2}}\label{propag_xt}.\end{equation}
In momentum space,\begin{eqnarray}
D_{R,L}^{\left(0\right)}\left(q,i\omega_{n}\right) & \equiv & -\int_{0}^{L}dx\, e^{-iqx}\int_{0}^{\beta}d\tau\, e^{i\omega_{n}\tau}D_{R,L}^{\left(0\right)}\left(x,\tau\right)\nonumber \\
 & = & \frac{\pm q}{i\omega_{n}\mp vq }.\label{eq:propagator}\end{eqnarray}

In order to calculate the dynamical structure factor defined in (\ref{eq:dynamical}),
we express the fluctuation of the spin operator in terms of the bosonic
field $\phi$. From (\ref{eq:WignerJordan}) and (\ref{eq:densityop}),
we have\begin{equation}
S_{j}^{z}\sim\sqrt{\frac{K}{\pi}}\partial_{x}\phi.\end{equation}
 In the continuum limit,\begin{eqnarray}
\chi\left(q,i\omega_{n}\right) & = & -\frac{K}{\pi}\int_{0}^{L}dx\, e^{-iqx}\int_{0}^{\beta}d\tau\, e^{i\omega_{n}\tau}\left\langle \partial_{x}\phi\left(x,\tau\right)\partial_{x}\phi\left(0,0\right)\right\rangle _{0}\nonumber \\
 & = & \frac{K}{2\pi}\, D^{\left(0\right)}\left(q,i\omega_{n}\right),\label{eq:chifree}\end{eqnarray}
 where $D^{\left(0\right)}(q,i\omega_{n})$ is the free boson propagator
(for the $\partial_{x}\phi$ field)\begin{equation}
D^{\left(0\right)}\left(q,i\omega\right)\equiv D_{R}^{\left(0\right)}\left(q,i\omega\right)+D_{L}^{\left(0\right)}\left(q,i\omega\right)=\frac{2vq^{2}}{\left(i\omega\right)^{2}-\left(vq\right)^{2}}.\end{equation}
It follows that the retarded correlation function is\begin{equation}
\chi^{ret}\left(q,\omega\right)=\frac{Kq}{2\pi}\left[\frac{1}{\omega-vq+i\eta}-\frac{1}{\omega+vq+i\eta}\right].\label{eq:chiretarded}\end{equation}
 Finally, using (\ref{eq:spectralfunction}), the dynamical structure
factor for the free boson model is ($q>0$)\begin{equation}
S^{zz}\left(q,\omega\right)=Kq\,\delta\left(\omega-vq\right).\label{eq:Szzfreeboson}\end{equation}
 The result in (\ref{eq:Szzfreeboson}) is analogous to (\ref{eq:deltapeak}).
Since the Luttinger model exhibits Lorentz invariance, $S^{zz}\left(q,\omega\right)$
is a delta function peak at the energy carried by the single boson with momentum
$q$. This solution should be asymptotically exact in the limit $q\rightarrow0$,
which means that any corrections to it must be suppressed by higher
powers of momentum. However, the free boson result misses many of
the features that the complete solution must have. For example, the
exact solution for the XX point suggests a broadening of the delta
peak with a width $\delta\omega_{q}\sim q^{2}$. Like in that case,
it is necessary to incorporate information about band curvature at
the Fermi level by keeping the quadratic term in the fermion dispersion
in order to get a finite width for $S^{zz}\left(q,\omega\right)$.
As we shall discuss in the next section, the problem is that such a
term is mapped via bosonization onto a boson-boson interaction term. Even though the interaction term is irrelevant, finite-order perturbation
theory in these operators leads to a singular frequency dependence
close to $\omega=vq$. It turns out that broadening the
delta function peak within a field theory approach is a not an easy task. A
complete solution that recovers the scaling $\delta\omega_{q}\sim q^{2}$
requires summing an infinite series of diagrams, as we will point
out in section \ref{sec:Broadening-of-the}. Another feature expected
for $S^{zz}\left(q,\omega\right)$ when $\Delta\neq0$ is a high-frequency
tail associated with multiple particle-hole excitations. This tail
can be calculated in the region $\delta\omega_{q}\ll\omega-vq\ll J$
by lowest-order perturbation theory in the fermionic interaction ($\propto \Delta$) starting from
a model of free fermions with quadratic dispersion \cite{pustilnik1}.
In section \ref{sec:High-frequency-tail} we obtain this result by
including fermionic interactions exactly (finite $\Delta$) and doing
perturbation theory in the band curvature terms.

\subsection{Irrelevant operators\label{sub:Irrelevant-operators-due}}

In order to go beyond the Luttinger model, we need to treat the irrelevant
operators that break Lorenz invariance. There are two sources of such
terms: band curvature terms, which are quadratic in fermions but involve
higher derivatives, and irrelevant interaction terms \cite{rozhkov}.
The first type appeared in (\ref{eq:kincontinuum}) and corresponds
to the $k^{2}$ term in the expansion of the fermion dispersion\begin{eqnarray}
\delta\mathcal{H}_{bc} & = & -\frac{1}{2m}\left(:\psi_{R}^{\dagger}\partial_{x}^{2}\psi_{R}^{\phantom{\dagger}}:\,+\,:\psi_{L}^{\dagger}\partial_{x}^{2}\psi_{L}^{\phantom{\dagger}}:\right).\label{eq:deltaH}\end{eqnarray}
 We derive the bosonized version of a general band curvature term
in the following way (see \cite{haldane}). We define the operator\begin{eqnarray}
F\left(x,\epsilon\right) & = & \psi_{R}^{\dagger}\left(x+\frac{\epsilon}{2}\right)\psi_{R}^{\phantom{\dagger}}\left(x-\frac{\epsilon}{2}\right)\nonumber \\
 & = & \sum_{k=0}^{\infty}\frac{1}{k!}\left(\frac{\epsilon}{2}\right)^{k}\partial_{x}^{k}\psi_{R}^{\dagger}\sum_{l=0}^{\infty}\frac{1}{l!}\left(-\frac{\epsilon}{2}\right)^{l}\partial_{x}^{l}\psi_{R}^{\phantom{\dagger}}\nonumber \\
 & = & \sum_{n=0}^{\infty}\left(-\frac{\epsilon}{2}\right)^{n}\psi_{R}^{\dagger}\partial_{x}^{n}\psi_{R}^{\phantom{\dagger}}\sum_{k=0}^{n}\frac{1}{k!(n-k)!}+\dots,\label{eq:Fexpansion}\end{eqnarray}
 where $\dots$ is a total derivative. Organizing by powers of $\epsilon$,
we can write\begin{equation}
F\left(x,\epsilon\right)=\sum_{n=0}^{\infty}\frac{\left(-1\right)^{n}}{n!}\,\epsilon^{n}F^{\left(n\right)}\left(x\right),\label{eq:defF_n}\end{equation}
 where \begin{equation}
F^{\left(n\right)}\left(x\right)=\psi_{R}^{\dagger}\partial_{x}^{n}\psi_{R}^{\phantom{\dagger}}.\end{equation}
 According to (\ref{eq:mandelstam}), we have\begin{equation}
\psi_{R}\sim\frac{1}{\sqrt{2\pi\alpha}}\, e^{-i\sqrt{2\pi}\phi_{R}}\sim\frac{1}{\sqrt{L}}\, e^{-i\sqrt{2\pi}\phi_{R}^{+}}e^{-i\sqrt{2\pi}\phi_{R}^{-}},\end{equation}
 where $\phi_{R}^{\pm}$ are the creation and annihilation parts of $\phi_{R}\left(x\right)=\phi_{R}^{+}\left(x\right)+\phi_{R}^{-}\left(x\right)$ and we have used the identity $e^{A+B}=e^{A}e^{B}e^{-\left[A,B\right]/2}$ with
\begin{equation}
\left[\phi_{R}^{-}\left(x\right),\phi_{R}^{+}\left(y\right)\right]\approx-\frac{1}{2\pi}\,\log\left[-\frac{2\pi i}{L}\left(x-y+i\alpha\right)\right],\end{equation}
 for large $L$. Then we express $F\left(x,\epsilon\right)$ in terms
of the bosonic fields\begin{equation}
F\left(x,\epsilon\right)=\frac{1}{L}e^{i\sqrt{2\pi}\phi_{R}^{+}\left(x+\epsilon/2\right)}e^{i\sqrt{2\pi}\phi_{R}^{-}\left(x+\epsilon/2\right)}e^{-i\sqrt{2\pi}\phi_{R}^{+}\left(x-\epsilon/2\right)}e^{-i\sqrt{2\pi}\phi_{R}^{-}\left(x-\epsilon/2\right)}.\end{equation}
 After normal ordering the operators, we can do the expansion in $\epsilon$  (dropping the normal ordering sign)\begin{eqnarray}
 &  & \psi_{R}^{\dagger}\left(x+\frac{\epsilon}{2}\right)\psi_{R}^{\phantom{\dagger}}\left(x-\frac{\epsilon}{2}\right)\nonumber \\
 & = & -\frac{1}{2\pi i\epsilon}\,\exp\left\{ i\sqrt{2\pi}\left[\phi_{R}\left(x+\frac{\epsilon}{2}\right)-\phi_{R}\left(x-\frac{\epsilon}{2}\right)\right]\right\} \nonumber \\
 & = & -\sum_{\ell=0}^{\infty}\frac{\left(2\sqrt{2\pi}i\right)^{\ell}}{2\pi i\epsilon\,\ell!}\sum_{\left\{ m_{j}\right\}}\frac{\ell !}{\prod_jm_j!}\left(\frac{\epsilon}{2}\right)^{\sum_{j}^{\prime}jm_{j}}\prod_{j=1,3,\cdots}\left(\frac{\partial_{x}^{j}\phi_{R}}{j!}\right)^{m_{j}}.\label{eq:expandderiv}\end{eqnarray}
 From (\ref{eq:defF_n}) and the coefficient of the $\epsilon^{n}$
term in (\ref{eq:expandderiv}), we have \begin{equation}
F^{\left(n\right)}\left(x\right)=\frac{\left(-1\right)^{n+1}n!}{2^{n+1}2\pi i}\sum_{\left\{ m_{j}\right\} }\frac{\left(2\sqrt{2\pi}i\right)^{\sum_{j}m_{j}}}{\prod_{j}\left(m_{j}!\right)}\,\prod_{j=1,3,\cdots}\left(\frac{\partial_{x}^{j}\phi_{R}}{j!}\right)^{m_{j}},\label{eq:F_nboson}\end{equation}
where the $m_{j}$'s obey the constraint $\sum_{j}jm_{j} = n + 1$. In
particular, for $n=2$ the sum in (\ref{eq:F_nboson}) contains only
two terms (either $m_{1}=3$, $m_{3}=0$ or $m_{1}=0$, $m_{3}=1$).
We get\begin{equation}
F^{\left(2\right)}\left(x\right)=\psi_{R}^{\dagger}\partial_{x}^{2}\psi_{R}^{\phantom{\dagger}}=\frac{\sqrt{2\pi}}{3}\,\left(\partial_{x}\phi_{R}\right)^{3}-\frac{1}{12\sqrt{2\pi}}\,\partial_{x}^{3}\phi_{R}.\end{equation}
 The last term is a total derivative and can be omitted from the Hamiltonian. Similar
expressions for the left-moving field $\phi_{L}$ are obtained straightforwardly
by using the symmetry under the parity transformation $x\rightarrow-x$, $R\rightarrow L$.
The bosonized version of the band curvature terms in (\ref{eq:deltaH})
is then\begin{equation}
\delta\mathcal{H}_{bc}=\frac{\sqrt{2\pi}}{6m}\left[\left(\partial_{x}\phi_{L}\right)^{3}-\left(\partial_{x}\phi_{R}\right)^{3}\right].\label{eq:d_xphi3}\end{equation}

We now rewrite $\delta \mathcal{H}_{bc}$ in terms of the right and left components
of the rescaled field $\phi$. Using (\ref{eq:phitilde}) and (\ref{eq:thetatilde}),
\begin{eqnarray}
\delta\mathcal{H}_{bc} & = & \frac{\sqrt{2\pi}}{6m}\left[\left(\frac{\partial_{x}\tilde{\theta}+\partial_{x}\tilde{\phi}}{\sqrt{2}}\right)^{3}-\left(\frac{\partial_{x}\tilde{\theta}-\partial_{x}\tilde{\phi}}{\sqrt{2}}\right)^{3}\right]\nonumber \\
 & = & \frac{\sqrt{\pi/K}}{6m}\int_{0}^{L}dx\,\left[3\left(\partial_{x}\theta\right)^{2}\partial_{x}\phi+K^{2}\left(\partial_{x}\phi\right)^{3}\right].\label{eq:deltaHphitheta}\end{eqnarray}
 Finally, using (\ref{eq:phi_LR}), we get (in accordance with \cite{Aristov})\begin{eqnarray}
\delta\mathcal{H}_{bc} & = & \frac{\sqrt{2\pi/K}}{6}\frac{3+K^{2}}{4m}\left[\left(\partial_{x}\varphi_{L}\right)^{3}-\left(\partial_{x}\varphi_{R}\right)^{3}\right]\nonumber \\
 &  & +\frac{\sqrt{2\pi/K}}{6}\frac{3(1-K^{2})}{4m}\left[\left(\partial_{x}\varphi_{L}\right)^{2}\partial_{x}\varphi_{R}-\left(\partial_{x}\varphi_{R}\right)^{2}\partial_{x}\varphi_{L}\right].\label{eq:deltaHbandcurv}\end{eqnarray}

Besides $\delta\mathcal{H}_{bc}$, we need to include the irrelevant
operators which arise from the expansion of the fermionic interaction
in the lattice spacing, as we encountered in (\ref{eq:expandinterac}).
In terms of $\varphi_{R,L}$, this contribution reads\begin{eqnarray}
\delta\mathcal{H}_{int} & = & \frac{J\Delta K^{3/2}}{3\sqrt{2\pi}}\sin(2k_{F})\left\{ \left[\left(\partial_{x}\varphi_{L}\right)^{3}-\left(\partial_{x}\varphi_{R}\right)^{3}\right]\right.\nonumber \\
 &  & \left.-3\left[\left(\partial_{x}\varphi_{L}\right)^{2}\partial_{x}\varphi_{R}-\left(\partial_{x}\varphi_{R}\right)^{2}\partial_{x}\varphi_{L}\right]\right\} .\label{eq:deltaHinteraction}\end{eqnarray}
Combining (\ref{eq:deltaHbandcurv}) and (\ref{eq:deltaHinteraction}),
we can write the irrelevant operators in the most general form \bea
\delta H&=&\frac{\sqrt{2\pi}}{6}\int dx\,\left\{ \eta_{-}\left[\left(\partial_{x}\varphi_{L}\right)^{3}-\left(\partial_{x}\varphi_{R}\right)^{3}\right]\right. \nonumber\\ & &\left.+\eta_{+}\left[\left(\partial_{x}\varphi_{L}\right)^{2}\partial_{x}\varphi_{R}-\left(\partial_{x}\varphi_{R}\right)^{2}\partial_{x}\varphi_{L}\right]\right\} .\label{eq:deltaHzetas}\eea
 To first order in $\Delta$, the coupling constants $\eta_{\pm}$
are given by\begin{eqnarray}
\eta_{-} & \approx & \frac{1}{m}\left(1+\frac{2\Delta}{\pi}\sin k_{F}\right),\label{eq:zetaminus}\\
\eta_{+} & \approx & -\frac{3\Delta}{\pi m}\sin k_{F}.\label{eq:zetaplus}\end{eqnarray}

\psfrag{eta+}{$\eta_+$}

\psfrag{eta-}{$\eta_-$}

\psfrag{Right}{$R$}

\psfrag{Left}{$L$}

\begin{figure}
\begin{center} \includegraphics[%
scale=0.5]{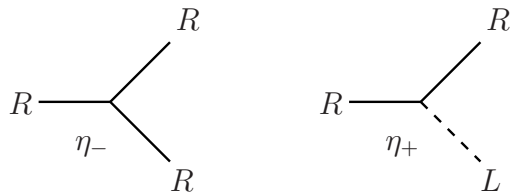}\end{center}

\caption{Interaction vertices in the low energy effective Hamiltonian. The
solid (dashed) lines represent propagators for right- (left-) moving
bosons $D_{R}^{\left(0\right)}$($D_{L}^{\left(0\right)}$).\label{cap:Interaction-vertices-in}}
\end{figure}

The perturbation $\delta H$ in (\ref{eq:deltaHzetas}) might as well
have been introduced phenomenologically in the effective Hamiltonian.
In fact, the dimension-three operators $(\partial_{x}\varphi_{R,L})^{3}$
are the leading irrelevant operators that are allowed by symmetry. They obey the parity symmetry $\varphi_L\rightarrow \varphi_R$, $x\rightarrow -x$, but not spin reversal (or particle-hole) $\varphi_{R,L}\rightarrow-\varphi_{R,L}$, which is absent for $h\neq0$. Such terms give rise to three-legged interaction
vertices which scale with powers of the momenta of the scattered bosons
(figure \ref{cap:Interaction-vertices-in}). They are responsible, for example, for corrections to the long distance
asymptotics of the correlation functions \cite{haldane}. Note that
as $\Delta\rightarrow0$ ($K\rightarrow1$), $\eta_{-}\rightarrow1/m$
while $\eta_{+}$ vanishes because there is no mixing between right
and left movers at the free fermion point. Moreover, the weak coupling expressions predict that both $\eta_{-}$
and $\eta_{+}$ vanish in the limit $h\rightarrow0$ ($m\rightarrow\infty$),
in which particle-hole symmetry is recovered. (See, however, figure \ref{etas} below.) For $h=0$ the leading
irrelevant operators are the dimension-four operators $\left(\partial_{x}\varphi_{R,L}\right)^{4},\left(\partial_{x}\varphi_{R}\right)^{2}\left(\partial_{x}\varphi_{L}\right)^{2}$
and the umklapp interaction $\cos(4\sqrt{\pi K}\phi$), which becomes
nonoscillating \cite{pereira}. 

The condition that a general model of the form $H_{LL}+\delta H$
be unitarily equivalent to free fermions up to dimension-four operators \cite{rozhkov} amounts to
imposing that the Bogoliubov transformation that diagonalizes $H_{LL}$
in the $R/L$ basis also diagonalizes the cubic operators in $\delta H$.
In our notation, this condition is expressed as $\eta_{+}=0$. That condition is not satisfied by the XXZ model except for the trivial case $\Delta=0$. However, the contributions from this extra ({\it i.e.}, not present for free fermions) dimension-three operator to $S^{zz}(q,\omega)$ are of $O(\eta_+^2)$, as we will discuss in section \ref{sec:High-frequency-tail}.

Similarly to what happens for $v$ and $K$, (\ref{eq:zetaminus})
and (\ref{eq:zetaplus}) should be regarded as weak-coupling expressions.
Again we can use the fact that the XXZ model is integrable and obtain
the exact (renormalized) values of $\eta_{\pm}$ by comparison with
Bethe Ansatz. In section \ref{sec:Determination-of-the} we will discuss
how to fix these coupling constants in order to obtain a parameter-free
theory.

\subsection{Determination of the renormalized coupling constants\label{sec:Determination-of-the}}

As mentioned in section \ref{sub:Irrelevant-operators-due}, the renormalized
parameters $\eta_{\pm}$ can be determined by comparison with exact
Bethe Ansatz results for infinite length. We will proceed by analogy with the calculation
for the zero-field case in \cite{lukyanov}. One difficulty is that
there are no analytical solutions of the Bethe Ansatz equations for
finite fields, so we must be satisfied with a numerical evaluation
of the parameters. In the following, we will relate $\eta_{\pm}$
to the coefficients of the expansion of $v$ and $K$ as functions
of the magnetic field, by comparing the corrections to the free boson
result for the free energy calculated in two different ways.

Let us consider the response to a small variation in the magnetic
field around a finite value $h_{0}$. In the limit $\delta h=h-h_{0}\ll1$,
such response is well described by the Luttinger model\begin{equation}
H=\int dx\,\left\{ \frac{v}{2}\left[\left(\partial_{x}\theta\right)^{2}+\left(\partial_{x}\phi\right)^{2}\right]-J\delta h\sqrt{\frac{K}{\pi}}\partial_{x}\phi\right\} ,\label{eq:1stapproach}\end{equation}
 where $v\left(h\right)$ and $K\left(h\right)$ are known exactly
from the Bethe Ansatz equations. For $h_{0}=0$, the cutoff-independent
terms of the free energy density according to field theory read\begin{equation}
f\left(h_{0}=0\right)\sim-\frac{\pi T^{2}}{6v}-\frac{K}{2\pi v}\left(J\delta h\right)^{2},\label{eq:freeenergy}\end{equation}
 where $v$ and $K$ are given by (\ref{eq:vBAzerofield}) and (\ref{eq:KBAzerofield}),
respectively. The magnetic susceptibility at zero temperature is $\chi=-J^{-2}\left.\left(\partial^{2}f/\partial h^{2}\right)\right|_{T=0}=K/\pi v$,
which is the familiar free boson result. For finite field $h_0\neq0$,
the free energy assumes some general form\begin{equation}
f\left(h_0\neq0\right)\sim-\frac{\pi T^{2}}{6v(h)}-C\left(h\right),\end{equation}
 and the $T=0$ susceptibility is obtained by\begin{equation}
\chi=-\frac{1}{J^{2}}\left.\left(\frac{\partial^{2}f}{\partial h^{2}}\right)\right|_{h,T=0}=-\frac{1}{J^{2}}\left.\left(\frac{\partial^{2}C}{\partial\left(\delta h\right)^{2}}\right)\right|_{h,T=0}=\frac{K(h)}{\pi v(h)},\end{equation}
 where the last identity holds for any Luttinger liquid. 

We would like to calculate the corrections to $f$ and $\chi$ that
involve higher powers of the perturbation $\delta h$. Our first approach
is to assume that the field dependence is already completely contained
in the definitions of $v(h)$ and $K(h)$, so that we can employ the
expansion\begin{eqnarray}
v(h) & = & v(h_{0})\left[1+a\,\delta h+O\left(\delta h^{2}\right)\right],\\
K(h) & = & K(h_{0})\left[1+b\,\delta h+O\left(\delta h^{2}\right)\right],\end{eqnarray}
 where the coefficients $a$ and $b$ can be extracted from the exact
$v$ and $K$ by linearizing the field dependence around $h=h_{0}$.
Consequently, the lowest-order correction to the free boson susceptibility
around $h=h_{0}$ is \begin{equation}
\chi=\frac{K(h_{0})}{\pi v(h_{0})}\left[1-\left(a-b\right)\delta h+O\left(\delta h^{2}\right)\right].\label{eq:susceptibility1}\end{equation}
 Likewise, the free energy at finite temperature must contain a term
of the form\begin{equation}
\delta f\sim a\frac{\pi\,\delta h\, T^{2}}{6v(h_{0})},\label{eq:deltaf1st}\end{equation}
 due to the field dependence of the velocity. Both $a$ and $b$ depend
on $h_{0}$ and the anisotropy $\Delta$. As an example, at the XX
point, $K=1$ for any value of the field, therefore $b(\Delta=0,h_{0})=0$.
From (\ref{eq:kF}), we have\begin{equation}
v_{F}=J\sin k_{F}=J\sqrt{1-h^{2}}\approx v_{F}(h_{0})-\frac{J^{2}h_{0}}{v_{F}(h_{0})}\,\delta h+O\left(\delta h^{2}\right),\label{eq:expand_vF(h)}\end{equation}
 so that we get\begin{equation}
a\left(\Delta=0,h_{0}\right)=-\frac{J^{2}h_{0}}{v_{F}^{2}(h_{0})}=\frac{\cos k_{F}}{\sin^{2}k_{F}}.\end{equation}

In our second approach, we take $v=v(h_{0})$ and $K=K(h_{0})$ to
be fixed and assume that the corrections to the free boson result
are generated by the irrelevant operators. We consider the effective
Hamiltonian $H=H_{LL}+\delta H$, with $\delta H$ defined in (\ref{eq:deltaHzetas}).
An equivalent Lagrangian formulation in imaginary time is \begin{equation}
\mathcal{L}=\mathcal{L}_{0}+\delta\mathcal{L},\end{equation}
\begin{eqnarray}
\mathcal{L}_{0} & = & \frac{\left(\partial_{\tau}\phi\right)^{2}}{2v}+\frac{v}{2}\left(\partial_{x}\phi\right)^{2}-J\delta h\sqrt{\frac{K}{\pi}}\partial_{x}\phi,\\
\delta\mathcal{L} & = & -\frac{A\sqrt{\pi}}{6v^{2}}\left(\partial_{\tau}\phi\right)^{2}\partial_{x}\phi+\frac{B\sqrt{\pi}}{6}\left(\partial_{x}\phi\right)^{3}+O\left(\eta_{\pm}^{2}\right),\end{eqnarray}
 where $A=3\eta_{-}+\eta_{+}$ and $B=\eta_{-}-\eta_{+}$. We shift
the field by $\phi\rightarrow\phi+\frac{J\delta h}{v}\sqrt{\frac{K}{\pi}}x$
to absorb the term linear in $\partial_{x}\phi$ and get\begin{eqnarray}
\mathcal{L}_{0} & = & \frac{\left(\partial_{\tau}\phi\right)^{2}}{2v}+\frac{v}{2}\left(\partial_{x}\phi\right)^{2}+\frac{K\left(J\delta h\right)^{2}}{2\pi v},\\
\delta\mathcal{L} & = & -\frac{A\sqrt{K}J\delta h}{6v^{3}}\left(\partial_{\tau}\phi\right)^{2}+\frac{B\sqrt{K}J\delta h}{2v}\left(\partial_{x}\phi\right)^{2}\nonumber \\
 &  & +\frac{BK^{3/2}\left(J\delta h\right)^{3}}{6\pi v^{3}}+\,\textrm{odd powers of }\phi.\end{eqnarray}
 We then calculate the free energy density from the partition function\begin{equation}
Z=\int\mathcal{D}\phi\,\exp\left\{ -\int_{0}^{\beta}d\tau\int_{0}^{L}dx\,\left(\mathcal{L}_{0}+\delta\mathcal{L}\right)\right\} ,\end{equation}
\begin{equation}
f=-\frac{T}{L}\,\ln Z\approx f_{0}+\frac{T}{L}\int_{0}^{\beta}d\tau\int_{0}^{L}dx\left\langle \delta\mathcal{L}\right\rangle ,\end{equation}
 where $f_{0}$ reproduces the free boson result \begin{equation}
f_{0}\sim-\frac{\pi T^{2}}{6v}-\frac{K}{2\pi v}\left(J\delta h\right)^{2},\end{equation}
 and $\left\langle \delta\mathcal{L}\right\rangle $ is the expectation
value of $\delta\mathcal{L}$ calculated with the unperturbed Hamiltonian.
In order to compute $\left\langle \delta\mathcal{L}\right\rangle $,
we need the finite temperature propagators \bea
\left\langle \partial_{x}\phi\left(x+\epsilon\right)\partial_{x}\phi\left(x\right)\right\rangle &=&-\frac{1}{v^{2}}\left\langle \partial_{\tau}\phi\left(x+\epsilon\right)\partial_{\tau}\phi\left(x\right)\right\rangle \nonumber\\&=&-\frac{1}{2\pi}\frac{\left(\pi T/v\right)^{2}}{\sinh^{2}\left(\pi T\epsilon/v\right)}.\eea
 Now we use the expansion $\sinh^{-2}\left(\pi T\epsilon/v\right)\approx\left(v/\pi T\epsilon\right)^{2}-1/3$
for $\epsilon\rightarrow0$ and drop the cutoff-dependent terms in
$\delta f$. The reason is that the latter simply renormalize the
corresponding terms in $f_{0}$ and have already been accounted for
in the renormalization of $v$ and $K$. The correction to the free
energy to first order in $A$ and $B$ becomes\bea
\delta f&=&\frac{T}{L}\int_{0}^{\beta}d\tau\int_{0}^{L}dx\left\langle \delta\mathcal{L}\right\rangle\nonumber\\ &\sim&\left(A+3B\right)\frac{\pi\,\sqrt{K}J\delta h\, T^{2}}{36v}+B\,\frac{K^{3/2}\left(J\delta h\right)^{3}}{6\pi v^{3}}.\label{eq:deltaf2nd}\eea
 The susceptibility obtained from $f_{0}+\delta f$ is \begin{equation}
\chi=-\frac{1}{J^{2}}\left.\frac{\partial^{2}\left(f_{0}+\delta f\right)}{\partial\left(\delta h\right)^{2}}\right|_{T=0}=\frac{K}{\pi v}-B\,\frac{K^{3/2}J\delta h}{\pi v^{3}}+O\left(\delta h^{2}\right).\end{equation}
 Comparing with the expression (\ref{eq:susceptibility1}), we can
identify\begin{equation}
a-b=\frac{\sqrt{K}J}{v^{2}}\, B=\frac{\sqrt{K}J}{v^{2}}\left(\eta_{-}-\eta_{+}\right).\label{eq:a-b}\end{equation}
 Besides, from the $\delta h\, T^{2}$ term in (\ref{eq:deltaf1st})
and (\ref{eq:deltaf2nd}), we have\begin{equation}
a=\frac{\sqrt{K}J}{6v^{2}}\left(A+3B\right)=\frac{\sqrt{K}J}{3v^{2}}\left(3\eta_{-}-\eta_{+}\right).\label{just_a}\end{equation}
 Finally, combining (\ref{eq:a-b}) and (\ref{just_a}) and writing
$a=v^{-1}\partial v/\partial h$ and $b=K^{-1}\partial K/\partial h$,
we find the formulas first presented in \cite{pereira}\begin{eqnarray}
J\eta_{-} & = & \frac{v}{K^{1/2}}\frac{\partial v}{\partial h}+\frac{v^{2}}{2K^{3/2}}\frac{\partial K}{\partial h},\label{eq:identity1}\\
J\eta_{+} & = & \frac{3v^{2}}{2K^{3/2}}\frac{\partial K}{\partial h}.\label{eq:identity2}\end{eqnarray}
 The above relations allow us to calculate the renormalized values
of $\eta_{\pm}$ once we have the field dependence of $v$ and $K$.
Notice that $\eta_{+}\propto\partial K/\partial h$ and as expected
vanishes at the XX point. On the other hand, $\eta_{-}$ remains finite
at $\Delta=0$ because $\partial v/\partial h\neq0$ and we recover
$\eta_{-}=(v_{F}/J)\partial v_{F}/\partial h=J\cos k_{F}=m^{-1}$.
It is also possible to check the validity of (\ref{eq:identity1})
and (\ref{eq:identity2}) explicitly in the weak coupling limit, using
the expressions for $v\left(\Delta\ll1,h\right)$ and $K\left(\Delta\ll1,h\right)$
in (\ref{eq:velocity}) and (\ref{eq:luttingerK}) as well as the
weak coupling expressions for $\eta_{\pm}$ in (\ref{eq:zetaminus})
and (\ref{eq:zetaplus}).

\section{Bethe Ansatz solution}\label{sec:BAsolution}

Although the Bethe Ansatz is first and foremost a method for
calculating the energy levels of an exactly solvable model
(readers who are unfamiliar with the subject 
are invited to consult standard textbooks, for example \cite{GaudinBOOK,KorepinBOOK,TakahashiBOOK}),
recent progress stemming from the Algebraic Bethe Ansatz means that 
we can now use it to make many nontrivial statements about dynamical quantities.  
Assuming that certain specific families of excited states carry the dominant part 
of the structure factor, we can delimit the
energy and momentum continua where we expect most of the correlation weight to be found,
and provide the specific lineshape of the structure factor both within this interval,
and further up within the higher-energy tail.  We start here by introducing the important aspects
of the Bethe Ansatz which we will make use of later on when studying the correspondence with
field theory results.

\subsection{Bethe Ansatz setup and fundamental equations}
As is well-known, an eigenbasis for the $XXZ$ chain (\ref{eq:XXZ}) on $N$ sites is obtained from the Bethe
Ansatz \cite{BetheZP71,OrbachPR112},
\begin{equation}
\Psi_M (j_1, ..., j_M) = \sum_{P} (-1)^{[P]} e^{i \sum_{a=1}^M k_{P_a} j_a - \frac{i}{2} \sum_{1 \leq a < b \leq M}
\phi(k_{P_a}, k_{P_b})}.
\label{BA_XXZ}
\end{equation}
Here, $M \leq N/2$ represents the number of overturned spins, starting from the reference state $|0\rangle = 
\otimes_{i=1}^N | \uparrow \rangle_i$ ({\it i.e.} the state with all spins pointing upwards in the $\hat{z}$
direction).  The total magnetization of the system along the $\hat{z}$ axis, $S^z_{tot} = N \sigma = \frac{N}{2} - M$
is conserved by the Hamiltonian.  $P$ represents a permutation of the integers $\{ 1, ..., M \}$ 
and $j_i$ are the lattice coordinates.  The quasi-momenta $k$ are parametrized in terms of
rapidities $\lambda$, 
\begin{equation}
e^{ik} = \frac{\sinh (\lambda + i\zeta/2)}{\sinh (\lambda - i\zeta/2)}, \hspace{1cm}
\Delta = \cos \zeta,
\end{equation}
such that the two-particle scattering phase shift becomes a function of the rapidity 
difference only, $\phi (k_a, k_b) = \phi_1 (\lambda_a - \lambda_b)$ with $\phi_1$ defined below.
An individual eigenstate is thus fully characterized by a set of
rapidities $\{ \lambda \}$, satisfying the quantization conditions (Bethe equations) 
obtained by requiring periodicity of the Bethe wavefunction (\ref{BA_XXZ}):
\begin{equation}
\phi_1 (\lambda_j) - \frac{1}{N} \sum_{k=1}^M \phi_2 (\lambda_j - \lambda_k) = 2\pi \frac{I_j}{N},
\hspace{1cm} j = 1, ..., M,
\label{BE_XXZ}
\end{equation}
in which $I_j$ are half-odd integers for $N-M$ even and integers for $N-M$ odd, and 
where we have defined the functions
\begin{equation}
\phi_n (\lambda) = 2 \arctan \left( \frac{\tanh(\lambda)}{\tan(n \zeta/2)} \right).
\end{equation}
The energy and momentum of an eigenstate are simple functions of its rapidities, 
\begin{eqnarray}
E = -\pi J \sin\zeta \sum_{j} a_1 (\lambda_j) - h S^z_{tot}, 
\nonumber \\
P = \pi M - \sum_j \phi_1 (\lambda_j) = \pi M - \frac{2\pi}{N} \sum_j I_j\label{energyBA}
\end{eqnarray}
in which
\begin{equation}
a_n (\lambda) = \frac{1}{2\pi} \frac{d}{d\lambda} \phi_n (\lambda) = \frac{1}{\pi} \frac{\sin (n \zeta)}{\cosh(2\lambda) - \cos(n\zeta)}.
\end{equation}

Each solution of the set of coupled nonlinear equations 
(\ref{BE_XXZ}) for sets of non-coincident rapidities represents an eigenstate
(if two rapidities coincide, the Bethe wavefunction (\ref{BA_XXZ}) formally vanishes).
The space of solutions is not restricted to real rapidities:  it has been known since Bethe's
original paper that there exist solutions having complex rapidities ('string' states), representing bound states
of magnons.  In fact, obtaining {\it all} wavefunctions from solutions to the Bethe equations 
(or degenerations thereof) remains to this day an open problem in the theory of integrable models.  
It is however possible to construct the vast majority of eigenstates using this procedure, allowing to
obtain reliable results for thermodynamic quantities and correlation functions.  In all our 
considerations in the present paper, we can and will restrict ourselves to real solutions to the Bethe equations.

\subsection{Ground state and excitations}\label{GSBA}
The simplest state to construct is the ground state, which 
is obtained by setting the quantum numbers $I_j$ to (we consider $N$ even from now on 
for simplicity)
\begin{equation}
I_j^{GS} = -\frac{M+1}{2} + j, \hspace{1cm} j = 1,...,M.
\end{equation}

\begin{figure}\begin{center}
\includegraphics[width=10cm]{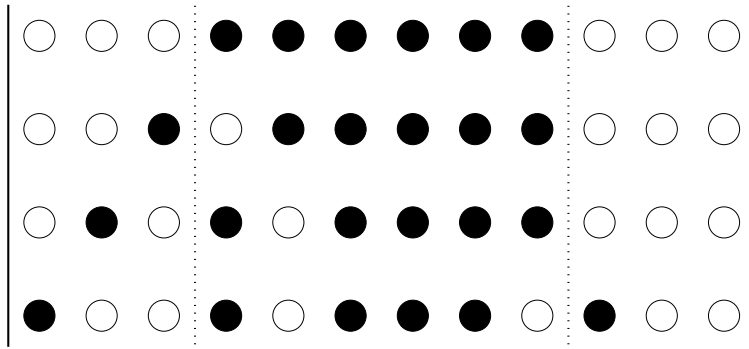}\end{center}
\caption{Representation of various quantum number configurations:  a black (empty) circle
represents an occupied (unoccupied) allowable quantum number (which here are taken to be half-odd integers;
the figure is centered on zero).  The top set represents the ground state configuration,
whereas the second and third from top represent two-particle excitations at different momenta, 
obtained by creating a particle-hole pair on the ground-state configuration.  The bottom
set is for a four-particle state obtained from two particle-hole pairs.  The dotted line
delimits the ground state interval, whereas the solid lines delimit the quantum numbers
for which real solutions to the Bethe equations can be obtained in this illustrative case (see main text).}
\label{Quantum_nrs}
\end{figure}

The simplest excited states which can be constructed at finite magnetic field are
obtained by introducing particle-hole excitations on the ground-state quantum number
distributions, see figure (\ref{Quantum_nrs}).  Since we limit ourselves to real solutions
to the Bethe equations, we require $|\lambda_j| < \infty$ and thus $|I| < I_{\infty}$,
where, from (\ref{BE_XXZ}),
\begin{equation}
I_{\infty} = \frac{N - M}{2} - (\frac{N}{2} - M) \frac{\zeta}{\pi}.
\end{equation}

The momentum of an excited state is simply
given by the left-displacement of the quantum numbers with respect to those in the ground state, 
$q = \frac{2\pi}{N} \delta_l I$, where $\delta_l I = \sum_j (I^{GS}_j - I_j)$.  
At a given fixed (small) momentum, we can thus construct $q N/2\pi$ two-particle states
by shifting the particle and hole quantum numbers, leaving their difference fixed.  Since the energy of these two-particle
states at fixed momentum are non-degenerate, this defines a two-particle continuum whose
characteristics will be studied later. Higher-particle states can
be similarly constructed and counted.  

The restriction to real rapidities and a single particle-hole pair therefore means that our
subsequent arguments will apply only to the region
$q < \mbox{Min} (2k_F, k_{\infty})$ where $k_F = \pi \frac{M}{N}$ and $k_{\infty}$
is given by the maximal displacement of the outermost quantum number, 
$k_{\infty} = 2\pi \frac{I_{\infty} - M/2}{N}$.
We can thus write our restriction as
\begin{equation}
q < \mbox{Min} \{\pi (1 - 2\sigma), 2\sigma (\pi - \zeta) \}\label{restrictionBA}
\end{equation}
in terms of the magnetization, noting in particular that the window of validity of
our arguments vanishes in the case of zero magnetic field.

For a finite chain with $N$ sites and $M$ overturned spins, the Hilbert space is
finite, and therefore so is the sum over intermediate states in the Lehmann 
representation for the structure factor (\ref{eq:lehmann}).  Each intermediate
state is obtained by solving the Bethe equations, the space of states being
reconstructed by spanning through the sets of allowable quantum numbers.  
The form factor of a local spin operator between the ground state and a
particular excited state is obtained from the Algebraic Bethe Ansatz as a
determinant of a matrix depending only on the rapidities of the eigenstates
involved \cite{KitanineNPB554,KitanineNPB567} even in the case of string states with
complex rapidities \cite{CauxJSTATP09003}.  This enables
to obtain extremely accurate results on the full dynamical spin-spin correlation
functions in integrable Heisenberg chains \cite{CauxPRL95,CauxJSTATP09003}.
We will make use of this method in what follows to compare results from
the Bethe Ansatz to field theory predictions for the structure factor at small momentum.

\section{Width of the on-shell peak\label{sec:Broadening-of-the}}

Linearizing the dispersion around the Fermi points is a key step for the bosonization
technique. By doing so all the particle-hole excitations with same momentum
$q\ll k_{F}$ become exactly degenerate and one can associate a particular
linear combination with a single-boson state \cite{haldane}. In this
approximation, the single boson state $\left|b\right\rangle \equiv a_{q}^{R\dagger}\left|0\right\rangle $
is the only state that couples to the ground state via $S_{q}^{z}$. The associated weight in $S^{zz}(q,\omega)$ is given by\begin{equation}
\left|\left\langle 0\left|S_{q>0}^{z}\right|b\right\rangle \right|^{2}=\frac{KqN}{2\pi}\left|\left\langle 0\left|a_{q}^{R}\right|b\right\rangle \right|^{2}=\frac{KqN}{2\pi}.\end{equation}
However, as we will see in section \ref{compareff}, the exact eigenstates in the Bethe Ansatz solution, whose energies are given by (\ref{energyBA}), are nondegenerate. In fact, most of the above
spectral weight is shared by $qN/2\pi$ two-particle states whose energies are
spread around $\omega=vq$. This is reminiscent of the exact solution
for the free fermion point in section \ref{sub:Exact-solution-for}. In the bosonic picture, on the other hand, the broadening $\delta\omega_{q}$
is related to a finite lifetime for the bosons of the Luttinger model.
Once band curvature is introduced via the irrelevant operators in
(\ref{eq:deltaHzetas}), the single boson is allowed to decay and
the coupling to the multiboson states lifts the previous degeneracy.
The fact that the irrelevant operators have the same scaling dimension
as in the noninteracting case suggests that for $\Delta\neq0$ the
width should also vanish as $q^{2}$ in the limit $q\rightarrow0$.
In this section we argue in favor of a $q^{2}$ scaling for $\delta\omega_{q}$
for all values of $\Delta$ in the gapless regime, \emph{as long as} $\eta_-\neq 0$, based on two different
approaches. First, we explain how the expansion of the bosonic
diagrams in the interaction vertex $\eta_{-}$, neglecting $\eta_{+}$,
coincides with the expansion of the free fermion result (\ref{eq:fermionbubble})
in powers of $1/m$. $\eta_{-}$ is then interpreted as a renormalized
inverse mass, in the sense that the width of the peak for $\Delta\neq0$
is given by $\delta\omega_{q}=|\eta_{-}|q^{2}$. Second, we derive from
the Bethe Ansatz equations an analytical expression for the width of
the two-particle continuum at finite fields and show that it coincides
with the field theory prediction for the width of $S^{zz}\left(q,\omega\right)$.
Finally, we confirm these results directly by analyzing the numerical
form factors calculated for finite chains of lengths up to 7000 sites.

\subsection{Width from field theory}\label{secwidthFT}

We saw that the width $\delta\omega_{q}$ is well defined for the
free fermion point, in which case $S^{zz}\left(q,\omega\right)$ has
sharp lower and upper thresholds $\omega_{L,U}\left(q\right)$. For
the interacting case, $S^{zz}\left(q,\omega\right)$ still vanishes below some finite lower threshold $\omega_{L}(q)$ at zero temperature due to simple kinematic constraints. However, the on-shell peak has to match a high-frequency tail somewhere around $\omega_{U}\left(q\right)$, hence the meaning of
an upper threshold is no longer clear. 

In their solution for weakly
interacting spinless fermions, Pustilnik \emph{et al}. \cite{pustilnik2}
found that $\omega_{U}$ has to be
interpreted as the energy at which the peak joins the high frequency
tail by approaching a finite value with an infinite slope. Although
it is actually possible that the singularity at $\omega_{U}\left(q\right)$
get smoothed out if one treats the decay of the {}``high-energy electron''
for a general model \cite{khodas}, the singularity may be protected
in integrable models such as the XXZ model.

Of course the situation is a lot simpler for models with no high-frequency
tail, where the dynamical structure factor is finite only within the
interval $\omega_{L}\left(q\right)<\omega<\omega_{U}\left(q\right)$.
Such is the case for the Calogero-Sutherland model \cite{SutherlandBOOK}. The absence of
a tail for $S\left(q,\omega\right)$ in the Calogero-Sutherland model
can be attributed to the remarkable property that the quasiparticles
are all right movers \cite{pustilnik3}. As we will discuss in section
\ref{sec:High-frequency-tail}, the $\eta_{+}$ term that mixes $R$
and $L$ in our low energy effective Hamiltonian (figure \ref{cap:Interaction-vertices-in}) is responsible for the high-frequency tail for $h\neq0$ because it allows for intermediate
states with two bosons moving in opposite directions, thus carrying
small momentum and high energy $\omega\gg vq$.

In contrast, the $\eta_{-}$ interaction  has matrix elements between multiboson states which contain only right movers. All these states have $\omega\approx vq$. Therefore $\eta_-$ must be related to the broadening of the on-shell peak. It has already been pointed out in \cite{rozhkov}
that the model with $\eta_{+}=0$ is equivalent to free fermions up
to irrelevant operators with dimension four and higher. For this case
one can write down an approximate expression for the dynamical structure
factor which misses more subtle features in the lineshape ({\it e.g.}, the
power law singularities at the thresholds) but accounts for the renormalization
of the width due to interactions. Even for models with nonzero $\eta_{+}$,
such as the XXZ model in the entire gapless regime, it is reasonable
to expect that $\delta\omega_{q}$, if well defined, will depend primarily
on the interaction between excitations created around the same Fermi
point. For that reason, we will neglect the $\eta_{+}$ interaction
in an attempt to derive an expression for the width of $S^{zz}\left(q,\omega\right)$
from the bosonic Hamiltonian. In the following we apply perturbation
theory in $\eta_{-}$ up to \emph{fourth order} and show that it recovers
the expansion of the logarithm for the density-density correlation
function. This fact has already been noticed in \cite{teber,Aristov} up to $O(\eta_{-}^{2})$. However, irrelevant interaction terms such as (\ref{eq:deltaHinteraction}) were
neglected in \cite{teber,Aristov}. Such terms are crucial to obtain the correct effective inverse mass, since the correction of first order in the fermionic interaction $\Delta$ stems from (\ref{eq:deltaHinteraction}).

For $\eta_{+}=0$, the Hamiltoninan $H_{LL}+\delta H$ decouples into
right and left movers. For excitations with $q>0$, we can consider
only right movers and work with\begin{equation}
\mathcal{H}_{R}=\frac{v}{2}\left(\partial_{x}\varphi_{R}\right)^{2}-\frac{\sqrt{2\pi}}{6}\eta_{-}\left(\partial_{x}\varphi_{R}\right)^{3}.\label{eq:Hbosonright}\end{equation}
The first attempt to broaden the delta function peak in $S^{zz}\left(q,\omega\right)$
would be to calculate the corrections to the propagator \begin{equation}
\chi\left(q,i\omega\right)=-\frac{K}{2\pi}\int_{-\infty}^{+\infty}dx\, e^{-iqx}\int_{0}^{\beta}d\tau\, e^{i\omega\tau}\left\langle T_{\tau}\partial_{x}\varphi_{R}\left(x,\tau\right)\partial_{x}\varphi_{R}\left(0,0\right)\right\rangle ,\end{equation}
by using perturbation theory in the cubic term. Unfortunately, any
finite order perturbation theory in $\eta_{-}$ breaks down near $\omega\approx vq$.
Even using the Born approximation, which sums an infinite series but
not all the diagrams, one finds that the self-energy to $O(\eta_{-}^{2})$
is divergent: $\textrm{Im }\Sigma\left(q,\omega\right)\sim\delta\left(\omega-vq\right)$
\cite{samokhin}. This is actually not surprising if we look at the exact
solution for the free fermion point. Expanding the positive-frequency
part of (\ref{eq:fermionbubble}) in powers of $1/m$, we get\begin{equation}
\chi\left(q,i\omega\right)=\frac{q}{2\pi w}\left[1+\frac{1}{3}\left(\frac{q^{2}/m}{2w}\right)^{2}+\frac{1}{5}\left(\frac{q^{2}/m}{2w}\right)^{4}+\dots\right],\label{eq:expandlog}\end{equation}
where $w\equiv i\omega-v_{F}q$. Stricly speaking, such expansion
is valid only for $\omega-v_{F}q\gg q^{2}/2m$. For $\omega\approx v_{F}q$,
the expansion in band curvature produces increasingly singular terms
that need to be summed up to produce the finite result in (\ref{eq:fermionbubble}).

\begin{figure}
\begin{center}\includegraphics[%
  scale=0.7]{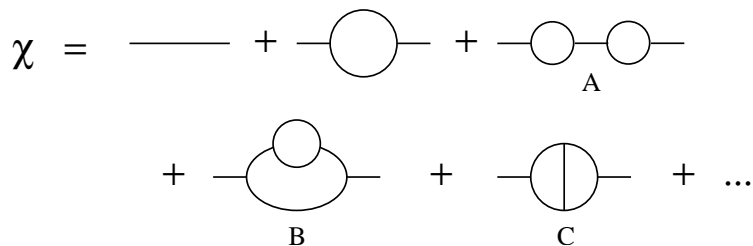}\end{center}

\caption{Perturbative diagrams up to fourth order in $\eta_{-}$.\label{cap:Perturbative-diagrams-up}}
\end{figure}

\begin{figure}
\begin{center}\includegraphics[%
  scale=0.65]{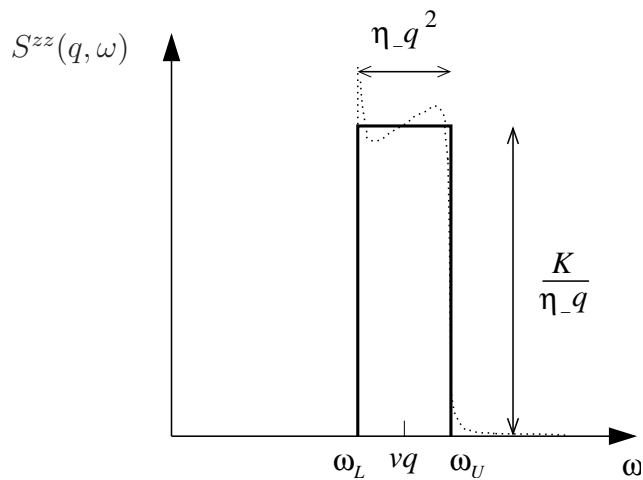}\end{center}

\caption{Lineshape in the approximation with the $\eta_{-}$ interaction only (solid line). The dotted line illustrates the expected true lineshape for small $\Delta$ (see section \ref{compareff}
). \label{cap:Lineshape-in-the}}
\end{figure}

In any case, it is legitimate to examine the expansion of bosonic
diagrams and ask whether it can at least reproduce the free fermion
result. We use the bare propagator $D_{R}^{\left(0\right)}\left(x,\tau\right)=\left\langle T_{\tau}\partial_{x}\varphi_{R}\left(x,\tau\right)\partial_{x}\varphi_{R}\left(0,0\right)\right\rangle _{0}$ in (\ref{propag_xt}),
with Fourier transform\begin{equation}
D_{R}^{\left(0\right)}\left(q,i\omega\right)=\frac{q}{w},\end{equation}
to calculate the expansion of $\chi_{R}\left(q,i\omega\right)$ up
to $O\left(\eta_{-}^{4}\right)$, as represented in figure \ref{cap:Perturbative-diagrams-up}.
The zeroth-order result is simply the same as in (\ref{eq:chiretarded})\begin{equation}
\chi^{\left(0\right)}\left(q,\omega\right)=\frac{Kq}{2\pi w}.\end{equation}
The $O(\eta_{-}^{2})$ correction is \begin{eqnarray}
\chi^{\left(2\right)}\left(q,i\omega\right) & = & -\frac{K}{2\pi}\int d^{2}x\, e^{-iqx+i\omega\tau}\int d^{2}x_{1}\int d^{2}x_{2}\frac{1}{2}\left(\frac{\sqrt{2\pi}}{6}\eta_{-}\right)^{2}\times\nonumber \\
 &  & \times\left\langle T_{\tau}\partial_{x}\varphi_{R}\left(x\right)\left[\partial_{x}\varphi_{R}\left(1\right)\right]^{3}\left[\partial_{x}\varphi_{R}\left(2\right)\right]^{3}\partial_{x}\varphi_{R}\left(0\right)\right\rangle \nonumber \\
 & = & \frac{K}{2\pi}\left[D_{R}^{\left(0\right)}\left(q,i\omega\right)\right]^{2}\Pi_{RR}\left(q,i\omega\right),\label{eq:2ndorderchi}\end{eqnarray}
where $\Pi_{RR}\left(q,i\omega\right)$ is the bubble with two right-moving
bosons\begin{eqnarray}
\Pi_{RR}\left(q,i\omega\right) & \equiv & -\pi\eta_{-}^{2}\int_{0}^{q}\frac{dk}{2\pi}\int_{-\infty}^{+\infty}\frac{d\nu}{2\pi}\, D_{R}^{\left(0\right)}\left(k,i\nu\right)D_{R}^{\left(0\right)}\left(q-k,i\omega-i\nu\right)\nonumber \\
 & = & \frac{\eta_{-}^{2}q^{3}}{12w}.\label{eq:PibubbleRR}\end{eqnarray}
Note that $\Pi_{RR}$ is singular at $\omega=vq$, which prevents
us from treating it as a self-energy. The origin of the singularity
is that the two right-moving bosons in the intermediate state always
carry energy $\omega=vq$, no matter how the momentum is distributed
between the pair. Substituting (\ref{eq:PibubbleRR}) back into (\ref{eq:2ndorderchi}),
we get\begin{equation}
\chi^{(2)}\left(q,i\omega\right)=\frac{Kq}{2\pi w}\,\frac{1}{12}\left(\frac{\eta_{-}q^{2}}{w}\right)^{2}.\label{chi2eta-}\end{equation}
To $O\left(\eta_{-}^{4}\right)$, there are three topologically distinct
diagrams (figure \ref{cap:Perturbative-diagrams-up}), which give
the following contributions\begin{eqnarray}
\chi_{A}^{\left(4\right)}\left(q,i\omega\right) & = & \frac{Kq}{2\pi w}\,\frac{1}{144}\left(\frac{\eta_{-}q^{2}}{w}\right)^{4},\nonumber \\
\chi_{B}^{\left(4\right)}\left(q,i\omega\right) & = & \frac{Kq}{2\pi w}\,\frac{1}{504}\left(\frac{\eta_{-}q^{2}}{w}\right)^{4},\label{eq:diagrams}\\
\chi_{C}^{\left(4\right)}\left(q,i\omega\right) & = & \frac{Kq}{2\pi w}\,\frac{1}{280}\left(\frac{\eta_{-}q^{2}}{w}\right)^{4}.\nonumber \end{eqnarray}
The coefficients for each diagram are nontrivial and result from both
combinatorial factors and integration over internal momenta (recall that the interaction vertex is momentum-dependent because of the derivatives in (\ref{eq:deltaHzetas})). Remarkably, all
the fourth-order diagrams have the same $q$ and $\omega$ dependence
with comparable amplitudes. We are not allowed to drop any of them
and there is no justification for the use of a self-consistent Born
approximation, for example \cite{samokhin}. Putting all the terms
together, we end up with the expansion %
\begin{equation}
\chi\left(q,i\omega\right)=\frac{Kq}{2\pi w}\left[1+\frac{1}{3}\left(\frac{\eta_{-}q^{2}}{2w}\right)^{2}+\frac{1}{5}\left(\frac{\eta_{-}q^{2}}{2w}\right)^{4}+\dots\right],\label{eq:serieseta-}\end{equation}
which is analogous to (\ref{eq:expandlog}) with the replacements
$1/m\rightarrow\eta_{-}$, $v_{F}\rightarrow v$ and an extra factor
of $K$. This proves that the expansion of bosonic diagrams reproduces
the expansion of the free fermion result up to fourth order in $1/m$.
Since there is no simple way to predict the prefactors of each diagram,
all we can do is to check this correspondence order by order in perturbation
theory. However, \emph{if we believe that the bosonic theory reproduces
the free fermion result to all orders in} $\eta_{-}$, we must conclude
that in the interacting case the series in (\ref{eq:serieseta-})
sums up to give the result\begin{equation}
\chi\left(q,i\omega\right)=\frac{K}{2\pi\eta_{-}q}\,\log\left[\frac{i\omega-vq+\eta_{-}q^{2}/2}{i\omega-vq-\eta_{-}q^{2}/2}\right],\label{dressedprop}\end{equation}
from which we obtain\begin{equation}
S^{zz}\left(q,\omega\right)=\frac{K}{|\eta_{-}|q}\theta\left(\frac{|\eta_{-}|q^{2}}{2}-\left|\omega-vq\right|\right).\end{equation}
 This result predicts that $S^{zz}\left(q,\omega\right)$ is finite
and flat within an interval of width \begin{equation}
\delta\omega_{q}=|\eta_{-}|q^{2}.\label{widtheta_-}\end{equation}
This lineshape (illustrated in figure \ref{cap:Lineshape-in-the})
is the exact one for the case of free fermions with quadratic dispersion.
The reason is simple: because the bosonization of the operator $\sim k^{2}c_{k}^{\dagger}c_{k}$
only generates the $\eta_{-}$ term, one could invert the problem
and refermionize the Hamiltonian (\ref{eq:Hbosonright}) to an effective
free fermion model with inverse mass $\eta_{-}$. In a more general
model, more irrelevant operator have to be added to the effective
Hamiltonian to reproduce details of the lineshape that are higher
order in $q$. For example, we expect the power-law singularities
present at $\omega_{L,U}$ for $\Delta\neq0$ \cite{pustilnik2} to
be associated with dimension-four operators such as $\left(\partial_{x}^{2}\varphi_{R}\right)^{2}$
and $\left(\partial_{x}\varphi_{R}\right)^{4}$ (with corrections
of $O(\eta_{+}^{2})$, see section \ref{sec:zerofield}). This means that if we write \begin{equation}S^{zz}\left(q,\omega\right)\equiv \frac{q}{\delta\omega_q}f\left(q,\frac{\omega-vq}{\delta \omega_q}\right),\end{equation}
the \emph{rescaled} function $f(q,x)$ approaches the flat distribution of figure \ref{cap:Lineshape-in-the}
in the limit $q\rightarrow0$. Finally, we note that this approximate
solution yields the same sum rules as the free boson result\begin{eqnarray}
\int_{0}^{\infty}d\omega\, S^{zz}\left(q,\omega\right) & = & Kq,\label{sumweight}\\
\int_{0}^{\infty}d\omega\,\omega S^{zz}\left(q,\omega\right) & = & vKq^{2},\end{eqnarray}
and also the magnetic susceptibility\begin{equation}
\chi=\chi\left(q=0\right)=\lim_{q\rightarrow0}\frac{1}{\pi}\int_{0}^{\infty}\frac{d\omega}{\omega}\, S^{zz}\left(q,\omega\right)=\frac{K}{\pi v},\end{equation}
independent of the value of $\eta_{-}$.%

\psfrag{Sqw}{$S^{zz}(q,\omega)$}
\psfrag{Kovereta}{$\frac{K}{\eta_-q}$}

\subsection{Width from Bethe Ansatz\label{WidthBA}}

The purpose of this section is to provide an analytical derivation of the
quadratic width formula (\ref{widtheta_-}), making use of standard methods associated
to the thermodynamic Bethe Ansatz, and assuming that single particle-hole type
excitations in the Bethe eigenstates basis carry the most important part of the
structure factor.  We first set our notations and underline certain
characteristics of the ground state of the infinite chain in a field which will prove to be useful
for our purposes.  We then
discuss particle-hole excitations in the thermodynamic limit, and obtain a relationship giving the width 
in terms of solutions of integral equations, simplifying to the conjectured field theory result
in the small momentum limit.

Let us begin by taking the thermodynamic limit $N \rightarrow \infty$ of the equations of
Section 4.  To do this, we first define
particle and hole densities as functions of the continuous variable $x = \frac{I}{N}$, 
\begin{equation}
\rho (x) = \frac{1}{N} \sum_{l \in \{ I \} } \delta (x - \frac{l}{N}), \hspace{1cm}
\rho^h (x) = \frac{1}{N} \sum_{m \notin \{ I \} } \delta (x - \frac{m}{N})
\end{equation}
in such a way that $\rho(x) + \rho^h(x) \rightarrow 1$ as $N \rightarrow \infty$.  We can also write
these in rapidity space by using the transformation rule for $\delta$ functions, so that the Bethe
equations become
\begin{equation}
\phi_1 (\lambda) - \int_{-\infty}^{\infty} d\lambda' \phi_2 (\lambda - \lambda') \rho(\lambda') = 2\pi x(\lambda)
\label{BE_cont1}
\end{equation}
where we view $x$ as an implicit function of $\lambda$.  Taking the derivative of this
with respect to $\lambda$ and using $\frac{dx(\lambda)}{d\lambda} = \rho(\lambda) + \rho^h(\lambda)$ yields
\begin{equation}
a_1 (\lambda) - \int_{-\infty}^{\infty} d\lambda' a_2 (\lambda - \lambda') \rho(\lambda') = \rho(\lambda) + \rho^h (\lambda),
\hspace{0.3cm} \lambda \in \mathbf{R}.
\label{deriv_BE}
\end{equation}
For the particular case of the ground state, the occupation density 
$\rho_{GS}(\lambda)$ is non-vanishing in a symmetric interval
$[-B, B]$, with $\rho^h_{GS}(\lambda)$ vanishing.  Outside of this interval, $\rho_{GS}$ vanishes but not $\rho^h_{GS}$.
$\lambda = \pm B$ therefore represent the two Fermi points in the rapidity distribution of the ground state,
which is obtained by solving 
\begin{equation}
\rho_{GS} (\lambda) + \int_{-B}^{B} d\lambda' a_2(\lambda - \lambda') \rho_{GS} (\lambda') = a_1 (\lambda), 
\hspace{0.3cm} \lambda \in [-B, B].
\label{rho_GS_eqn}
\end{equation}
The magnetic field dependence is encoded in the constraint
\begin{equation}
\int_{-B}^B d\lambda \rho_{GS} (\lambda) = \frac{M}{N} = \frac{1}{2} - \sigma
\label{constraint_GS}
\end{equation}
where $\sigma$ is the field-dependent average magnetization per site along the $z$ axis.
These two equations determine $B$ and $\rho_{GS}$, and therefore also $\rho^h_{GS}$.
We can write a formal solution as follows.
Let us define the inverse operator $L(\lambda, \lambda')$, $\lambda, \lambda' \in [-B, B]$, 
inverse of the kernel in (\ref{rho_GS_eqn}) in the sense that
\begin{eqnarray}
\int_{-B}^B d\lambda' [\delta(\lambda - \lambda') + L(\lambda, \lambda')] [\delta(\lambda' - \bar{\lambda}) + a_2 (\lambda' - \bar{\lambda})]
= \delta (\lambda - \bar{\lambda}).
\end{eqnarray}
This operator is symmetric, $L(\lambda, \lambda') = L(\lambda', \lambda)$, unique and analytic 
in its domain of definition \cite{YangPR150}.  In particular, the definition implies the identity
\begin{eqnarray}
\fl
a_2 (\lambda -\bar{\lambda}) + L(\lambda, \bar{\lambda}) + \int_{-B}^B d\lambda' L(\lambda, \lambda') a_2 (\lambda' - \bar{\lambda}) = 0,
\hspace{0.3cm} \lambda, \bar{\lambda} \in [-B, B].
\end{eqnarray}
In terms of this operator, we have the explicit solution of equation (\ref{rho_GS_eqn}) for the ground state distribution,
\begin{eqnarray}
\rho_{GS} (\lambda) = \left\{ \begin{array}{cc}
\int_{-B}^B d\lambda' [\delta(\lambda - \lambda') + L(\lambda, \lambda')] a_1 (\lambda') & |\lambda| \leq B, \\
0 & |\lambda| > B.
\end{array} \right.
\label{rho_GS_soln}
\end{eqnarray}
Knowing $\rho_{GS}$ then yields $\rho_{GS}^h$ from (\ref{deriv_BE}), namely
\begin{eqnarray}
\rho_{GS}^h (\lambda) = \left\{ \begin{array}{cc}
0 & |\lambda| \leq B, \\
a_1(\lambda) - \int_{-B}^B d\lambda' a_2 (\lambda - \lambda') \rho_{GS} (\lambda') & |\lambda| > B.
\end{array} \right.
\end{eqnarray}
The ground state can also be obtained from the thermodynamic Bethe Ansatz formalism \cite{YangPR150}
in the following way.  Given distributions $\rho(\lambda)$ and $\rho^h(\lambda)$, the free energy 
$f = (E - T S)/N$ is written to leading order in $N$ as
\begin{eqnarray}
\fl
f = -\frac{h}{2} + \int_{-\infty}^{\infty} d\lambda \left[ \varepsilon_0 \rho 
- T (\rho + \rho^h) \ln (\rho + \rho^h) + T \rho \ln \rho + T \rho^h \ln \rho^h \right]
\end{eqnarray}
in which we have suppressed the $\lambda$ functional arguments and defined the bare energy
\begin{equation}
\varepsilon_0 (\lambda) = h - \pi J \sin \zeta a_1 (\lambda).
\label{bare_energy}
\end{equation}
Introducing the quasi-energy $\varepsilon (\lambda) = T \ln \frac{\rho^h(\lambda)}{\rho(\lambda)}$, 
the condition of thermodynamic equilibrium $\delta F = 0$ under the constraint of the Bethe equations 
(\ref{deriv_BE}) then gives 
after standard manipulations \cite{YangPR150} (taking the limit $T \rightarrow 0$, so here and
in what follows, $\varepsilon (\lambda)$ is for the ground state configuration)
\begin{eqnarray}
\varepsilon(\lambda) + \int_{-B}^{B} d\lambda' a_2 (\lambda - \lambda') \varepsilon(\lambda')
= \varepsilon_0 (\lambda), \hspace{0.5cm} \lambda \in ]-\infty, \infty[.
\label{quasienergy}
\end{eqnarray} 
In particular, we have that
\begin{equation}
\varepsilon (\pm B) = 0, \hspace{0.5cm} \varepsilon (\lambda) \leq 0 (> 0) \mbox{for} \lambda \in (\notin) [-B, B].
\end{equation}
Similarly to (\ref{rho_GS_soln}), 
we can also solve for $\varepsilon (\lambda) = \varepsilon^{-} (\lambda) + \varepsilon^{+} (\lambda)$
with $\varepsilon^{\pm} (\lambda) \geq (\leq) 0$ using the inverse integral kernel:
\begin{eqnarray}
\varepsilon^{-} (\lambda) = \left\{ \begin{array}{cc}
\int_{-B}^B d\lambda' [\delta(\lambda - \lambda') + L(\lambda, \lambda')] \varepsilon_0 (\lambda') & |\lambda| \leq B, \\
0 & |\lambda| > B,
\end{array} \right.  \\
\varepsilon^{+} (\lambda) = \left\{ \begin{array}{cc}
0 & |\lambda| \leq B, \\
\varepsilon_0 (\lambda) - \int_{-B}^B d\lambda' a_2 (\lambda - \lambda') \varepsilon^{-} (\lambda') & |\lambda| > B.
\end{array} \right.
\end{eqnarray}
The free energy simplifies to
\begin{equation}
f = -\frac{h}{2} + \int_{-B}^B d\lambda a_1 (\lambda) \varepsilon (\lambda).
\end{equation}
The magnetic equilibrium condition $\frac{\partial F}{\partial h} = 0$ then is
\begin{eqnarray}
\int_{-B}^{B} d\lambda a_1(\lambda) \frac{\partial \varepsilon (\lambda)}{\partial h} = \frac{1}{2}.
\end{eqnarray}
By defining the dressed charge $Z(\lambda)$ as solution to
\begin{equation}
Z(\lambda) + \int_{-B}^{B} d\lambda' a_2 (\lambda - \lambda') Z(\lambda') = 1,
\label{dressedcharge}
\end{equation}
which we can solve as
\begin{equation}
Z (\lambda) = 1 + \int_{-B}^B d\lambda' L(\lambda, \lambda'),
\end{equation}
we have the identity
$Z(\lambda) = \frac{\partial \varepsilon (\lambda)}{\partial h}$ by making use of
(\ref{bare_energy}) and (\ref{quasienergy}).  The Luttinger parameter $K$ is given
by the square of the dressed charge at the Fermi boundary (see {\it e.g.} \cite{KorepinBOOK}), 
\begin{equation}
K = Z^2(-B).
\end{equation}

The magnetic field dependence of the Fermi boundary $B$ can be obtained by taking the $h$ derivative of
(\ref{quasienergy}):
\begin{equation}
\int_{-B}^B d\lambda' [\delta(\lambda -\lambda') + a_2 (\lambda - \lambda')] \frac{\partial \varepsilon(\lambda')}{\partial B} = \frac{\partial h}{\partial B}.
\end{equation}
Since $\varepsilon(-B) = 0$, we have 
\begin{equation}
\frac{\partial \varepsilon(\lambda)}{\partial \lambda} |_{\lambda = -B} = 
\frac{\partial \varepsilon(\lambda)}{\partial B} |_{\lambda = -B}
\end{equation}
and therefore
\begin{equation}
\frac{\partial h}{\partial B} = \frac{\varepsilon' (-B)}{Z(-B)}.
\label{dh_dB}
\end{equation}
The magnetization is
\begin{eqnarray}
\sigma = -\frac{\partial f}{\partial h} 
= \frac{1}{2} - \int_{-B}^B d\lambda a_1 (\lambda) \frac{\partial \varepsilon(\lambda)}{\partial h}
= \frac{1}{2} - \int_{-B}^B d\lambda a_1 (\lambda) Z (\lambda).
\end{eqnarray}
To get the susceptibility, we start from
\begin{equation}
\frac{\partial \sigma}{\partial B} = -\int_{-B}^B d\lambda a_1(\lambda) \frac{\partial Z(\lambda)}{\partial B} - a_1 (B)Z(B) - a_1 (-B) Z(-B).
\end{equation}
The integral equation for the dressed charge (\ref{dressedcharge}) gives
\bea
\frac{\partial Z(\lambda)}{\partial B} &=& -\int_{-B}^B d\lambda' [\delta (\lambda - \lambda') + L(\lambda, \lambda')]\nonumber \\& &\times
[Z(B) a_2 (\lambda' - B) + Z(-B) a_2 (\lambda' + B)]
\eea
which yields after simple manipulations and use of symmetry
\begin{equation}
\frac{\partial \sigma}{\partial B} = -2 \rho_{GS}(-B) Z(-B).
\end{equation}
The susceptibility is therefore given by
\begin{equation}
\chi = \frac{\partial \sigma}{\partial h} = \frac{\partial B}{\partial h} \frac{\partial \sigma}{\partial B} 
= -2 \frac{\rho_{GS} (-B) Z^2 (-B)}{\varepsilon'(-B)}.
\label{chi_1}
\end{equation}
This expression will be related to the Fermi velocity after discussing elementary excitations 
(see equation (\ref{Zsq_pivfchi})).

Finally, we will need the slope of the ground state rapidity distribution at the Fermi boundary,
$\frac{\partial \rho_{GS} (\lambda)}{\partial B}|_{-B}$.  From the integral equation for $\rho_{GS}$, we can write
\begin{equation}
\frac{\partial \rho_{GS} (\lambda)}{\partial B} 
= \rho_{GS} (-B) [L(\lambda, B) + L(\lambda, -B)].
\end{equation}
This can be related to the derivative of the dressed charge by using the representation
\begin{equation}
\frac{\partial Z(\lambda)}{\partial B} = L(\lambda, B) + L(\lambda, -B) + \int_{-B}^B d\lambda' \frac{\partial L(\lambda, \lambda')}{\partial B}.
\end{equation}
From the definition of $L(\lambda, \lambda')$, we can show that
\begin{equation}
\frac{\partial L(\lambda, \lambda')}{\partial B} = L(\lambda, B) L(\lambda', B) + L(\lambda, -B) L(\lambda', -B)
\end{equation}
and therefore
\begin{equation}
\frac{\partial Z(\lambda)}{\partial B} = [L(\lambda, B) + L(\lambda, -B)] Z(-B),
\end{equation}
finally yielding
\begin{equation}
\frac{\partial \rho_{GS} (\lambda)}{\partial B}|_{-B} = \frac{\rho_{GS} (-B)}{Z(-B)} \frac{\partial Z(\lambda)}{\partial B}|_{-B}.
\label{drhodB}
\end{equation}
We will make use of these identities later, while relating the width of the
two-particle continuum to field-dependent physical quantities.

Let's now construct an excited state over the finite-field ground state 
by generating a single particle-hole pair.  That is, we select
a quantum number $I_p \notin \{I^{GS} \}$ associated to a particle and $I_h \in \{ I^{GS} \}$
associated to a hole, and write the excited state densities in $x$ space as
\begin{eqnarray}
\rho (x) = \rho_{GS} (x) + \frac{1}{N} \delta (x - \frac{I_p}{N}) - \frac{1}{N} \delta (x - \frac{I_h}{N}), \nonumber \\
\rho^h (x) = \rho^h_{GS} (x) - \frac{1}{N} \delta (x - \frac{I_p}{N}) + \frac{1}{N} \delta (x - \frac{I_h}{N}),
\end{eqnarray}
with once again $\rho(x) + \rho^h(x) \rightarrow 1$ as $N \rightarrow \infty$.
We can again map to rapidity space, with $\lambda_p \leq -B$ and $|\lambda_h| \leq B$. 
%
Upon creating such a particle-hole pair, the induced distribution $\rho(\lambda)$ will be only very
slightly shifted (order $1/N$) as compared to the ground state one (for $\lambda \neq \lambda_p, \lambda_h$).  
We therefore define a backflow function
$K (\lambda; \lambda_p, \lambda_h) \sim \mbox{O}(N^0)$ as
\begin{equation}
\rho(\lambda) = \rho_{GS} (\lambda) + \frac{1}{N} \left[ K(\lambda; \lambda_p, \lambda_h)
+ \delta (\lambda - \lambda_p) - \delta(\lambda - \lambda_h) \right].
\end{equation} 
By subtracting the equations for the ground state from those of the excited state, 
the backflow function is shown to obey the constraint
\begin{equation}
K(\lambda; \lambda_p, \lambda_h) + \int_{-B}^B d\lambda' a_2 (\lambda - \lambda')
K(\lambda'; \lambda_p, \lambda_h) = -a_2 (\lambda - \lambda_p) + a_2 (\lambda - \lambda_h)
\end{equation}
for $\lambda \in [-B, B]$, with $K = 0$ outside of this domain.  
We can again formally solve for $K$ by applying the inverse integral operator $1 + L$, 
\begin{eqnarray}
\fl
K (\lambda; \lambda_p; \lambda_h) = -a_2 (\lambda - \lambda_p) - \int_{-B}^B d\lambda' L(\lambda, \lambda') a_2 (\lambda' - \lambda_p)
- L(\lambda, \lambda_h).
\label{K_as_L}
\end{eqnarray}
In terms of this kernel, 
the energy of the excited state is
\begin{eqnarray}
\fl
E \!-\! E_{GS} \!=\! N \int_{-\infty}^{\infty} d\lambda \varepsilon_0 \left[ \rho - \rho_{GS} \right]
= \varepsilon_0 (\lambda_p) - \varepsilon_0 (\lambda_h) + \int_{-B}^B d\lambda \varepsilon_0 (\lambda) K (\lambda; \lambda_p, \lambda_h), 
\end{eqnarray}
which can be rewritten after basic manipulations as ($|\lambda_p| > B$ and $|\lambda_h| < B$)
\begin{equation}
E - E_{GS} = \varepsilon (\lambda_p) - \varepsilon (\lambda_h).
\end{equation}
Similarly, the momentum of the excited state is
\begin{eqnarray}
P - P_{GS} &=& -\phi_1 (\lambda_p) + \phi_1 (\lambda_h) - \int_{-B}^B d\lambda \phi_1 (\lambda)
K (\lambda; \lambda_p, \lambda_h).
\end{eqnarray}

\begin{figure}
\begin{center}\includegraphics[width=9cm]{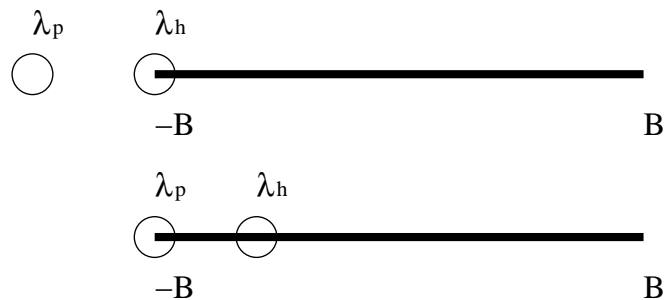}\end{center}
\caption{Highest and lowest energy two-particle excited states at fixed momentum.  The straight line represents 
the interval $\lambda \in [-B, B]$ within which the ground-state rapidity $\rho_{GS}(\lambda)$ is nonvanishing.
$\lambda_p$ and $\lambda_h$ respectively represent the positions of the particle and hole rapidities for 
the highest (top) and lowest (bottom) two-particle excited states at a fixed value of momentum.}
\label{excitations_fig}
\end{figure}

Single particle-hole pairs as described above constitute a set of two-particle 
excitations labeled by the particle and hole rapidities $\lambda_p$ and $\lambda_h$.
This continuum is well-defined and spanned by the intervals $\lambda_p \in ~]-\infty, -B]$,
$\lambda_h \in [-B, B]$.  Assuming that the mapping
from $(\lambda_p, \lambda_h)$ to $(\omega, q)$ is
one-to-one and onto and that the particle dispersion
curvature is greater than the hole one (this monotonicity
assumption will be discussed further in section \ref{compareff}), the highest energy state at a given
fixed momentum $q$ will be given by the choice $\lambda_p = \lambda_p(q)$,
$\lambda_h = -B$, where $\lambda_p(q)$ is solution to
\begin{equation}
q = -\phi_1 (\lambda_p(q)) + \phi_1 (-B) + \int_{-B}^B d\lambda \phi_1 (\lambda) 
K(\lambda; \lambda_p(q); -B).
\label{lambdap_k}
\end{equation}
Similarly, the lowest energy state will correspond to the choice $\lambda_p = -B$,
$\lambda_h = \lambda_h(q)$, where $\lambda_h(q)$ is solution to
\begin{equation}
q = \phi_1 (\lambda_h(q)) - \phi_1 (-B) - \int_{-B}^B d\lambda \phi_1 (\lambda) 
K(\lambda; -B; \lambda_h(q)).
\label{lambdah_k}
\end{equation}
As discussed in Section 4, 
this continuum is well-defined ({\it i.e.} finite real solutions to both (\ref{lambdap_k}) and
(\ref{lambdah_k}) can be found) as long as $q \leq \mbox{Min}(2k_F, k_{\infty})$,
with $2k_F = \pi (1 - 2\sigma)$ and $k_{\infty} = 2\sigma (\pi - \zeta)$.
This is illustrated in Figure (\ref{excitations_fig}).
The width of the two-particle continuum defined by these excitations
will thus be given by the energy difference between these two limiting configurations, namely 
\begin{eqnarray}
W(q) = \varepsilon (\lambda_p(q)) + \varepsilon(\lambda_h(q)) - 2\varepsilon (-B)
= \varepsilon (\lambda_p(q)) + \varepsilon(\lambda_h(q))
\end{eqnarray}
where we have used $\varepsilon(\pm B) = 0$.
These functions are exact in the thermodynamic limit, in the sense that they
allow at least in principle to obtain the exact function $W(q)$ for the momentum
region where these excitations are defined.  These coupled equations 
unfortunately cannot be solved explicitly at nonzero magnetic field (where $B$ is finite).
We can however obtain analytical results in the small momentum limit, where these
excitations always exist in a finite region at finite field.

At small momentum, we can expand the width at fixed magnetic field as
\begin{equation}
W = q W^{(1)} + q^2 W^{(2)} + O(q^3)
\label{width_small_k}
\end{equation}
with coefficients given explicitly by
\bea
W^{(1)} &=& 
\frac{\partial}{\partial q} (\varepsilon (\lambda_p(q)) + \varepsilon(\lambda_h(q))) |_{q = 0},\\
W^{(2)} &=& 
\frac{1}{2} \frac{\partial^2}{\partial q^2} 
(\varepsilon (\lambda_p(q)) + \varepsilon(\lambda_h(q))) |_{q = 0}.
\eea
Let us treat the linear term first.  Considering that (\ref{lambdah_k}) also defines a
function $q(\lambda_h)$, we can rewrite the hole contribution to the coefficient as 
\begin{equation}
\frac{\partial}{\partial q} \varepsilon (\lambda_h(q)) |_{q = 0} = \frac{\frac{\partial \varepsilon (\lambda_h)}{\partial \lambda_h}|_{\lambda_h = -B}}
{\frac{\partial q(\lambda_h)}{\partial \lambda_h}|_{\lambda_h = -B}}.
\end{equation}
The denominator is obtained from (\ref{lambdah_k}) as
\begin{eqnarray}
\frac{\partial q}{\partial \lambda_h} = 2\pi a_1 (\lambda_h) - \int_{-B}^B d\lambda \phi_1 (\lambda) \frac{\partial K(\lambda; -B; \lambda_h)}{\partial \lambda_h}
= 2\pi \rho_{GS} (\lambda_h)
\end{eqnarray}
where we have used (\ref{K_as_L}), the symmetry of $L$ and partial integration.
This contribution is by definition related to the field-dependent Fermi velocity, namely
\begin{equation}
\frac{\partial}{\partial q} \varepsilon (\lambda_h) |_{q = 0} = \frac{1}{2\pi} \lim_{\lambda \rightarrow -B^+} 
\frac{{\varepsilon^-}' (\lambda)}{\rho_{GS} (\lambda)} \equiv -v.
\label{Fermi_velocity}
\end{equation}
In particular, this allows us to relate the susceptibility to the Fermi velocity and the dressed charge
using relation (\ref{chi_1}),
\begin{equation}
Z^2 (-B) = \pi v \chi.
\label{Zsq_pivfchi}
\end{equation}
For the particle contribution to the linear term, 
we find similarly that $\frac{\partial q}{\partial \lambda_p} = -2\pi \rho_{GS}^h(\lambda_p)$.
Since $\lim_{\lambda \rightarrow -B^-} \rho_{GS}^h (\lambda) = \lim_{\lambda \rightarrow -B^+} \rho_{GS} (\lambda)$, 
we also have 
$\frac{\partial}{\partial q} \varepsilon (\lambda_p) |_{q = 0} = \frac{-1}{2\pi} 
\lim_{\lambda \rightarrow -B^-} \frac{{\varepsilon^+}' (\lambda)}{\rho_{GS}^h (\lambda)} = v$
since $\varepsilon$ is smooth around this point.
Therefore, in the momentum expansion (\ref{width_small_k}) for the width, the linear term vanishes:
\begin{equation}
W^{(1)} = \frac{\partial}{\partial q} (\varepsilon(\lambda_p(q)) + \varepsilon(\lambda_h(q))) |_{q = 0} = 0.
\end{equation}

The width therefore depends at least quadratically on momentum.  To compute the coefficient of the quadratic term, 
we first note that given a function $\lambda(q)$ and its inverse $q(\lambda)$, the chain rule allows us to write
\begin{equation}
\left.\frac{\partial^2}{\partial q^2} \varepsilon (\lambda(q))\right|_{q = 0} = \left[ \left.\frac{\partial q}{\partial \lambda}\right|_{-B} \right]^{-2}
 \left( \left.\frac{\partial^2 \varepsilon (\lambda)}{\partial \lambda^2} \right|_{-B} - \left.\frac{\partial^2 q}{\partial \lambda^2}\right|_{-B} 
\left.\frac{\partial \varepsilon (\lambda)}{\partial \lambda}\right|_{-B} \right).
\label{d2e_dk2}
\end{equation}
From (\ref{lambdap_k}) and (\ref{lambdah_k}), we have that the particle and hole parts are related through
\bea
\left.\frac{\partial q (\lambda_p)}{\partial \lambda_p} \right|_{\lambda_p = -B} &=& -\left. \frac{\partial q (\lambda_h)}{\partial \lambda_h} \right|_{\lambda_h = -B},\\
\left.\frac{\partial^2 q (\lambda_p)}{\partial \lambda_p^2} \right|_{\lambda_p = -B} &=& - \left.\frac{\partial^2 q (\lambda_h)}{\partial \lambda_h^2} \right|_{\lambda_h = -B},
\eea
so using (\ref{d2e_dk2}) for $\lambda_p(q)$ and $\lambda_h(q)$, we obtain that the quadratic coefficient of the width 
can be simplified to
\begin{eqnarray}
\fl
W^{(2)} = \left.\frac{1}{2}\frac{\partial^2}{\partial q^2} (\varepsilon (\lambda_p(q)) + \varepsilon(\lambda_h(q)))\right|_{q = 0} 
= \left[ \left.\frac{\partial q}{\partial \lambda} \right|_{-B} \right]^{-2}
\left.\frac{\partial^2 \varepsilon (\lambda)}{\partial \lambda^2} \right|_{-B}.
\label{W2_coeff_1}
\end{eqnarray}
While this expression for the width is an end in itself, 
it is much more enlightening to relate it to more physical quantities
by making use of the identities derived earlier.  Starting from
$\frac{\partial^2 \varepsilon (\lambda)}{\partial \lambda^2} |_{-B} = \frac{\partial^2 \varepsilon (\lambda)}{\partial B^2} |_{-B}$
and using (\ref{Fermi_velocity}) together with (\ref{drhodB}) and (\ref{Zsq_pivfchi}), we get
\begin{equation}
\left.\frac{\partial^2 \varepsilon (\lambda)}{\partial \lambda^2} \right|_{-B}
= -2\pi \rho_{GS} (-B) \left( \frac{3}{2} \frac{\partial v}{\partial B} + \frac{1}{2} \frac{v}{\chi}\frac{\partial \chi}{\partial B}\right).
\end{equation}
Putting this in (\ref{W2_coeff_1}) and making use of (\ref{dh_dB}), (\ref{Fermi_velocity}) and (\ref{Zsq_pivfchi}) again, 
this finally gives
\begin{equation}
W^{(2)} = \sqrt{\frac{v}{\pi \chi}} \left[ \frac{3}{2} \frac{\partial v}{\partial h} + \frac{1}{2} \frac{v}{\chi} \frac{\partial \chi}{\partial h}\right].
\end{equation}
Since we have the identity $K = Z^2(-B) = \pi v \chi$, this coincides with (\ref{eq:identity1}). It also reduces to the formula derived in \cite{pereira} for $\Delta\ll 1$ by linearizing the Bethe Ansatz equations. While our derivation was done for the anisotropic chain in the gapless regime, 
the same calculation can be performed for the isotropic antiferromagnet by simply
using the appropriate scattering kernels in the Bethe equations.  This result is
however limited to chains with finite magnetization, in view of the fact that the region of validity 
of the excitations we have used to compute the width collapses to zero when the
field vanishes.

\subsection{Comparison with numerical form factors} \label{compareff}

In order to compare the field theory results with the dynamical structure factor for finite chains, we first fix the parameters of the bosonic model introduced in section \ref{sub:The-free-boson}. We do that by calculating $v(\Delta,h)$ and $K(\Delta,h)=\pi v(\Delta,h)\chi(\Delta,h)$ numerically using the Bethe Ansatz integral equations in the thermodynamic limit. $\eta_-$ and $\eta_+$ are obtained by linearizing the field dependence of $v$ and $K$ around some fixed $h_0$ and using (\ref{eq:identity1}) and (\ref{eq:identity2}). As examples, we consider three values of the anisotropy, $\Delta=0.25$, $\Delta=0.75$ and the Heisenberg point $\Delta=1$, at a fixed magnetization per site $\sigma=-0.1$.
Table \ref{cap:Parameters-for-the} lists the values of the important parameters
(we set $J=1$). Note that $b$ is negative for $\sigma<0$
($m>0$) because $K$ decreases as we approach half-filling \cite{giamarchi}. 
\begin{table}

\caption{Parameters for the low-energy effective model for $\Delta=0.25$, $\Delta=0.75$ and $\Delta=1$  and finite magnetic field $h_{0}$ (in all cases
the magnetization per site is $\sigma=-0.1$). \label{cap:Parameters-for-the}}

\begin{indented}
\item[]\begin{tabular}{@{}cccccccc}
\br 
$\Delta$&
 $h_{0}$&
 $v$&
 $K$&
 $a$&
 $b$&
 $\eta_{-}$&
 $\eta_{+}$ \\
\mr
0.25&
 -0.414&
 1.087&
 0.871&
 0.306&
-0.050&
 0.356&
-0.095\\
0.75&
 -0.652&
 1.313&
 0.699&
 0.271&
-0.145&
 0.409&
-0.449 \\
1&
-0.791&
1.399&
0.639&
0.256&
-0.188&
0.397&
-0.690\\
\br
\end{tabular}\end{indented}

\end{table}

\begin{figure}
\begin{center}\includegraphics[%
  scale=0.5]{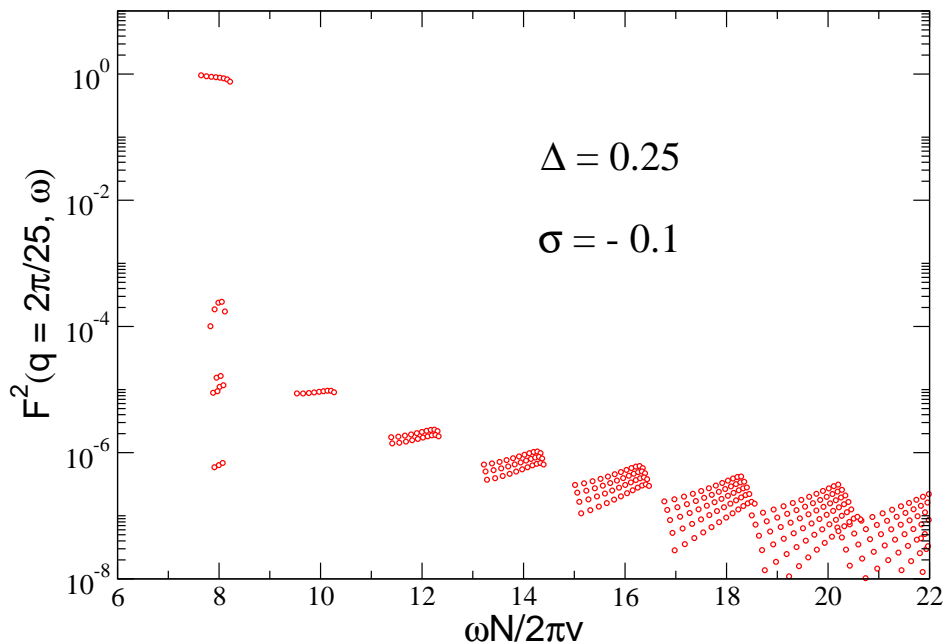}\end{center}

\caption{Numerical form factors squared (transition probabilities) for states with momentum $q=2\pi/25$,
for a chain with $N=200$ sites, anisotropy $\Delta=0.25$, magnetization
per site $\sigma=-0.1$. The energies of the eigenstates are rescaled
by the level spacing of the bosonic states predicted by field theory.
The on-shell states are the ones at $\omega N/2\pi v=qN/2\pi=8$.
\label{cap:Numerical-form-factors}}
\end{figure}
As mentioned in section \ref{GSBA}, we are able to calculate the exact transition probabilities $F^2(q,\omega)\equiv \left|\langle0\left|S_q^z\right|\alpha\rangle\right|^2$ for finite chains by means of the Algebraic Bethe Ansatz \cite{KitanineNPB554, KitanineNPB567,CauxJSTATP09003}. Figure \ref{cap:Numerical-form-factors} illustrates a typical result obtained for finite anisotropy and finite magnetic field. In contrast with the free fermion case, we observe two main differences when we turn on the fermion interaction $\Delta$: First, the form factors for the two-particle (on-shell) states become $\omega$-dependent; second, the form factors for multiparticle states are now finite and account for a finite spectral weight extending up to high energies. For the four-particle states (two particle-hole pairs), we expect $\left\langle 0\left|S_{q}^{z}\right|\alpha\right\rangle \sim O\left(\Delta\right)$,
but this is not true near $\omega\approx vq$ where perturbation theory
in the interaction diverges \cite{pustilnik2}. Figure
\ref{cap:Numerical-form-factors} suggests that
most of the exact form factors evolve smoothly from the XX point, except close to the lower and upper thresholds. If that is the case, the two-particle states still carry most of the spectral weight. In the thermodynamic limit, $F^2(q,\omega)$ has to be combined with the density of states factor \begin{equation}D(q,\omega)=\frac{2\pi}{N}\sum_{\alpha} \delta(\omega-E_{\alpha}+E_{GS}),\label{DOS}
\end{equation}to define the lineshape of $S^{zz}(q,\omega)$ (see (\ref{eq:lehmann})). 

We can count the states at each energy
level of the finite system in the Bethe Ansatz the same way we count states for weakly interacting fermions. For example, in figure \ref{cap:Numerical-form-factors} we see $n\equiv qN/2\pi=8$ two-particle states with $F^2\sim O(1)$. One can also verify that for $n=8$ there are 14 states with two right-moving particle-hole pairs (of the form $c^{\dagger}_{p_1+q_1,R}c^{\phantom{\dagger}}_{p_1,R}c^{\dagger}_{p_2+q_2,R}c^{\phantom{\dagger}}_{p_2,R}\left| 0\right\rangle$) and no states with
three or more pairs. The 14 on-shell states
with $F^2\left(q,\omega\right)<10^{-3}$ in figure
\ref{cap:Numerical-form-factors} are all four-particle states. Furthermore,
for small $\Delta$ the main contribution to the high-frequency tail ($\ell\equiv \omega N/2\pi v>n$) is due to states containing
two particle-hole excitations created around the two different Fermi
points \cite{pustilnik1}. If the momenta of the pairs at the right and left branches are $q_{1}=2\pi n_{1}/N>0$ 
and $q_{2}=2\pi n_{2}/N<0$, such that $n_{1}=(\ell+n)/2$ and $n_{2}=-(\ell-n)/2$, then
the number of such states is given by $\left|n_{1}\times n_{2}\right|=(\ell^{2}-n^{2})/4$.
This is in agreement with the counting of states in figure \ref{cap:Numerical-form-factors}.
We also find much smaller form factors for states with three particle-hole pairs (not shown in the figure).

\begin{figure}
\begin{center}\includegraphics[%
  scale=0.5]{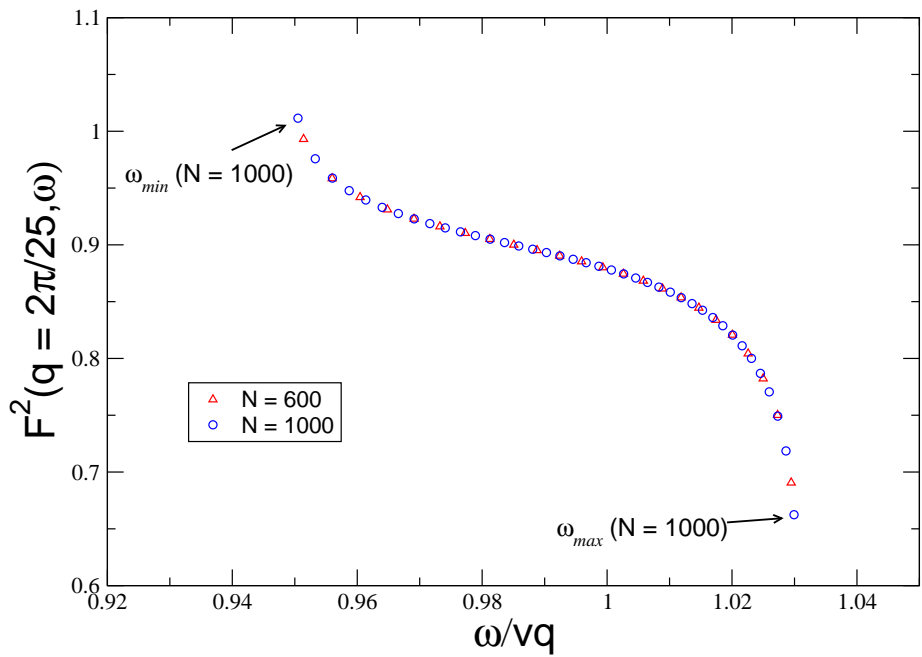}\end{center}

\caption{Form factors squared for the two-particle states for two values of system
size $N$ (we set $q=2\pi/25$, $\Delta=0.25$ and $\sigma=-0.1$). The points
seem to collapse on a single curve, showing very little size dependence. The minimum and maximum energies converge to the thresholds
of the two-particle continuum when $N\rightarrow\infty$. \label{cap:zoominonpeak}}
\end{figure}

We now focus on the two-particle states inside the peak, with $\omega\approx vq$. 
If we seek only these states with dominant form factors it is possible to reach
much larger system sizes (we go up to 7000 sites). The number of two-particle states is always $n=qN/2\pi$. Figure \ref{cap:zoominonpeak}
shows $F^2(q,\omega)$
 for a fixed value of $q=2\pi/25$ and two different
system sizes. %

\begin{figure}
\begin{center}\includegraphics[%
  scale=0.5]{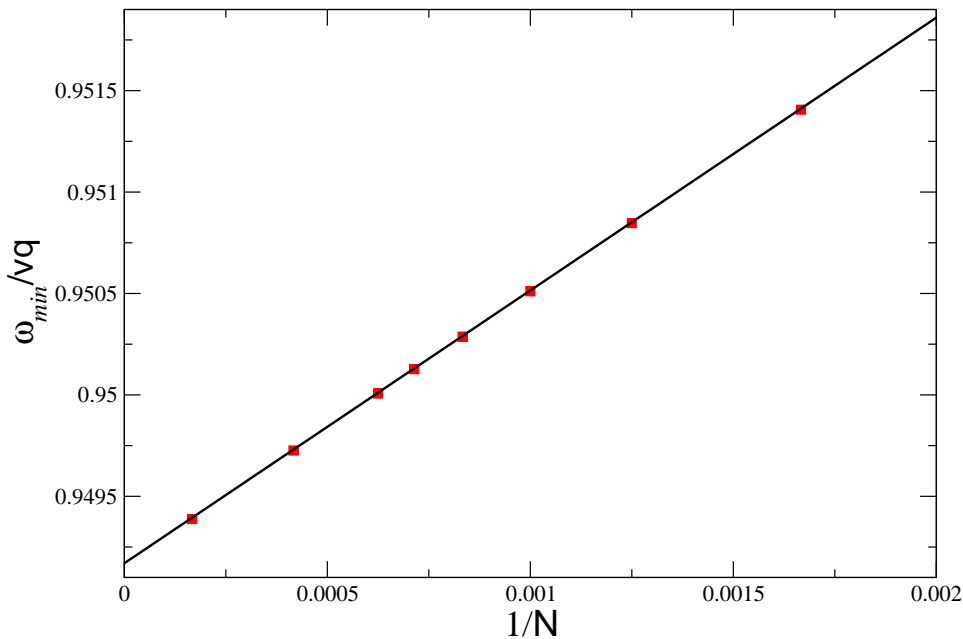}\end{center}

\caption{Finite size scaling of the minimum energy for two-particle states
with $q=2\pi/25$, $\Delta=0.25$ and $\sigma=-0.1$.\label{cap:Finite-size-scaling}}
\end{figure}

\begin{figure}
\begin{center}\includegraphics[%
  scale=0.5]{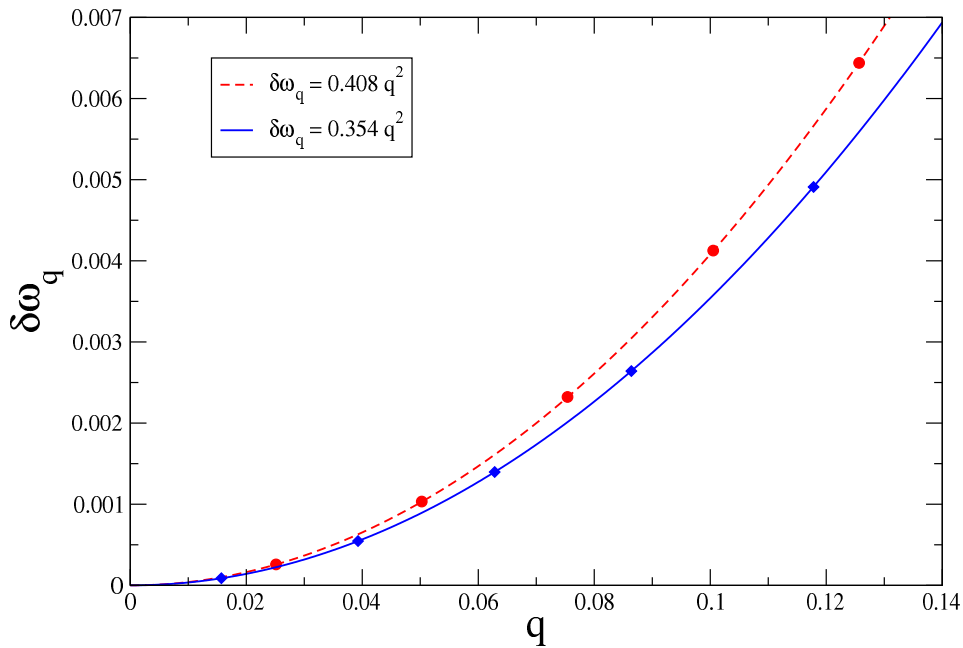}\end{center}

\caption{Width of the on-shell peak (based on the two-particle contribution)
as a function of momentum $q$ for $\sigma=-0.1$ and two values of
anisotropy: $\Delta=0.25$ (blue diamonds) and $\Delta=0.75$ (red
circles). The lines are the best fit to the data. \label{cap:Width-of-the}}
\end{figure}

We extract $\delta\omega_{q}$ from the numerical form factors as
follows. We see from figure \ref{cap:zoominonpeak} that the separation between energy levels inside the peak is of order $\delta\omega_q/N$ and decreases from $\omega_L(q)$ to $\omega_U(q)$. As $N$ increases, the maximum and minimum energies $\omega_{max,min}(N)$ converge to fixed values which  we identify as the thresholds of the
two-particle continuum. Figure \ref{cap:Finite-size-scaling} shows
the finite size scaling of the minimum energy for $\Delta=0.25$, $\sigma=-0.1$ and
$q=2\pi/25$. The same $N^{-1}$ dependence is observed for the
maximum energy. We use this scaling to determine the lower and upper
thresholds $\omega_{L,U}\left(q\right)$ in the thermodynamic limit for several values of $q$. %
We then calculate the width $\delta\omega_{q}=\omega_{U}\left(q\right)-\omega_{L}\left(q\right)$.
As expected, we find that $\delta\omega_{q}=q^{2}/m^*$ for small
$q$ (figure \ref{cap:Width-of-the}). Table \ref{eta_width} compares the coefficients $1/m^*_{\tiny{FIT}}$ obtained by
fitting the data with the predicted values of $\eta_{-}$ taken 
from Table \ref{cap:Parameters-for-the}. The perturbative result in (\ref{eq:zetaminus}) is also shown for comparison. The agreement supports our formula for the width in the strongly interacting (finite $\Delta$) regime. Note that $\eta_-$ is a nonmonotonic function of $\Delta$.

\begin{table}

\caption{Effective inverse mass, defined as the coefficient of the $q^2$ scaling of the width $\delta\omega_q$. The data are for $\sigma=-0.1$ and anisotropy parameters $\Delta=0.25, 0.75, 1$. \label{eta_width}}
\begin{indented}
\item[] \begin{tabular}{@{}cccc}
\br 
$\Delta$&
 $1/m^*_{\tiny{FIT}}$&
 $\eta_-$&
 $\frac{1}{m}\left(1+\frac{2\Delta}{\pi}\sin k_F\right)$\\
\mr
0.25&
 0.354&
 0.356&
0.356\\
0.75&
 0.408&
 0.409&
0.449 \\ 
1&
0.396&
0.397&
0.496\\
\br

\end{tabular}\end{indented}

\end{table}

\begin{figure}
\begin{center}\includegraphics[%
  scale=0.5]{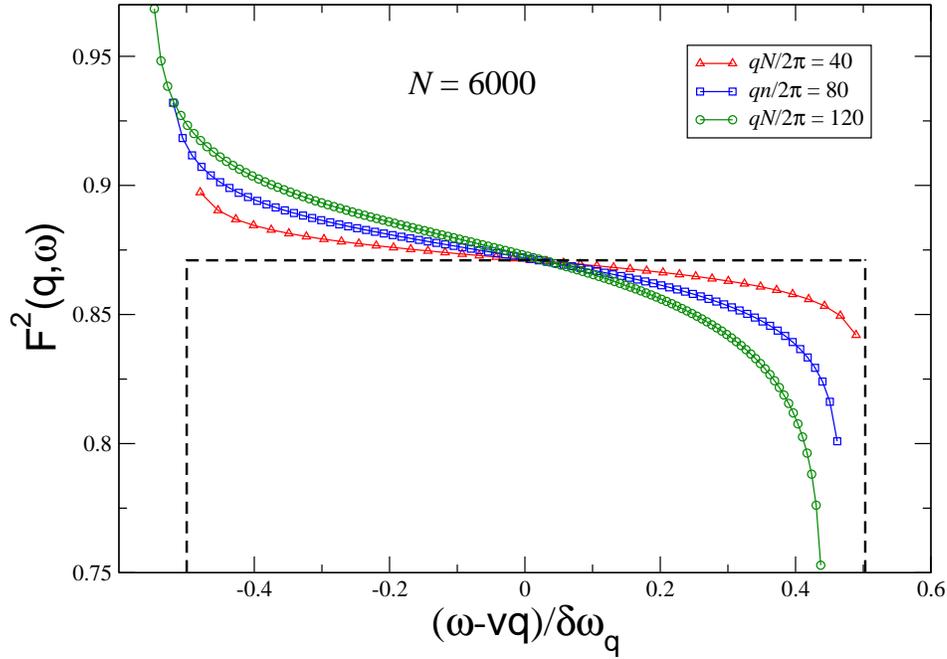}\end{center}

\caption{Frequency dependence of the form factors squared for $N=6000$, $\Delta=0.25$,
$\sigma=-0.1$, and three values of momentum. The dashed line represents the field theory prediction $F^{2}\left(q,\omega\right)=K\approx0.871$, as in figure \ref{cap:Lineshape-in-the}.
\label{cap:Frequency-dependence-of}}
\end{figure}

In figure \ref{cap:Frequency-dependence-of} we confirm that, despite the enhancement (suppression) near the lower (upper) threshold, $F^2(q,\omega)$ converges to the constant value $F^2(q,\omega)=K$
in the limit $q\rightarrow0$, as expected from the box-like shape shown in figure \ref{cap:Lineshape-in-the} (see however the subtleties about the thermodynamic limit in section \ref{sec:finitesizescal}). This is in agreement with the fact that the exponents of the singularities
at the edges are linear in $q$ for $h\neq0$ \cite{pustilnik2}.
Notice that $F^2(q,\omega)$ (and therefore $S^{zz}\left(q,\omega\right)$) is {\it not} a scaling function
of $\left(\omega-vq\right)/\delta\omega_{q}$.%

\begin{figure}
\begin{center}\includegraphics[%
  scale=0.5]{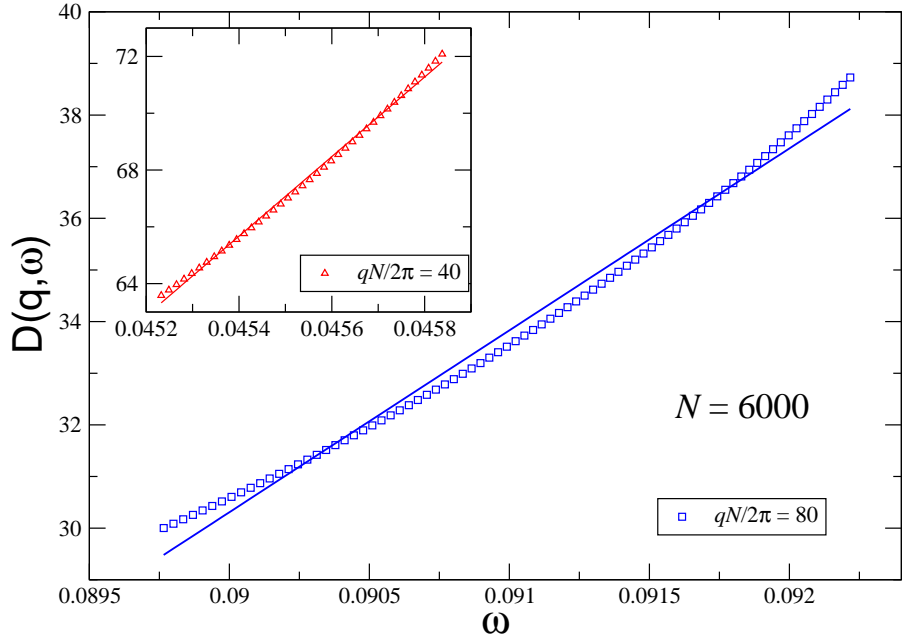}\end{center}

\caption{Density of states $D(q,\omega)$ for the two-particle states obtained using (\ref{diffE_j}). As in figure \ref{cap:Frequency-dependence-of}, we use $N=6000$, $\Delta=0.25$ and $\sigma=-0.1$. The main graph is for $qN/2\pi=80$. The inset shows the density of states for a smaller value of momentum, $qN/2\pi=40$. The solid lines are meant to illustrate the deviation of $D(q,\omega)$ from the linear dependence in $\omega$. \label{figDOS}}
\end{figure}

\begin{figure}
\begin{center}\includegraphics[%
  scale=0.5]{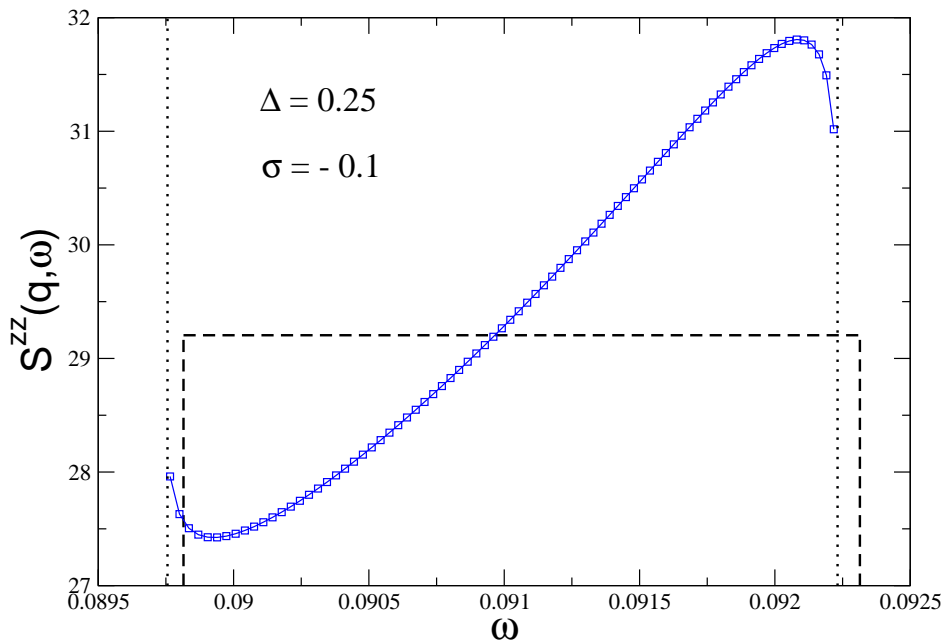}\end{center}

\caption{Lineshape of $S^{zz}(q,\omega)$ estimated from the two-particle states ($\mathcal{S}^2_N(q,\omega)$ in the notation of section \ref{sec:finitesizescal}). For this graph $\Delta=0.25$, $\sigma=-0.1$, $N=6000$ and $qN/2\pi=80$. The dashed line is the
flat distribution of figure \ref{cap:Lineshape-in-the}. The dotted lines are the exact boundaries of the two-particle continuum in the thermodynamic limit.  \label{truelineshape}}
\end{figure}

If the density of states $D(q,\omega)$ for the two-particle states were constant, the extrapolation of figure \ref{cap:zoominonpeak} to the thermodynamic limit would be representative of the lineshape of $S^{zz}(q,\omega)$. This would be exactly the case if the exact energies $E_{\alpha}-E_{GS}$ could be written as the sum of the energies of particles and holes with parabolic dispersion (as in a Galilean-invariant system, {\it e.g.} the Calogero-Sutherland model \cite{pustilnik3}). This is also the case considered in \cite{pustilnik2}. In our case $D(q,\omega)$ does vary inside the peak because of the cubic terms in the dispersion of the particles in the Bethe Ansatz. For large enough $N$ we can include the density of states factor (\ref{DOS}) if we rescale $F^2(q,\omega)$ by the separation between energy levels inside the peak \begin{equation}S^{zz}(q,\omega)=D(q,\omega) F^2(q,\omega)\approx \frac{2\pi}{N}\frac{F^2(q,\omega)}{E_{j+1}-E_j},\label{diffE_j}\end{equation}
where $E_j$ and $E_{j+1}$ are the energies of the two-particles states, ordered in energy, with $E_j-E_{GS}=\omega$. The approximate density of states calculated this way is illustrated in figure \ref{figDOS}. The resulting lineshape is shown in figure \ref{truelineshape}. This lineshape should be contrasted with the free fermion result in figure \ref{freefermion}. The exact boundaries of the two-particle continuum (dotted line in figure \ref{truelineshape}) are actually shifted to lower energies relatively to the prediction $\omega_{U,L}(q)=vq\pm\eta_-q^2/2$ (dashed lines) because of the cubic term in the exact dispersion, which was neglected in the field theory approach. Notice that there appears to be a peak at the exact lower threshold of the two-particle continuum. The result of Pustilnik \textit{et al.} predicts that there is actually a power-law singularity at $\omega_L(q)$, which is related to the physics of the X-ray edge problem \cite{pustilnik2}. We do not attempt to study the singularity in the form factors in this paper (see discussion in section \ref{sec:finitesizescal}). Interestingly, however, the density of states competes with the energy dependence of the form factors, leading to a minimum in $S^{zz}(q,\omega)$ above $\omega_L(q)$ and a rounded peak below $\omega_U(q)$. In the limit $q\ll \cot k_F$ we can linearize the density of states for the two-particle states\begin{equation}
D(q,\omega)\approx\frac{2\pi/N}{E_{j+1}-E_j}\approx \frac{1}{\eta_-q}\left[1+\frac{\tilde{\gamma}q}{\eta_-}\frac{\omega-vq}{\delta\omega_q}\right],
\end{equation}
where $\tilde{\gamma}$ is a fitting parameter analogous to $\gamma$ in (\ref{definevmgamma}) for the free fermion model. The inset of figure \ref{figDOS} shows the density of states for a smaller value of $q=2\pi(40/6000)$. We have checked that $D(q,\omega)$ becomes more linear and $\tilde{\gamma}$ converges to a finite value as $q$ decreases. For $\Delta=0.25$ and $\sigma=-0.1$ we estimate $\tilde{\gamma}\approx 1.11$, which is larger than the value for free fermions $\gamma=\sin(2\pi/5)\approx 0.951$. Combining this density of states with the power-law singularity proposed in \cite{pustilnik2}, the behavior near the lower threshold is described by the function\begin{equation}
S^{zz}(q,\omega)\approx \frac{K}{\eta_-q}\left[1-\frac{\tilde{\gamma}q}{2\eta_-}+\frac{\tilde{\gamma}q}{\eta_-}\frac{\omega-\omega_L(q)}{\delta\omega_q}\right]\left[\frac{\omega-\omega_L(q)}{\delta\omega_q}\right]^{-\mu_q},
\end{equation}
where $\mu_q$ is the exponent of the X-ray edge singularity. The position of the minimum is then \begin{equation}\frac{\omega^*-\omega_L(q)}{\delta\omega_q}\approx\frac{\eta_-\mu_q}{\tilde{\gamma} q},\label{minimumS}\end{equation}
for $\tilde{\gamma}q/\eta_-\ll 1$ and $\mu_q\ll 1$. Since $\mu_q\propto q$ for small $q$, the right-hand side of (\ref{minimumS}) becomes constant in the limit $q\rightarrow 0$. In this sense, the X-ray edge singularity and the energy dependence of the density of states are effects of the same order in $q$. We notice that the difference $\Delta S^{zz}$, defined between the maximum and the minimum of $S^{zz}(q,\omega)$, converges to a finite value as $q\to 0$ (as it did for free fermions). The precise value depends on both the density of states $D(q,\omega)$ and the frequency dependence of $F^2(q,\omega)$ (which is approximately linear with a negative slope for $|\omega-vq|\ll \delta\omega_q$). As a result, the slope of $S^{zz}(q,\omega)$ near the center of the peak diverges as $1/q^2$ as $q\to 0$. This is a rather singular dependence of the lineshape on $\tilde{\gamma}$ and is potentially important for systems in which the dispersion is not exactly parabolic ({\it e.g.} due to band mixing in semiconductor quantum wires).
\begin{figure}
\begin{center}\includegraphics[%
  scale=0.5]{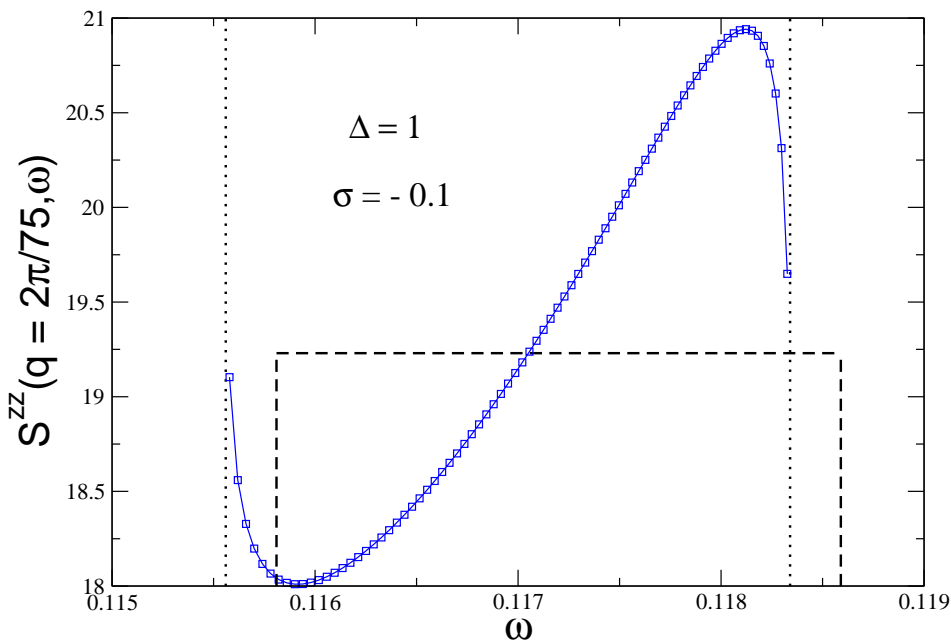}\end{center}

\caption{Lineshape for the Heisenberg chain at finite field ($\Delta=1$, $\sigma=-0.1$, $N=6000$ and $qN/2\pi=80$). Lines and symbols are represented as in figure \ref{truelineshape}.
\label{lineshapeDelta1}}
\end{figure}

Figure \ref{lineshapeDelta1} shows the lineshape for the isotropic point $\Delta=1$ and the same values of $\sigma$ and $q$ used in figure \ref{truelineshape}. In comparison with the weak coupling value $\Delta=0.25$, there is an enhancement of the singularities near the lower and upper thresholds. The shift of the peak to lower energies (another ``$q^3$ effect") is also more pronounced, but the width is very well described by the field theory formula (prefactor given in table \ref{eta_width}).

\begin{figure}
\begin{center}\includegraphics[%
  scale=0.30]{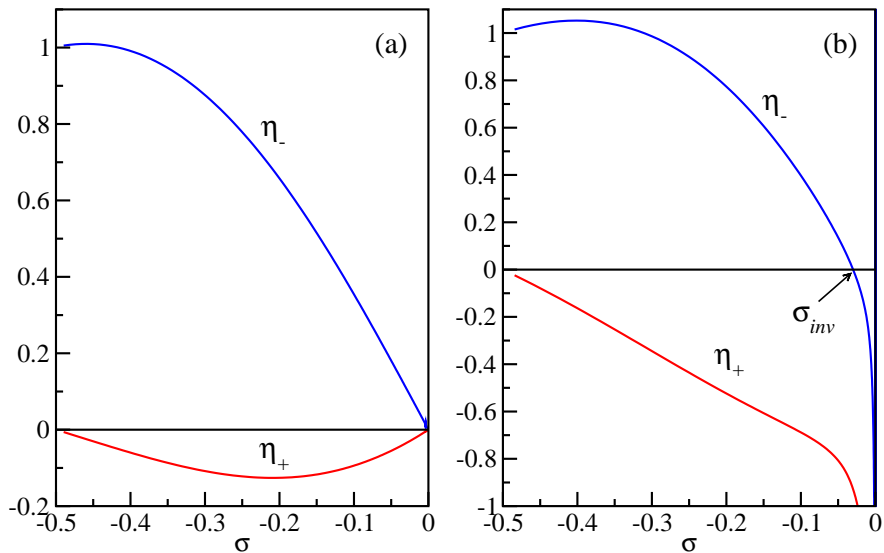}\end{center}

\caption{Parameters $\eta_\pm$ for the low energy effective Hamiltonian as a function of the magnetization $\sigma<0$ for two values of anisotropy: (a) $\Delta=0.25$; (b) $\Delta=1$. For $\sigma>0$, we have $\eta_\pm(-\sigma)=-\eta_\pm(\sigma)$.
\label{etas}}
\end{figure}

Finally, let us comment on the validity of the $q^2$ scaling for the width as a function of magnetic field. Figure \ref{etas} shows the dependence of the coupling constants of the irrelevant operators on the magnetization $\sigma$ for $\Delta=0.25$  and $\Delta=1$.  From the field theory standpoint, we expect that the $q^2$ scaling is valid as long as  $\eta_-q^2\ll vq$ (the peak is narrow) and $\eta_{\pm}q^2\gg \tilde{\gamma} q^3$ (the cubic terms yield the leading correction to the free boson result and we can drop operators with dimension four and higher in the effective Hamiltonian). For $\Delta=0.25$, we see that $\eta_{\pm}$ follow the behavior predicted by the weak coupling expressions (\ref{eq:zetaminus}) and (\ref{eq:zetaplus}), vanishing at $\sigma=0$. In this case, $\eta_{\pm}$ are always of $O(1)$. The restrictions are similar to the ones for the approximation (\ref{eq:Szzquadratic}) for the dynamical structure factor of the XX model, namely $q\ll k_F$ and $q\ll \cot k_F$ (which becomes $q\ll \pi \sigma$ for small $\sigma$). On the other hand, for $\Delta=1$ we find that $|\eta_{\pm}|\to \infty$ as $\sigma\to 0$. This is a direct consequence of formulas (\ref{eq:identity1}) and (\ref{eq:identity2}) in the strong coupling regime. It is known that the magnetic susceptibility at small fields is given by $\chi(h)\sim \textrm{const} + C_1 h^2+C_2 h^{8K-4}$, where $C_{1,2}$ are constants \cite{TakahashiBOOK,SirkerBortz}. The exponent $8K-4$ is a manifestation of the Umklapp scattering term at zero magnetic field.  As a result, $\partial\chi/\partial h$ diverges as $h\to 0$ for $K<5/8$ or $\Delta >\cos(\pi/5)\approx 0.81$. In other words, the Luttinger parameter has an infinite slope at $h=0$ (see \cite{affleckPRB60} for the isotropic case). Since $\partial K/\partial h$ and $\partial v/\partial h$ have opposite signs, $\eta_-$ goes through zero for a finite value of $\sigma$. Therefore, we predict that $\delta\omega_q$ is a nonmonotonic function of $\sigma$ for $\Delta > \cos(\pi/5)$ and $|\sigma|\ll 1$.  
\begin{figure}
\begin{center}\includegraphics[%
  scale=0.5]{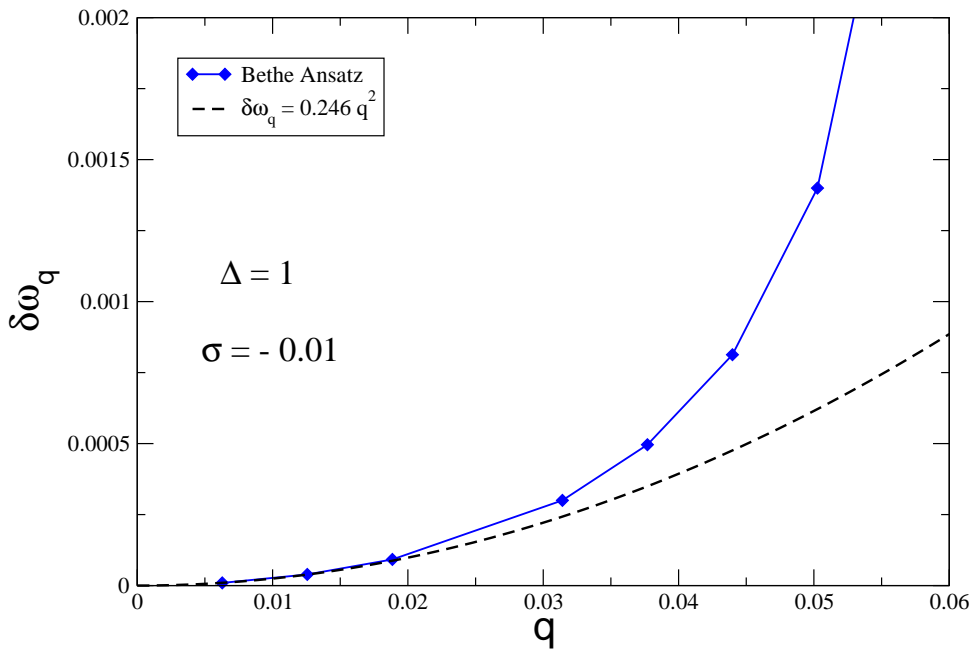}\end{center}

\caption{Width $\delta\omega_q$ for $\Delta=1$ and $\sigma=-0.01$ (in the region where $\eta_-<0$). The blue diamonds represent the width defined as the difference between the maximum and minimum energies of the two-particle states calculated in the Bethe Ansatz, extrapolated to the thermodynamic limit. For $q\ll \pi|\sigma|$, we recover the behavior $\delta\omega_q=|\eta_-| q^2$, with $|\eta_-| \approx 0.246$ (dashed line).
\label{width001}}
\end{figure}
The sign change in $\eta_-$ is reflected in the Bethe Ansatz data as the inversion of the ordering of the energies of the two-particle states as a function of hole momentum. For $\Delta=1$, the ``inversion point" where $\eta_-=0$ occurs at $|\sigma_{inv}|\approx 0.030$. At this point we observe that $\delta\omega_q\propto q^3$ for $q\ll \pi\sigma$.  However, $\eta_+\sim O(1)$, so the lineshape defined by the two-particle states must be different from the one at zero field, where there is also a $q^3$ scaling (see section \ref{sec:zerofield}). We have also confirmed that the $q^2$ scaling holds in the region where $\eta_-<0$ and $q\ll \pi|\sigma|$ (figure \ref{width001}). In this regime we find that the sign change of $\eta_-$ is accompanied by the inversion of the lineshape of $S^{zz}(q,\omega)$: The form factors appear to vanish at the lower threshold and are peaked near the upper threshold (with a possible divergence at $\omega_U$) (figure \ref{invertedshape}). There is no maximum or minimum near the edges in this case. In order to understand this result, we recall that a \emph{converging X-ray edge} is possible in strongly interacting systems. An important point is that the exponents of the X-ray edge singularities calculated in \cite{pustilnik2}, which predict a diverging X-ray edge, are valid only to first order in the interactions. Second order corrections, which usually have the opposite sign because of the orthogonality catastrophe, tend to kill the singularity at the lower edge \cite{mahan}. 
\begin{figure}
\begin{center}\includegraphics[%
  scale=0.5]{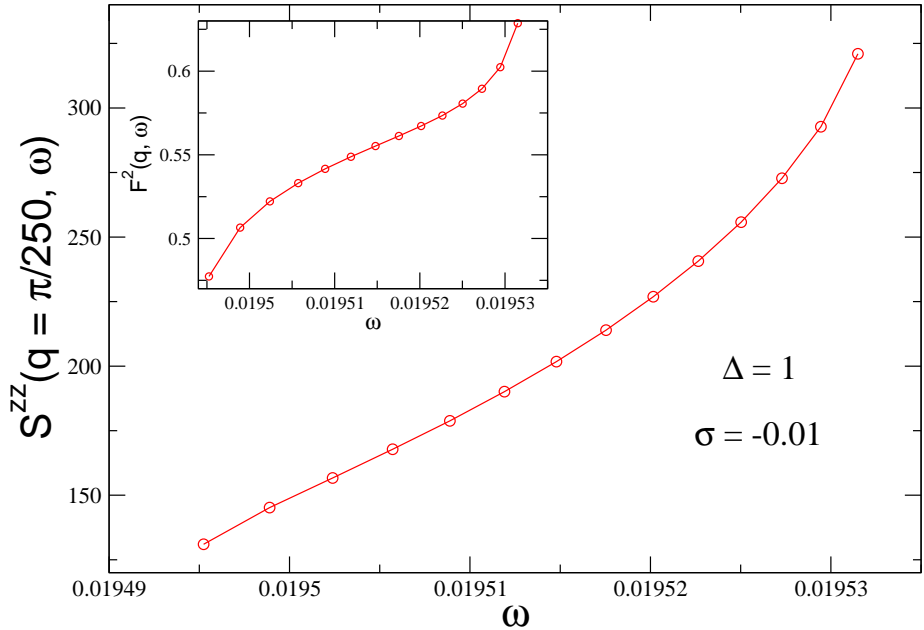}\end{center}

\caption{Lineshape of $S^{zz}(q,\omega)$ for magnetization below the inversion point, {\it i.e.} $|\sigma|<\sigma_{inv}$ ($\Delta=1$, $\sigma=-0.01$, $N=7000$ and $qN/2\pi=14$). This value of $q$ is in the domain where $\delta\omega_q\sim q^2$ (see figure \ref{width001}), but the lineshape is inverted. In contrast with figure  \ref{cap:zoominonpeak}, the form factors (shown in the inset) are peaked at the upper threshold of the two-particle continuum. 
\label{invertedshape}}
\end{figure}
For even smaller values of $\sigma$, the divergence of $\eta_-$ seems to be consistent with the Bethe Ansatz results. For $|\sigma|<\sigma_{inv}$, the width increases as $|\sigma|$ decreases at least down to $\sigma=-0.001$, the lowest magnetization we were able to analyze. In the limit $\sigma\to 0$ and $|\eta_{\pm}|\gg 1$, we expect for the isotropic point (using the results of \cite{BogoliubovNPB,affleckPRB60})\be
\frac{\eta_+}{3}\to \eta_- \to \frac{J}{8\sqrt{2}\sigma\ln|\sigma_0/\sigma|},\label{divergingetas}
\ee
where $\sigma_0=\sqrt{32/\pi e}$.  According to the conditions $\tilde{\gamma}q^2\ll \eta_-q\ll v$, the field theory result which predicts the $q^2$ scaling for a small fixed $q$ breaks down both near the inversion point $\sigma_{inv}$ and for $\sigma\to 0$. In the limit $\sigma \rightarrow 0$, as mentioned in section
4.2, the set of allowable quantum numbers for the single
particle-hole excitations becomes empty, as the $I_{\infty}$
quantum number tends to $N/2$ (meaning that the particle
part becomes trapped at the Fermi surface), and this family
of excitations disappears.  The vanishing field two-particle continuum
at nonvanishing momentum is then obtained from considering the
next simplest excitations, which are states having two holes
(spinons) within the ground state configuration together with a single
negative parity one-string (or, for the XXX chain, an infinite
rapidity).  At finite but small field, the contributions from these  
states
dominates $S^{zz} (q, \omega)$ for $q \gg \pi \sigma$ and allows
to smoothly recover the zero field behavior.  A full
discussion of all the possible lineshapes as a function of
$\Delta$ and $\sigma$ together with the characterization of
the dominant families of excitations is accessible from the results
of \cite{CauxJSTATP09003}, but is beyond the scope of the
present paper.

\section{High-frequency tail\label{sec:High-frequency-tail}}

We now turn to the calculation of $S^{zz}\left(q,\omega\right)$ in
the frequency range $\gamma_{q}\ll\omega-vq\ll J$, where
finite order perturbation theory is expected to be valid. This off-shell
spectral weight is possible because the $\eta_{+}$ interaction allows
for two-boson intermediate states with total momentum $q=q_{1}+q_{2}$
but energy $\omega=v\left|q_{1}\right|+v\left|q_{2}\right|>v\left|q\right|$
if $\textrm{sign}(q_{1})=-\textrm{sign}(q_{2})$. In other words,
the incoming boson can decay into one right-moving and one left-moving
boson, which together can carry small momentum but high energy $\omega\gg v\left|q\right|$.
In the limit $\Delta\ll1$, this is equivalent to a state with two
particle-hole pairs created around the two different Fermi points
\cite{pustilnik1}. In this sense, our $\eta_{+}$ is analogous to
the $U_{q}$ interaction in \cite{pustilnik2}. We should stress that,
although the tail carries a small fraction of the spectral weight of
$S^{zz}\left(q,\omega\right)$, it is important for response functions
that depend on the overlap of two spectral functions, {\it e.g.} the drag
resistivity in the fermionic version of the problem \cite{pustilnik1}. In our formalism the calculation of the tail provides a direct quantitative
check of the accuracy of the low energy effective model against the
form factors calculated by Bethe Ansatz.

\subsection{Field theory prediction}\label{fieldtail}

\begin{figure}
\begin{center}\includegraphics[%
  scale=0.9]{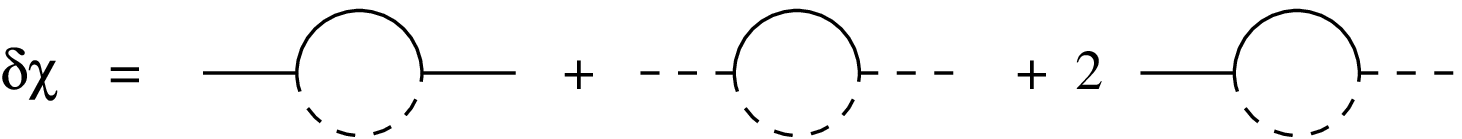}\end{center}

\caption{Diagrams at $O(\eta_{+}^{2})$ for the calculation of the tail.\label{cap:Diagrams-at-O}}
\end{figure}

The lowest-order correction to $\chi\left(q,i\omega_{n}\right)$ due
to the $\eta_{+}$ interaction is\begin{equation}
\delta\chi\left(q,i\omega_{n}\right)=-\int_{0}^{L}dx\, e^{-iqx}\int_{0}^{\beta}d\tau\, e^{i\omega\tau}\delta\chi\left(x,\tau\right),\end{equation}
 where $\delta\chi\left(x,\tau\right)$ is the correlation function
in real space given by\begin{eqnarray}
\delta\chi\left(x,\tau\right) & = & \frac{K}{\pi}\frac{1}{2}\left(\frac{\sqrt{2\pi}}{6}\eta_{+}\right)^{2}\int d^{2}x_{1}\int d^{2}x_{2}\,\nonumber \\
 &  & \times\left\langle \partial_{x}\phi\left(x\right)\left[\left(\partial_{x}\varphi_{L}\left(1\right)\right)^{2}\partial_{x}\varphi_{R}\left(1\right)-\left(R\leftrightarrow L\right)\right]\right.\nonumber \\
 &  & \times\left.\left[\left(\partial_{x}\varphi_{R}\left(2\right)\right)^{2}\partial_{x}\varphi_{L}\left(2\right)-\left(R\leftrightarrow L\right)\right]\partial_{x}\phi\left(0\right).\right\rangle \label{eq:deltachi}\end{eqnarray}
 This corresponds to the diagrams in figure \ref{cap:Diagrams-at-O}.
$\delta\chi$ can be factored in the form \begin{equation}
\delta\chi\left(q,i\omega\right)=\frac{K}{2\pi}\left[D_{R}^{\left(0\right)}\left(q,i\omega\right)+D_{L}^{\left(0\right)}\left(q,i\omega\right)\right]^{2}\Pi_{RL}\left(q,i\omega\right),\label{eq:definebubble}\end{equation}
 where $\Pi_{RL}\left(q,i\omega\right)$ is the bubble with right-
and left-moving bosons\begin{eqnarray}
\Pi_{RL}\left(q,i\omega\right)&=&-\frac{2\pi\eta_{+}^{2}}{9}\int_{-\infty}^{+\infty}dx\, e^{-iqx}\int_{0}^{\beta}d\tau\, e^{i\omega\tau}D_{R}^{\left(0\right)}\left(x,\tau\right)D_{L}^{\left(0\right)}\left(x,\tau\right)\nonumber\\
&=& -\frac{2\pi\eta_{+}^{2}}{9}\int_{-\infty}^{+\infty}\frac{dk}{2\pi}\int_{-\infty}^{+\infty}\frac{d\nu}{2\pi}D_{R}^{\left(0\right)}\left(k,i\nu\right)\nonumber\\ & & \qquad \times D_{L}^{\left(0\right)}\left(q-k,i\omega-i\nu \right).\end{eqnarray}
After integrating over frequency, we get\bea
\Pi^{ret}_{RL}\left(q,\omega\right)&=&-\frac{\eta_+^2}{9}\left[\int_0^{\Lambda}dk\, \frac{k(q+k)}{\omega+vq+2vk+i\eta}\right.\nonumber\\& &\left. +\int_q^{\Lambda}dk\, \frac{k(q-k)}{\omega+vq-2vk+i\eta}\right],
\eea
where $\Lambda\sim k_F$ is a momentum cutoff. Note that the real part of $\Pi^{ret}_{RL}$ is ultraviolet-divergent, but the imaginary part is not. The integration over the internal momentum yields\begin{equation}\Pi^{ret}_{RL}(q,\omega)=-\frac{\eta_+^2}{9}\left\{\frac{\Lambda^2}{2v}-\frac{\omega^2-v^2 q^2}{8v^3}\log\left[\frac{(vq)^2-(\omega+i\eta)^2}{4v^2\Lambda^2}\right]\right\}.\label{Pireal}\end{equation}
 Finally, using Eqs. (\ref{eq:definebubble}) and (\ref{eq:spectralfunction}),
we find that the high-frequency tail of $S^{zz}\left(q,\omega\right)$
is given by\begin{equation}
\delta S^{zz}\left(q,\omega\right)=\frac{K\eta_{+}^{2}q^{4}}{18v}\,\frac{\theta\left(\omega-vq\right)}{\omega^{2}-v^{2}q^{2}}.\label{eq:high-w_tail}\end{equation}
 This is the same $\omega^{-2}$ dependence obtained for weakly interacting
fermions with parabolic dispersion \cite{pustilnik1}. Since
the small parameter is $\eta_{+}\sim\Delta/m$, we approach the perturbative
regime either by $\Delta\rightarrow0$ or $m\rightarrow\infty$ (more
precisely, $q/mv\rightarrow0$). In this limit, our result (\ref{eq:high-w_tail})
agrees with equation (19) of \cite{pustilnik1} if we use (\ref{eq:zetaplus})
and $U\left(q\right)=\left(\Delta/2\right)\cos q$.

Since our model predicts that $\delta S^{zz}\left(q,\omega\gg vq\right)\sim O(\eta_{+}^{2})$,
one interesting consequence is that there will be no tail in $S\left(q,\omega\right)$
for models where the Luttinger parameter $K$ is independent of particle
density, since then $\eta_{+}=0$ according to (\ref{eq:identity2}).
This is the case for the Calogero-Sutherland model, where $K$ is
a function of the amplitude of the long-range interaction only \cite{KawakamiPRL67}. 

The divergence of the high-frequency tail of $\delta S^{zz}\left(q,\omega\right)$
as $\omega\rightarrow vq$ confirms that the on-shell region is not
accessible by our standard perturbation theory in the band curvature
terms. The matching of the tail to the on-shell peak at $\omega_U(q)$ is a complicated problem that has only been addressed in the regime $\Delta\ll 1$ (see \cite{pustilnik2}). The $(\omega-vq)^{-1}$ divergence in (\ref{eq:high-w_tail}) comes from the frequency dependence of the external legs in the diagrams of figure \ref{cap:Diagrams-at-O}. It is easy to see that if the bosonic propagators are replaced by the ``dressed'' propagator (all orders in $\eta_-$) given by (\ref{dressedprop}), the singularity at the upper threshold $\omega_U(q)$ becomes only logarithmic. This supports the picture that the $\eta_+$ interaction only modifies the shape of the on-shell peak very close to the edges. We expect that $\eta_+$ will contribute to the exponent of the singularity at the edges, since the exponent $\mu_q$ derived in \cite{pustilnik2} picks up corrections of second order in the interaction between right and left movers, {\it i.e.} $O(\eta_+^2)$. As discussed in section \ref{secwidthFT}, we believe that $\eta_+$ does not affect the width to $O(q^2)$. Evidence for that is that the perturbation theory in $\eta_+$ (second order given by (\ref{Pireal})) does not generate terms with the same $q$ and $\omega$ dependence as in (\ref{chi2eta-}) and (\ref{eq:diagrams}). If the frequency dependence is regularized in the peak region by summing the perturbation theory in $\eta_-$, the diagrams involving $\eta_+$ are always suppressed by higher powers of $q$ because of simple kinematics. Inside the peak the energy of the left moving boson that is put on shell when taking the imaginary part of $\chi(q,\omega)$ (as in the ``unitarity condition'' method used in \cite{Aristov}) has to be of order $\delta\omega_q=\eta_-q^2$ or smaller, which constrains the phase space for the internal momenta.

\subsection{Comparison with numerical form factors}
\begin{figure}
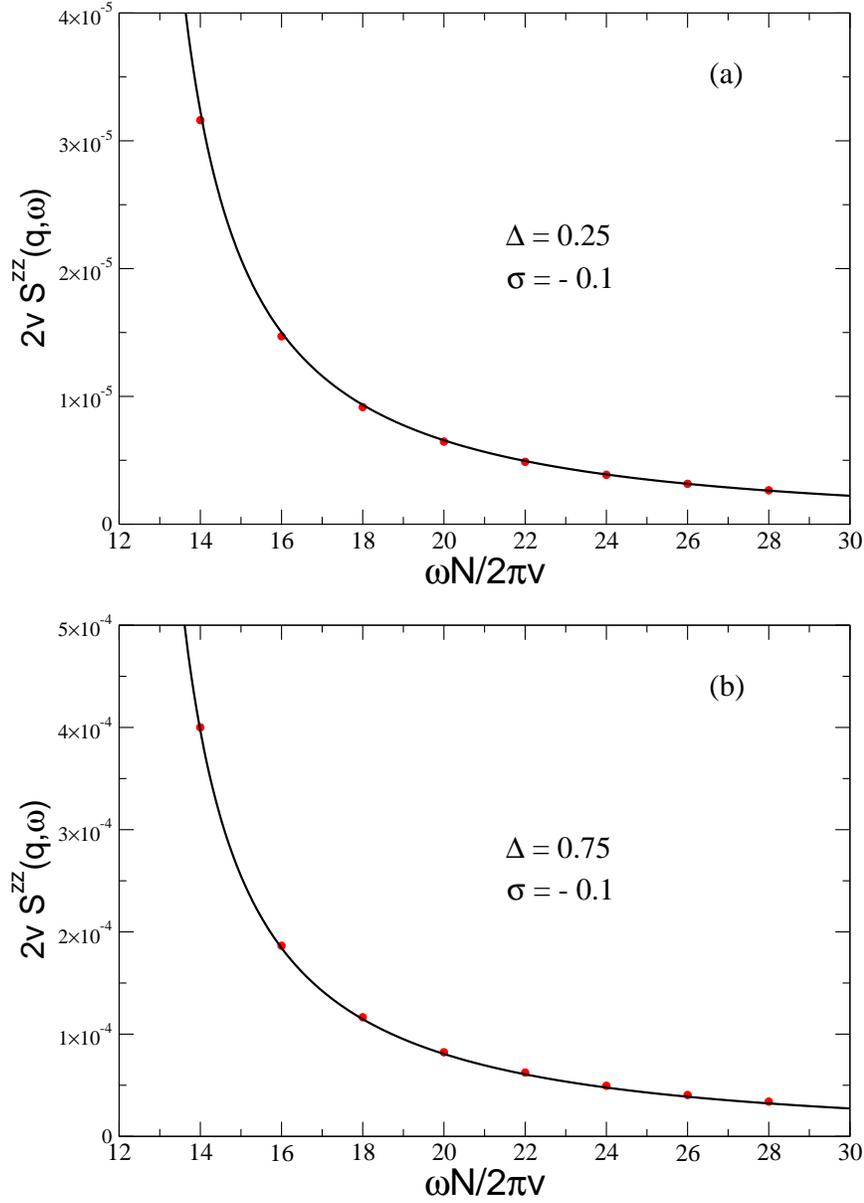

\begin{center}\includegraphics[%
  scale=0.45]{sqw_figure19a.eps}\end{center}
\begin{center}
\includegraphics[%
  scale=0.45]{sqw_figure19b.eps}\end{center}

\caption{Tail of $S^{zz}\left(q,\omega\right)$ for $q=2\pi/50$ and $\delta\omega_{q}\ll\omega-vq\ll J$.
The red dots represent the sum of the numerical $F^2(q,\omega)$ identified
with each energy level predicted by field theory ({\it c.f.} figure \ref{cap:Numerical-form-factors}). The solid line is the field theory result 
(\ref{eq:F(N,M)_fieldtheory}). The chain length is $N=600$. (a) $\sigma=-0.1$, $\Delta=0.25$; (b) $\sigma=-0.1$,
$\Delta=0.75$. \label{cap:Tail}}
\end{figure}

For a finite system with size $N$, the result for $\delta S^{zz}\left(q,\omega\right)$
must be expressed in terms of the transition probabilities $F^2(q,\omega)$. If the intermediate
bosons carry momenta $q_{1,2}=2\pi n_{1,2}/N$, such that $q_{1}+q_{2}=q\equiv2\pi n/N$,
the energy levels are given by the sum of their individual energies
$\omega=v\left|q_{1}\right|+v\left|q_{2}\right|$, {\it i.e.},\begin{equation}
\omega_{\ell}=\frac{2\pi v\ell}{N}\qquad,\qquad\ell=n+2,n+4,\dots\,.\label{eq:energylevels}\end{equation}
 Thus field theory predicts a uniform level spacing $4\pi v/N$ above
the mass shell. It is easy to verify (by simply replacing the integrals
by sums in momentum space) that $\delta S^{zz}\left(q,\omega\right)$
for the finite system can be written as \begin{equation}
\delta S^{zz}\left(q=\frac{2\pi n}{N},\omega\right)=\frac{2\pi}{N}\sum_{\mathcal{\ell}}F^{2}\left(q,\omega\right)\delta\left(\omega-\omega_{\mathcal{\ell}}\right),\end{equation}
 where $F^{2}\left(q,\omega\right)=\left|\left\langle 0\left|S_{q}^{z}\right|\alpha\right\rangle \right|^{2}$,
with $\left|\alpha\right\rangle $ a two-boson intermediate state,
is the transition probability for the state with energy $\omega_{\ell}$ and is given by\begin{equation}
F^{2}\left(q,\omega\right)=2v\,\delta S^{zz}\left(q,\omega\right)=\frac{4\pi^{2}K\eta_{+}^{2}}{9v^{2}N^{2}}\, \frac{n^{4}}{\mathcal{\ell}^{2}-n^{2}}.\label{eq:F(N,M)_fieldtheory}\end{equation}

We compare our field theory prediction with the form factors calculated
numerically for a chain with $N=600$ sites. We take $q=2\pi/50$ ($n=12$) and the previous values $\sigma=-0.1$ and $\Delta=0.25$ or $\Delta=0.75$ (for which the parameters are shown in table \ref{cap:Parameters-for-the}). As we saw in figure \ref{cap:Numerical-form-factors}, the energies of the eigenstates calculated by BA are actually scattered around the values of $\omega_{\ell}$ predicted
in (\ref{eq:energylevels}).
The broadening becomes comparable with the level spacing $4\pi v/N$
when $\mathcal{\ell}\approx30$ ($\omega\approx0.4J$). Again the number of states agrees with a picture of multiple particle-hole
excitations based on perturbation theory in the interaction. These
features are not predicted by the bosonization approach. In order to make the comparison
with (\ref{eq:F(N,M)_fieldtheory}), we group the form factors that
can be identified with a given energy level $\omega_{\mathcal{\ell}}$
and plot the total  $F^{2}\left(q,\omega\right)$
as a function of the integers $\mathcal{\ell}=\omega N/2\pi v$. We
emphasize that for very large $\ell$ we expect deviations from the
lowest-order field theory result due to the effect of more irrelevant operators
we have neglected. The results are shown in figure \ref{cap:Tail}.

\section{The zero field case}\label{sec:zerofield}
So far we have focused on the dynamical structure factor  at finite magnetic field, which is somewhat analogous to interacting fermions with parabolic dispersion. One may then ask whether the field theory calculations can be applied to the case $h=0$ ($k_F=\pi/2$). Let us first review what is known for the free fermion point $\Delta=0$. In this case $S^{zz}(q,\omega)$ is still given by (\ref{eq:szzcosine}), but the thresholds of the two-particle (two-spinon in the Bethe Ansatz solution) continuum are given by \begin{eqnarray}
\omega_{L}\left(q\right) & = & J\sin q,\label{h=0_lowerbound}\\
\omega_{U}\left(q\right) & = & 2J\sin\frac{q}{2}.\label{h=0upperbound}\end{eqnarray} As a result, $S^{zz}(q,\omega)$ develops a square root divergence at the upper threshold $\omega_U(q)$. The width now scales like $q^3$ for small $q$\begin{equation}
\delta\omega_q\approx \frac{J q^3}{8}. \label{q3scaling}\end{equation}
A crossover from $q^2$ to $q^3$ is observed as we decrease the magnetic field (or, equivalently, increase $q\ll k_F$) so as to violate (\ref{restrictFT}) or (\ref{restrictionBA}). The result (\ref{q3scaling}) is also obtained by keeping the leading correction to the linear dispersion around $k_F$\begin{equation}
\epsilon_{k}^{R,L}\approx\pm \left(v_{F}k-\frac{\gamma k^{3}}{6}+\dots\right)\label{dispersionh=0},\end{equation}
where $\gamma=v_F=J$. Bosonizing the band curvature term according to (\ref{eq:F_nboson}), we find \begin{equation}
\delta\mathcal{H}_{bc}=-\frac{\pi\gamma}{12}:\left(\partial_x\phi_R\right)^4:-\frac{\gamma}{24}:\left(\partial_x^2\phi_R\right)^2:+(R\rightarrow L),\end{equation}
which can be rewritten as\begin{equation}
\delta\mathcal{H}_{bc}=-\frac{\pi\gamma}{12}:\left(\partial_x\phi_R\right)^2:\, :\left(\partial_x\phi_R\right)^2:+(R\rightarrow L)\label{OPEquartic},\end{equation}
as follows from the operator product expansion of (\ref{OPEquartic}).

In the interacting case we also have to keep track of the irrelevant interaction terms, including the Umklapp interaction in (\ref{eq:interaction}). The general form for the leading irrelevant operators for zero field is \cite{lukyanov,pereira}\begin{eqnarray}\delta \mathcal{H}&=&\frac{\pi\zeta_-}{12}\left[:\left(\partial_x\varphi_R\right)^2:\,:\left(\partial_x\varphi_R\right)^2:+:\left(\partial_x\varphi_L\right)^2:\,:\left(\partial_x\varphi_L\right)^2:\right]\nonumber\\
& & +\frac{\pi\zeta_+}{2}\,:\left(\partial_x\varphi_R\right)^2:\,:\left(\partial_x\varphi_L\right)^2:+\frac{\lambda_1}{2\pi} \cos(4\sqrt{\pi K}\phi)+\dots,\label{deltaH_h=0}
\end{eqnarray}
where the dots stand for higher dimensional local counterterms. The coupling constants to first order in $\Delta$ can be obtained from the bosonization of the band curvature term and the irrelevant interaction terms. We find \begin{equation}\zeta_-\approx -J\left(1+\frac{\Delta}{\pi}\right),\hspace{0.5cm}
\zeta_+\approx-\frac{\Delta J}{\pi},\hspace{0.5cm}
\lambda_1\approx \frac{\Delta J}{\pi}.\end{equation}
The exact coupling constants for finite $\Delta$ can be taken from \cite{lukyanov}  \begin{eqnarray}
\zeta_-&=&-\frac{v}{4\pi K}\frac{\Gamma\left(\frac{6K}{4K-2}\right)\Gamma^3\left(\frac{1}{4K-2}\right)}{\Gamma\left(\frac{3}{4K-2}\right)\Gamma^3\left(\frac{2K}{4K-2}\right)},\\
\zeta_+&=&-\frac{v}{2\pi}\tan\left(\frac{\pi K}{2K-1}\right),\label{zeta+h=0}\\
\lambda_1&=&-\frac{4v\,\Gamma(2K)}{\Gamma(1-2K)}\left[\frac{\Gamma\left(1+\frac{1}{4K-2}\right)}{2\sqrt{\pi}\Gamma\left(1+\frac{K}{2K-1}\right)}\right]^{4K-2},
\end{eqnarray}
where $v$ and $K$ are given by (\ref{eq:vBAzerofield}) and (\ref{eq:KBAzerofield}), respectively.

One important point is that the other possible type of dimension-four operator $\left(\partial_x\varphi_R\right)^3\partial_x\varphi_L+R\leftrightarrow L$ is absent from the effective Hamiltonian for the XXZ model. We see this directly when calculating the coupling constants to first order in $\Delta$, but we can also show that it remains true for finite $\Delta$  by imposing the constraint that the XXZ model is integrable \cite{pereira}. Integrability implies the existence of nontrivial conserved quantities, the simplest one of which is the energy current operator $J^E=\sum_j j^E_j$ given by \cite{zotos, brenig}\bea 
J^E&=&J^2\sum_j\left[S_{j-1}^yS_j^zS_{j+1}^x-S_{j-1}^xS_j^zS_{j+1}^y+\Delta (S_{j-1}^xS_j^yS_{j+1}^z-S_{j-1}^zS_j^yS_{j+1}^x)\right.\nonumber\\
& &\left.+\Delta (S_{j-1}^zS_j^xS_{j+1}^y-S_{j-1}^yS_j^xS_{j+1}^z)\right].\label{JElattice}
\eea
The latter is defined by the continuity equation of the energy density at zero field \be
j^E_{j+1}-j^E_j=- \partial_t \mathcal{H}_j = i [\mathcal{H}_j,H], 
\ee
where $H=\sum_j \mathcal{H}_j$ is the Hamiltonian (\ref{eq:XXZ}) with $h=0$. One can then verify that  $J^E$ is conserved in the sense that $[J^E,H]=0$. 

Let us now look at the corresponding quantity in the low energy effective model. In the general case, we consider the Hamiltonian density $\mathcal{H}=\mathcal{H}_{LL}+\delta \mathcal{H}+\delta \mathcal{H}_3$, where $\delta H$ is given by (\ref{deltaH_h=0}) and we also add the interaction \be
\delta \mathcal{H}_3 = \pi \zeta_3  \left[\left(\partial_x\varphi_R\right)^3\partial_x\varphi_L+\left(\partial_x\varphi_L\right)^3\partial_x\varphi_R\right].
\ee
We obtain the energy current operator from the continuity equation in the continuum limit\be
\partial_x j^E(x) = -\partial_t \mathcal{H}(x)=i\int dy \left[\mathcal{H}(x),\mathcal{H}(y)\right].\label{continuityJE}
\ee
The energy current operator for the Luttinger model (with $\zeta_{\pm,3}=\lambda_1=0$) takes the form \bea
J^E_0&=&\int dx \, j^E_0(x)= \frac{v^2}{2}\int dx \left[\left(\partial_x \varphi_R\right)^2-\left(\partial_x \varphi_L\right)^2\right]\nonumber\\
&=&-v^2\int dx\, \partial_x\phi \partial_x\theta.
\eea
This coincides with the spatial translation operator of the Gaussian model \cite{fujimoto}. A nontrivial consequence of the conservation law arises when we consider the corrections to $J^E$ due to the irrelevant operators. We keep corrections up to operators of dimension four. Using (\ref{continuityJE}), we find $J^E=J^E_0+\delta J^E$ with \cite{pereira}\bea
\delta J^E&=&\pi v\int dx \left\{\frac{\zeta_-}{3} \left[\left(\partial_x \varphi_R\right)^4-\left(\partial_x \varphi_L\right)^4\right]\right. \nonumber \\& &\left. +2\zeta_3\left[\left(\partial_x \varphi_R\right)^3 \partial_x \varphi_L-\left(\partial_x \varphi_L\right)^3 \partial_x \varphi_R\right]\right\}.
\eea
Note that there are no first-order corrections to $J^E$ associated with the $\zeta_+$ interaction or the Umklapp scattering $\lambda_1$. (The case of the Umklapp perturbation was discussed in \cite{saito}).
The conservation of $J^E$ up to dimension-four operators implies\be
[J^E,H]=[J^E_0,H_{LL}]+[J^E_0,\delta H]+[J^E_0,\delta H_3]+[\delta J^E,H_{LL}]=0.
\ee
Since $J^E_0$ is conserved in the Luttinger model, we have $[J^E_0,H_{LL}]=0$. In fact, $J^E_0$ commutes with any local operator of the form $\int dx \, O(x)$ under periodic boundary conditions  \cite{fujimoto}. As a result, $[J^E_0,\delta H]=[J^E_0,\delta H_3]=0$ as well. We are left with the condition that the commutator $[\delta J^E,H_{LL}]$ vanishes. This is automatically satisfied by the contribution from the $\zeta_-$ term because it does not mix $R$ and $L$ and $
\left[\left(\partial_x \varphi_R\right)^4-\left(\partial_x \varphi_L\right)^4,H_{LL}\right]$ is a total derivative. We then have\bea
[\delta J^E,H_{LL}]&=& \pi v^2\zeta_3 \int dx \int dx^{\prime}\times\nonumber\\
& & \left[\left(\partial_x \varphi_R\right)^3 \partial_x \varphi_L-\left(\partial_x \varphi_L\right)^3 \partial_x \varphi_R\, ,\left(\partial_{x^\prime} \varphi_R\right)^2+\left(\partial_{x^\prime} \varphi_L\right)^2\right]\nonumber\\
&=&4\pi iv^2\zeta_3 \int dx \left[\left(\partial_x \varphi_R\right)^3 \partial^2_x \varphi_L+\left(\partial_x \varphi_L\right)^3 \partial^2_x \varphi_R\right].
\eea
Therefore, $[J^E,H]=0\Leftrightarrow \zeta_3=0$. 

This argument also applies to the finite field case. The model is still integrable for $h\neq 0$. Although the relevant quantity for thermal transport is now a linear combination of the energy current and the spin current operator (which is not conserved for the XXZ model), the energy current operator given by (\ref{JElattice}) commutes with the Hamiltonian (\ref{eq:XXZ}) for all values of $h$ \cite{brenig,sakai}. The corresponding conserved quantity in the low energy theory is the current operator $J^E$ obtained from the effective Hamiltonian at zero field, which has no dependence on the coupling constants $\eta_\pm$. Clearly, $[J^E_0,\delta H(h\neq 0)]=0$ for $\delta H(h\neq 0)$ given by (\ref{eq:deltaHzetas}), so integrability poses no constraints on the coupling constants $\eta_\pm$.

We have checked that $\zeta_3\neq 0$ for a nonintegrable model obtained by adding to the XXZ model the following next-nearest neighbour interaction\be
\delta H_{nnn} = J\Delta^\prime \sum_j S_j^zS_{j+2}^z, 
\ee
which is mapped by bosonization onto
\be
\delta H_{nnn}=J\Delta^\prime \int dx \left[ -\frac{3}{\pi}\left(\partial_x\tilde{\phi}\right)^2+\frac{16}{3}\left(\partial_x\tilde{\phi}\right)^4+\dots\right]. \label{intbreak}
\ee
The first  term in (\ref{intbreak}) is quadratic in the bosons and modifies the velocity and the Luttinger parameter of the Luttinger model. The second term is the irrelevant operator. To first order in $\Delta$ and $\Delta^\prime$, we find that it gives rise to a $\zeta_3$ term in the Hamiltonian, which is given by\be
\delta H_{nnn}\sim -\frac{29}{6}J\Delta^\prime \left[\left(\partial_x\varphi_R\right)^3\partial_x\varphi_L+\left(\partial_x\varphi_L\right)^3\partial_x\varphi_R\right].
\ee
This shows that, unlike the XXZ model, a low energy effective model describing a nonintegrable model must in general contain the $\zeta_3$ interaction.

This result establishes a connection between integrability and the field theory approach, by means of a restriction on the coupling constant of a band curvature type operator in the low energy effective model. More generally, if we keep more irrelevant operators in the effective Hamiltonian, integrability should manifest itself as a fine tuning of the coupling constants and the absence of certain perturbations. This connection may be important for understanding the role of integrability in the transport properties of one-dimensional systems \cite{zotos}.

With $\zeta_3=0$, only $\zeta_+$ and $\lambda_1$ mix right and left movers. We can apply second order perturbation theory in these interactions to calculate two contributions to the high-frequency tail in the frequency range $\delta\omega_q\ll \omega-vq\ll J$ \cite{pereira}.  For a finite chain with $N$ sites and fixed momentum $q=2\pi n/N$, the $\zeta_+$ operator gives rise to intermediate states with discrete energies $\omega_\ell=2\pi v\ell/N$, $\ell=n+2,n+4, \dots$. In the thermodynamic limit, the contribution to the tail is
\be
\delta S^{zz}_{\zeta_+}(q,\omega)=\frac{K(\zeta_+/v)^2}{192v}q^2\left(\frac{\omega^2-v^2q^2}{v^2}\right)\theta(\omega-vq).\label{tail_zeta+}
\ee
The states generated by the Umklapp operator have energies $\omega_\ell=2\pi v(\ell+4K)/N$, $\ell=n,n+2, \dots$ . For $4\pi v/N\gg\delta\omega_q$ it is easy to separate this contribution from the $\zeta_+$ one because of the shift in the energy levels by the noninteger factor $4K$. The corresponding contribution to the tail is \be
\delta S^{zz}_{\lambda_1}(q,\omega)=\frac{2\lambda_1^2K^2}{\Gamma^2(4K)}(2v)^{3-8K}q^2\left(\omega^2-v^2q^2\right)^{4K-3}\theta(\omega-vq).\label{betas_umklapp}
\ee The derivation of equations (\ref{tail_zeta+}) and (\ref{betas_umklapp}), as well as the result for the finite system, is presented in the appendix.

The result in (\ref{tail_zeta+}) and (\ref{betas_umklapp}) shows that the Umklapp operator (dimension $4K$) yields the dominant contribution to the tail near $\omega\sim vq$. For $0<\Delta<1/2$ ($3/4<K<1$), the next-leading contribution is given by $\zeta_+$ (dimension 4). For $1/2<\Delta<1$ ($1/2<K<3/4$) it is important to include in the effective Hamiltonian the operator\be
\delta\mathcal{H}_{\lambda_2}=\lambda_2 \partial_x\theta \cos(4\sqrt{\pi K}\phi),
\ee
which is a descendant of the Umklapp operator and is allowed by all symmetries. This operator has dimension $4K+1$ and is less irrelevant than $\zeta_+$ for $K<3/4$. Another reason to include $\lambda_2$ is that for $K\leq 3/4$ the exact amplitude $\zeta_+$ in (\ref{zeta+h=0}) diverges at the points $K=1/2+1/(4n)$, $n\geq 1$. This divergence has been discussed in the context of corrections to the bulk and boundary susceptibility of the open XXZ chain \cite{SirkerBortz}. There it was found that the susceptibility as a function of magnetic field or temperature has a correction of first order in $\zeta_+$. However, the corrections at any given order of $h$ or $T$ are always finite because the divergences of $\zeta_+$ at the points $K=1/2+1/(4n)$ are cancelled by the contribution from the Umklapp operator and the cancellation gives rise to logarithm corrections. In our case the tail (\ref{tail_zeta+}) is of order $\zeta_+^2$, so it must be cancelled by a more irrelevant operator with a $K$-dependent dimension. Note also that the $O(\lambda_1^2)$ term in (\ref{betas_umklapp}) does not diverge. Therefore, in order to recover a finite high-frequency tail for all values of $0<\Delta<1$ it is necessary that the amplitude of the terms generated by $\lambda_2$ also diverge (have poles) at the above values of $K$. By simple power counting, we expect that the divergence of the $\zeta_+^2$ term at $K=1/2+1/(4n)$ will be cancelled by the term of second order in $\lambda_2$ and $2(n-1)$-th order in $\lambda_1$, which scales like $\delta S^{zz}\sim \lambda_1^{2(n-1)}\lambda_2^2 q^2(\omega^2-v^2q^2)^{n(4K-2)}$. For  $n(4K-2)=1$, this term has the same $q$ and $\omega$ dependence as in (\ref{tail_zeta+}) and the cancellation is thus possible. In principle it is possible to determine the amplitude $\lambda_2$ from the Bethe Ansatz, but that would require  solving the Wiener-Hopf equations to higher orders than was done in \cite{SirkerBortz}.

Computing the broadening $\delta\omega_q$ for $h= 0$ from bosonization is much more challenging. Since $\zeta_-$ is the only vertex present at the free fermion point, the naive expectation is that we could derive the renormalization of the width at zero field by summing all orders of $\zeta_-$, as we did for $\eta_-$ in section \ref{secwidthFT}. Although we now have to deal with a four-legged vertex, which introduces three-boson intermediate states, the calculation of the lowest order diagrams is not much harder than the finite field case. However, the fundamental difference is that for $h=0$ the broadening has to be produced by dimension-four operators and is therefore of the same order of $q$ as the changes in the lineshape ({\it i.e.}, the density of states factor and the singularities near the thresholds). That implies that the lineshape of $S^{zz}(q,\omega)$ for $\Delta\neq 0$ cannot be approximated by the free fermion result in the limit $q\rightarrow 0$. Therefore it is not clear what the expansion of bosonic diagrams should sum up to.

\begin{figure}
\begin{center} \includegraphics[%
  scale=0.5]{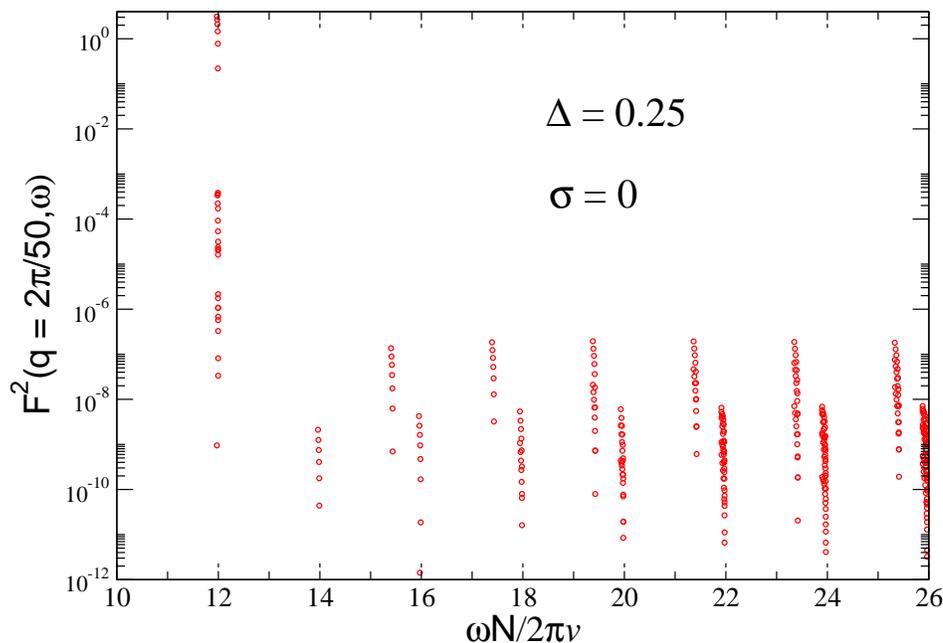}\end{center}

\caption{Numerical form factors squared for $\Delta=0.25$ and zero field. The chain length is $N=600$ and the momentum is set to $q=2\pi/50$.   \label{figffh=0}}
\end{figure}

Figure \ref{figffh=0} shows the form factors for $\Delta=0.25$ and $h=0$ calculated numerically by Bethe Ansatz for a chain of $N=600$ sites. In agreement with the field theory prediction, the states in the high-frequency tail cluster around the energy levels $\omega N/2\pi v=\ell$ (corresponding to the $\zeta_+$ contribution) and $\omega N/2\pi v=\ell+4K\approx \ell+3.45$ (the $ \lambda_1$ contribution). The comparison between the Bethe Ansatz data and the field theory results for the tail at zero field was done in \cite{pereira}, confirming the validity of (\ref{tail_zeta+}) and (\ref{betas_umklapp}) for $\delta\omega_q\ll \omega - vq\ll J$. An 
important consequence of integrability for the lineshape at zero field is that for $\Delta<1/2$ ($K>3/4$) the tail decreases as $\omega \to vq$. A finite $\zeta_3$ interaction would produce a contribution to the tail that diverges as $\omega\to vq$, similarly to the finite field case (see appendix). There is no such contribution in the Bethe Ansatz data for small $\Delta$. 

\begin{figure}
\begin{center} \includegraphics[%
  scale=0.5]{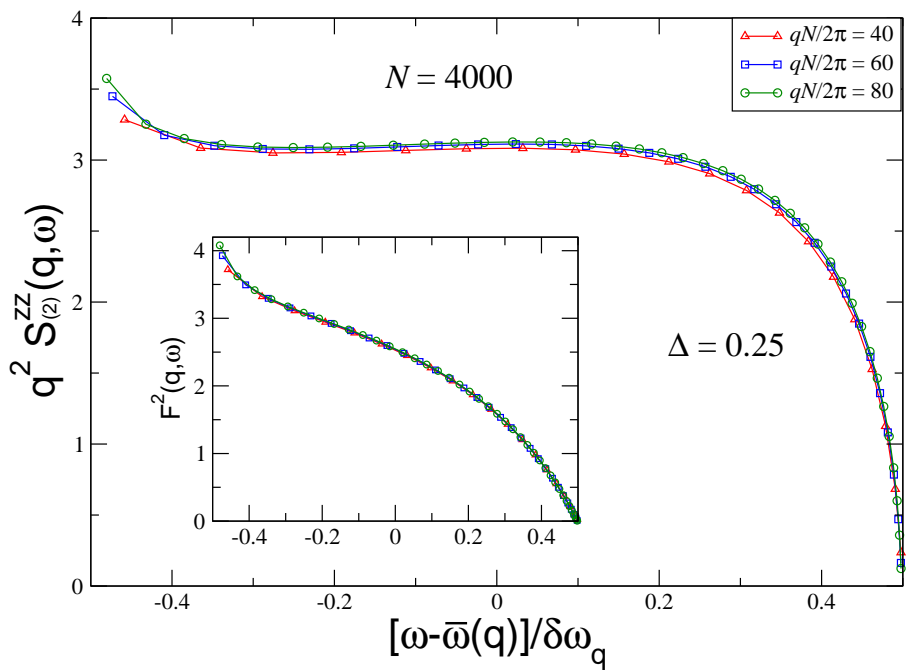}\end{center}

\caption{Two-spinon contribution to $S^{zz}(q,\omega)$ at zero field for $N=4000$, $\Delta=0.25$ and three values of $q$. The number of two-spinon states is given by $qN/4\pi$. Inset: transition probabilities $F^2(q,\omega)$ used to calculate $S^{zz}(q,\omega)$. We denote $\delta\omega_q=\omega_U(q)-\omega_L(q)$ and $\bar{\omega}(q)=[\omega_U(q)+\omega_L(q)]/2$. \label{figH0}}
\end{figure}

To study the peak region we can focus on the two-spinon states only (with $\omega\approx vq$ and form factors of $O(1)$) and reach lengths up to $N=4000$. We see that $F^2(q,\omega)$ is dominated by the two-spinon contribution, except very close to the upper threshold, where that contribution vanishes (inset of figure \ref{figH0}). Unlike the finite field case, the rescaled $F^2(q,\omega)$ does not become flat in the limit $q\to 0$. The density of states for the two-spinon states is known exactly \begin{equation}
D(q,\omega)=\frac{1}{\sqrt{\omega_U^2(q)-\omega^2}},\end{equation}where $\omega_U(q)=2v\sin(q/2)$. The two-spinon contribution to $S^{zz}(q,\omega)$ for zero field and $\Delta=0.25$ obtained by multiplying $F^2(q,\omega)$ by the above density of states is shown in figure \ref{figH0}.

We know from the exact solution for the two-spinon dynamical structure factor at the Heisenberg point $\Delta=1$ that there is a square-root divergence (with a logarithmic correction) at $\omega_L(q)$ and that the same contribution vanishes at $\omega_U(q)$  \cite{karbach}. Such behavior is completely opposite to what happens at the free fermion point (see \cite{muller}).  Numerical results suggest that the exponents change smoothly from $\Delta=0$ to $\Delta=1$, with spectral weight being transferred from the upper threshold to the lower threshold as $\Delta$ increases \cite{CauxJSTATP09003,SatoJPSJ73,CauxPRL95}.

In addition, the renormalization of $\delta\omega_q$ defined as the width of the two-particle continuum in the Bethe Ansatz solution is known exactly. The thresholds of the two-spinon continuum for $0<\Delta<1$ are a simple generalization of (\ref{h=0_lowerbound}) and (\ref{h=0upperbound}), with $J$ replaced by the renormalized velocity $v$ given by (\ref{eq:vBAzerofield}) \cite{KorepinBOOK,BiegelJPA36}. As a result, the width for finite $\Delta$ is\begin{equation}
\delta\omega_q\approx\frac{vq^3}{8}.\end{equation}
Since $v\approx J(1+2\Delta/\pi)$ for $\Delta\ll 1$, the above expression is different from the renormalization of $\zeta_-$. Therefore, the renormalization of $\delta\omega_q$ by interactions is \emph{not} given by $\zeta_-$.

A proper treatment of the dimension-four operators which allows to predict the lineshape of $S^{zz}(q,\omega)$ at zero field remains an open question.

\section{Sum rules and finite size effects}\label{sec:sumrules}

\subsection{Sum rules}
\label{SR} We checked the accuracy of the Bethe Ansatz data
by calculating the following sum rules \begin{equation}
I\left(q\right)\equiv\int_0^\infty \frac{d\omega}{2\pi}\, S^{zz}(q,\omega)=\frac{1}{N}\langle S^{z}_q S^{z}_{-q}\rangle\label{SR1}\end{equation}
 and \begin{equation}
L\left(q\right)\equiv\int_0^\infty \frac{d\omega}{2\pi}\,\omega S^{zz}(q,\omega)=-2\frac{\langle H_{xy}\rangle}{N}\sin^{2}\frac{q}{2}\;.\label{SR2}\end{equation}
These sum rules can be expressed in terms of sums over the form factors calculated by Bethe Ansatz for finite chains as \begin{eqnarray}I_{BA}(q)=\frac{1}{N}\sum_\alpha \left|\left\langle 0\left|S_{q}^{z}\right|\alpha\right\rangle \right|^{2}, \\ L_{BA}(q) = \frac{1}{N}\sum_\alpha \left(E_\alpha-E_{GS}\right) \left|\left\langle 0\left|S_{q}^{z}\right|\alpha\right\rangle \right|^{2}.\end{eqnarray} 

The identity in (\ref{SR2}) is a consequence of the \textit{f}-sum rule. The first moment sum rule $L(q)$ can then be calculated exactly by using the BA result for $\langle H_{xy}\rangle / N=2\langle S_j^{x} S_{j+1} ^{x}\rangle=e_0-\partial e_0/\partial \Delta$, where $e_0$ is the ground state energy per site \cite{Jimbo}.
Since there are no exact results for $I(q)$, we first compare $I_{BA}$ with the lowest-order field theory result  in (\ref{sumweight}) \begin{equation}
I(q)\approx I_{FT}(q)=\frac{K}{2\pi}q.\label{SR3}\end{equation}
This should be a reasonably good approximation for small $q$.

The other possibility is to calculate the static correlation function
by DMRG. In Table \ref{XX} we show DMRG results for periodic XXZ chains. The results
used the standard DMRG finite system method \cite{dmrg, dmrg2, dmrg3}, but with extra noise
terms added to the density matrix to speed convergence in 
the number of sweeps for the more difficult periodic boundaries case \cite{singlesite}.
We see that for these measurements the finite size effects are very small for $N=100$,
and that the truncation error depends significantly on $N$.
We can obtain results for $I(q)$ to an accuracy of $10^{-6}$ or $10^{-7}$ by
using the $m=2400$ results for $N=100$. Finite size corrections for larger $N$ appear
to be roughly the same size.

\begin{table}

\caption{DMRG results for $I(q)$, for $\Delta = 0.25$ and zero field ($\sigma=0$). The truncation error is $\varepsilon$
and $m$ is the number of states kept per block. Between 10 ($m=1200$) and 14 ($m=2400$)
sweeps were performed.\label{XX}}

\begin{indented}
\item[]\begin{tabular}{@{}ccccc}
\br 
$N$&
 $I(2 \pi/50)$&
 $I(2 \pi/25)$&
 $\varepsilon$&
 $m$\\
\mr
50 &	0.017237038&	0.034518387&	$7.5\times10^{-13}$&		1200\\
100&	0.017237380&	0.034518764&	$9.4\times10^{-11}$&		1200\\
100&    0.017237138&	0.034518522&	$2.1\times10^{-11}$&		1600\\
100&	0.017237109&	0.034518505&    $1.5\times10^{-12}$&		2400\\
200&	0.017237364&	0.034518657&	$1.2\times10^{-10}$&		2400\\
\br
\end{tabular}\end{indented}

\end{table}

A comparison between the sum rules obtained for the BA data and the
values expected from the equations above is shown in tables \ref{tab1} and \ref{tab2}. In all the cases shown here the Bethe Ansatz agrees with the DMRG and the exact results to better than $0.1\%$. 

\begin{table}[h]

\caption{Sum rules for $\Delta=0.25$ and zero field ($\sigma=0$). First sum rule: Results
for the BA data, $I_{BA}$, with a chain length of $N=400$ compared
with the field theory approximation (\ref{SR3}), $I_{FT}$, and results
from DMRG, $I_{DMRG}$, for a chain with $N=100$ sites. Second sum
rule: Results for the same BA data, $L_{BA}$, compared to the exact
result (\ref{SR2}), $L_{\mbox{\tiny exact}}$.\label{tab1}}

\begin{indented}
\item[]\begin{tabular}{@{}cccccc}
\br
 &
$I_{BA}$&
$I_{FT}$&
$I_{DMRG}$&
$L_{BA}$&
$L_{exact}$\\
\mr
 $q=\frac{2\pi}{50}$&
 0.017236 &
 0.017228 &
 0.017237 &
 0.002498 &
 0.002498 \\
$q=\frac{2\pi}{25}$&
 0.034513 &
 0.034457 &
 0.034518 &
 0.009952 &
 0.009953 \\
\br
\end{tabular}\end{indented}

\caption{Same sum rules as in table \ref{tab1}, but for finite field ($\sigma=-0.1$).\label{tab2}}

\begin{indented}
\item[]\begin{tabular}{@{}cccccc}
\br
 &
$I_{BA}$&
$I_{FT}$&
$I_{DMRG}$&
$L_{BA}$&
$L_{exact}$\\
\mr
 $q=\frac{2\pi}{50}$&
 0.017418 &
 0.0174 &
 0.017419 &
 0.002376 &
 0.002378\\
$q=\frac{2\pi}{25}$&
 0.034880 &
 0.0348 &
 0.034883 &
 0.009468 &
 0.009474\\
\br
\end{tabular}\end{indented}
\end{table}

\subsection{Size dependence of the $n$-particle contributions to $S(q,\omega)$
  within Bethe ansatz\label{sec:finitesizescal}} 
For any finite $N$, $S(q,\omega )$ is a sum of $\delta$-functions. However,
for $N\gg 1$, we expect to be able to approximate it by a continuous function
of $\omega$. One way this can be done is by ``binning''; i.e. we can define:
\begin{equation}{\mathcal S}(q,\omega )\equiv {2\pi/N\over \Delta \omega}\sum_{\alpha}\,' 
\left|\left\langle 0\left|S_{q}^{z}\right|\alpha\right\rangle \right|^{2}
,\label{binnedS}\end{equation}
where the sum is restricted to energies such that \begin{equation}
\omega -\Delta \omega /2<E_\alpha <\omega +\Delta \omega /2.\end{equation}
The bin size is chosen so that $2\pi/N \ll \Delta \omega/v \ll 2\pi$. For a fixed $N$, ${\mathcal S}$ is defined in (\ref{binnedS}) for discrete values of $q$. However, an extra binning could be defined for the wavectors as well, so that $q$ could be held fixed as we increase $N$.  Alternatively, we can use (\ref{binnedS}) and increase $N$ by an integer factor to reach the limit $N\rightarrow \infty$ in such a way that the ratio $qN/2\pi$ is always integer. In the large $N$ limit ${\mathcal S}(q,\omega )$ becomes a smooth nonzero function 
normalized so that:
\begin{equation}{1 \over N} \sum_q\int {d\omega \over 2\pi}{\mathcal S}(q,\omega )={1\over 4}.\end{equation}
We can define an approximation to this function, for large but finite $N$, 
${\mathcal S}_N(q,\omega )$, using a fixed (small) bin size. Furthermore, we 
can decompose this function into the $2n$-particle contributions: \footnote{Here, we call `particle' an excitation obtained
in the Bethe Ansatz by adding or removing a
quantum number from the ground state.  For the
sake of simplicity, we treat bound states in
a simplified manner and include their contribution
in the appropriate term (i.e.:  a two-string bound
state counts as two excitations).}
\begin{equation} {\mathcal S}_N(q,\omega )=\sum_{n=1}^\infty {\mathcal S}_N^{2n}(q,\omega ).\end{equation}
It is convenient to discuss the finite size behavior in terms of these binned
functions. While ${\mathcal S}_N(q,\omega )$ must have a finite limit as $N\to
\infty$, it is possible that each single contribution ${\mathcal
  S}^{2n}_N(q,\omega )$ vanishes. The question whether or not this is the case has
been addressed for the isotropic antiferromagnet in zero magnetic field. Here
it is known that two-spinon \cite{BougourziPRB54,BougourziPRB57} as well as the
four-spinon \cite{CauxJSTATP12013} contributions are finite in the thermodynamic
limit. For the $\Delta=0$ case, on the other hand, the two-spinon contribution
to the transverse structure factor $S^{+-}(q,\omega)$ vanishes in the
thermodynamic limit \cite{karbach}.

\begin{figure}
\begin{center}\includegraphics[%
  scale=0.5]{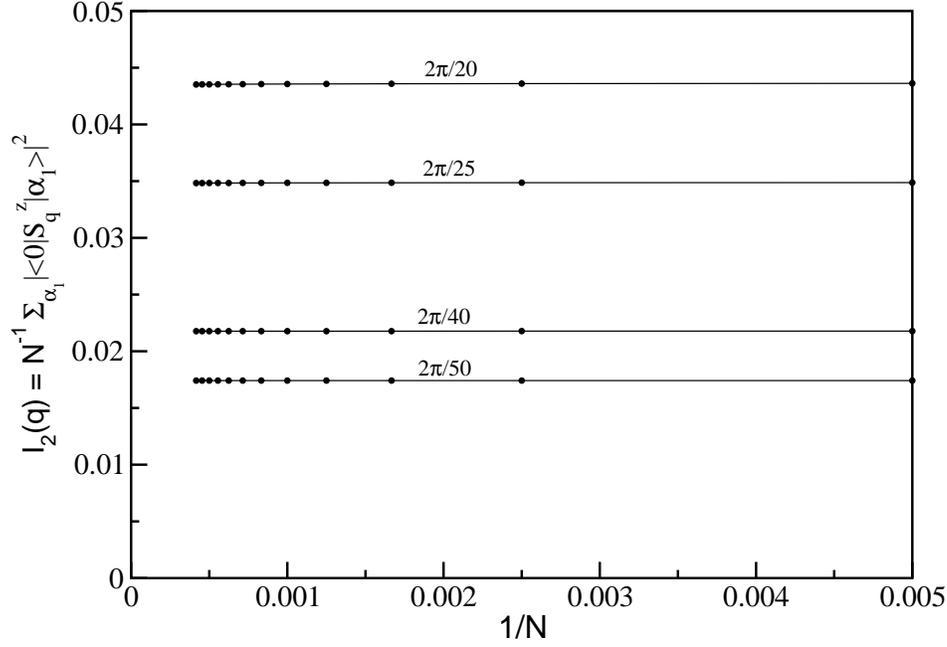}\end{center}

\caption{Sum of $F^2(q,\omega)$ for all single particle-hole as a function
of inverse length for different momenta $q=2\pi/20,\cdots,2\pi/50$.
Here $\Delta=0.25$ and $\sigma=-0.1$.\label{cap:Sum-of-all}}
\end{figure}

\begin{figure}
\begin{center} \includegraphics[%
  scale=0.5]{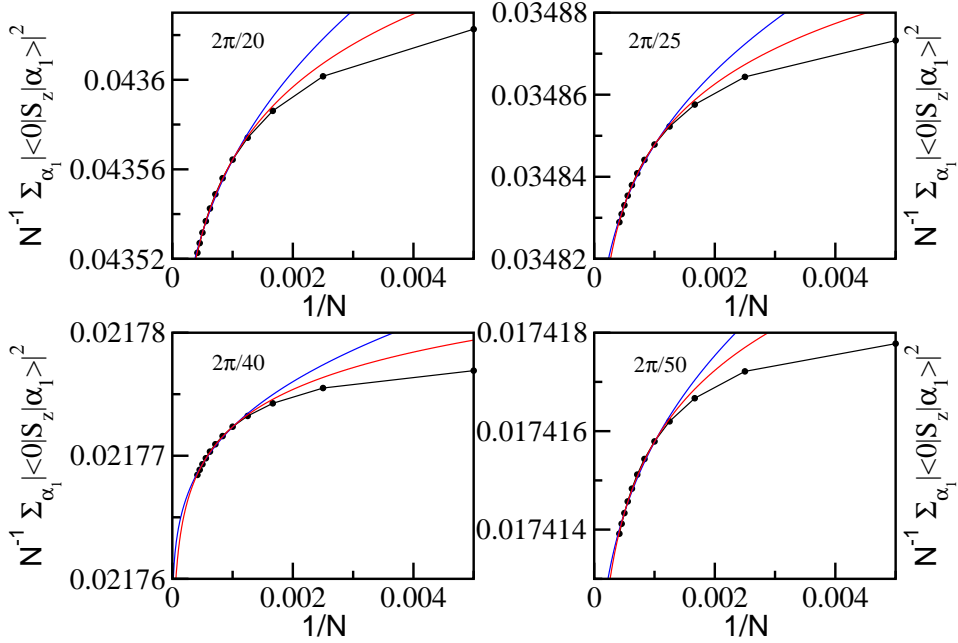}\end{center}

\caption{The same as figure \ref{cap:Sum-of-all}. The blue curves are fits
$\sim a+b/\ln(1/L)$ with fit parameters a,b. The fit parameters (for
decreasing $q$) are given by $a=0.04321,0.034687,0.021739,0.0174$ and
$b=-0.00245,-0.00111,-0.00023,-0.00011$. The red curves are fits
$\sim a\cdot x^{b}$ with fit parameters $a=0.0439,0.035,0.0218,0.0174$
and $b=0.00108,0.00061,0.00020,0.00012$.
\label{cap:The-same-as}}
\end{figure}

For the $XXZ$ chain in a finite magnetic field the form factors can only be
calculated numerically and the following discussion about the finite size
behavior of ${\mathcal S}^{2n}_N(q,\omega )$ has to be based on these numerical
data. In figure \ref{cap:Sum-of-all} we show the sum of all two-particle form
factors squared, denoted as $I_2(q)$, as a function of inverse length for
different fixed momenta $q$ and chain lengths up to $N=2600$. On this scale it
is hardly possible to detect any finite size effects at all. If we plot each
curve for a fixed $q$-value separately on a finer scale, however, we see that
the two-particle contribution is decreasing slightly with increasing length
(see figure \ref{cap:The-same-as}). By using extended precision arithmetics we
checked that this slow decrease of the two-particle contribution with
increasing length {\it is not} a numerical artefact. Based on the fits shown
in figure \ref{cap:The-same-as} it is impossible to decide whether the
two-particle contribution decreases (possibly logarithmically) to a finite value or
vanishes with a power law. If it does vanish with a power law, the exponent is
extremely small and apparently also $q$ dependent. 

Although we do not have a detailed explanation for the observed size
dependence, the following scenario seems plausible: Throughout the
gapless regime, two-particle (defined in the Bethe Ansatz sense of shifted
quantum numbers; these do not correspond to two-particle states in the field
theory) contributions vanish in the thermodynamic limit at finite field, but
are finite at zero magnetic field.  At finite field (in contrast to zero
field), there is room in the choice of quantum number configurations to
accommodate multiparticle excitations yielding solutions to Bethe equations in
terms of real rapidities. Within the gapless regime, backflows produced by
finite numbers of excitations at the Fermi boundary vanish. With the possibility of adding particles without modifying the state substantially, it is probable that the
contributions from 2, 4 and any finite number of excitations vanish in the thermodynamic limit.
 We can expect to get a good approximation for
the {\it exact} thermodynamic result by summing over {\it finite} numbers of
excitations for a {\it finite} chain, but for very large values of $N$ more and more families of states have to be included in the sum. In other words, although the
correlation weight distribution among families of excited states (which is not
an observable) might show some highly nontrivial size dependence, the full
correlation function (the true observable here), which is obtained by summing
over all states, shows a much weaker size dependence.  This is supported by the
analysis in \cite{KitanineNPB567} of the elementary blocks of the correlation
functions for the finite and infinite chains.  In the gapless regime all blocks
(from which any correlation function can be obtained) differ from their
thermodynamic limits by terms of order $1/N$.  

This scenario is corroborated by the agreement found when comparing the
numerical Bethe ansatz results to the field theory formulas for the
high-frequency tail as well as by the evaluation of the sum rules in the
previous subsection. The Bethe ansatz data in figure \ref{cap:Tail} correspond
to four-particle contributions for a chain of length $N=600$. 6, 8, ... and
higher particle contributions have been neglected. So the four-particle
contribution for a chain of length $N=600$ is apparently a good approximation
to the thermodynamic limit result.  If each n-particle contribution does
indeed vanish in the thermodynamic limit this suggests that it happens in such
way that $\mathcal{S}_N(q,\omega)$ does not change significantly in the tail
region. We call this the ``compensation scenario'' because when ${\mathcal
  S}^4_N$ starts to go to zero, ${\mathcal S}^6_N$ comes in and ``compensates"
${\mathcal S}^4_N$.  Similar compensations occur with ${\mathcal S}^8_N$,
${\mathcal S}^{10}_N$, etc.

Concerning the peak region, we also believe that ${\mathcal S}^2_N$ as shown
in figure \ref{truelineshape} is a good approximation of the lineshape in the
thermodynamic limit. It nearly saturates the sum rules indicating that even
for $N=6000$ the contribution of higher particle states are negligible.
Furthermore, the width of the peak obtained from ${\mathcal S}^2_N$ in section
\ref{WidthBA} does agree with the field theory result in section
\ref{secwidthFT}. If ${\mathcal S}^2_N$ indeed vanishes for $N\to\infty$ it
again seems to get compensated by higher particle excitations in such a way
that the lineshape does not change significantly. A notable exception might
occur very close to $\omega_{L,U}(q)$ where the methods of \cite{pustilnik2}
predict a very singular form for ${\mathcal S}(q,\omega )$ caused by large
numbers of excitations very close to the Fermi energy. It is difficult
to determine whether or not $S^2_N(q,\omega )$ approaches this form at large
$N$ and it seems to be possible that ${\mathcal S}_N(q,\omega )$ could
continue to change near $\omega_{L,U}(q)$, becoming more singular, out to very
large $N$.

\section{Conclusion}

Based on a low-energy effective theory which includes the leading (band
curvature type) irrelevant operators we have 
studied the longitudinal dynamical structure factor $S^{zz}(q,\omega)$ for the
$XXZ$ spin-1/2 chain in a magnetic field. By comparing with results for free
fermions we have conjectured a method to sum up the entire perturbation series
in one of these irrelevant operators allowing us to obtain an approximation for the shape of the
peak of $S^{zz}(q,\omega)$ which is valid for small $q$ and is non-perturbative  in the interaction strength (anisotropy $\Delta$). Besides the velocity $v$ and Luttinger parameter $K$, the important parameters to determine the lineshape are the coupling constants $\eta_{\pm}$ of the leading (dimension-three) irrelevant operators, which we relate to derivatives of $v$ and $K$ with respect to the magnetic field. A
summation of the entire series is necessary because perturbation theory in
the band curvature terms is divergent on shell, $\omega\sim vq$, although
these operators are formally irrelevant. Our field theory approach is valid in the regime $\tilde{\gamma}q^2\ll\eta_-q\ll v$, where $\tilde{\gamma}$ is of order of the coupling constants of the next-leading (dimension-four) irrelevant operators which we neglected for the case of a finite magnetic field. The result is a box shaped peak with width $\delta\omega_q=|\eta_-|q^2$ and height $K/|\eta_-|q$, similar to the exact $S^{zz}(q,\omega)$ for the XX model (free fermion point). 

The field-theoretical results for the width of the peak are supported by Bethe Ansatz calculations. Since the $XXZ$ model is integrable, we used the Bethe Ansatz equations in the thermodynamic limit to determine the parameters $\eta_{\pm}(\Delta,\sigma)$ numerically, so that the low-energy effective theory and the obtained results for
$S^{zz}(q,\omega)$ are {\it parameter free}.  We
have shown that the width of the peak obtained in field theory agrees with the
analytically calculated width of the two-particle continuum in the Bethe
Ansatz. Furthermore, we have demonstrated that the form factors obtained numerically by Bethe Ansatz approach the flat distribution predicted by field
theory for $q\to 0$. Applying our results to the strongly interacting  case (large $\Delta$), we found that for $\Delta >\cos(\pi/5)\approx 0.81$ the parameter $\eta_-$ goes through zero for a finite value of the magnetic field. At the ``inversion point" where $\eta_-(\Delta,\sigma_{inv})=0$ the $q^2$ scaling breaks down and the width of the two-particle continuum scales like $\delta\omega_q\sim q^3$. The $q^2$ scaling is recovered for $0<|\sigma|<\sigma_{inv}$. As a result, the width $\delta\omega_q$ is a non-monotonic function of $\sigma$, with a minimum at the inversion point.

The power-law singularities found in \cite{pustilnik2}
for $\Delta\ll 1$ near the lower and upper thresholds $\omega_L(q),\,
\omega_U(q)$ are not captured by our calculations. Within the
effective low-energy theory these singularities seem to be related to higher
dimension operators.  Within the Bethe Ansatz, on
the other hand, it is not clear if these singularities can be obtained by
considering only the form factors for two-particle states. It seems possible that the finite
size effects near these boundaries are complicated and form factors for
multi-particle excitations at very large system sizes have to be studied. Nevertheless, for finite chains and small $\Delta$ the behavior of the dominant form factors for the two-particles states   agrees qualitatively with the result of \cite{pustilnik2}. However, taking into account the energy dependence of the density of states leads to a maximum and a minimum of $S^{zz}(q,\omega)$ inside the two-particle continuum.  In the strongly interacting regime $\Delta > \cos(\pi/5)$, we have found that for $|\sigma|<\sigma_{inv}$ the dynamical structure factor exhibits a rather distinct lineshape,  reminiscent of a converging X-ray singularity at the lower threshold.

We also showed that in the interacting case $S^{zz}(q,\omega)$
has a high-frequency tail $\delta S^{zz}(q,\omega)$. Within the effective
theory this tail is related to an irrelevant operator (with coupling constant $\eta_+$) mixing excitations at
the right and left Fermi points. Contrary to the calculation for the on-shell
region, this term can be treated in finite-order perturbation theory for
$\delta\omega_q\ll \omega - vq\ll J$ and we find that the tail for finite field decays as
$\delta S^{zz}(q,\omega)\sim q^4/(\omega^2-v^2q^2)$. This result is again
supported by numerical calculations based on the Bethe ansatz.

We have proposed that the integrability of the XXZ model is manifested in the low-energy effective Hamiltonian at the order of the dimension-four, band curvature type operators. The conservation of the energy current operator imposes that the interaction denoted as $\zeta_3$ is absent. This has consequences for the lineshape of $S^{zz}(q,\omega)$ at zero magnetic field, since a nonzero $\zeta_3$ would change the behavior of the tail near the upper threshold of the two-particle continuum.

One promising test of our theory would be to measure, by means of inelastic neutron scattering experiments, the width of the peak as a function of $q$ and $h$ (equations (\ref{widtheta_-}) and (\ref{eq:identity1})). In some spin-1/2 compounds it is experimentally possible to go all the way up to the saturation field. The main limitation is the low intensity of the signal for small-$q$ scattering. However, one important point that may facilitate the experiment is that in the transverse channel the low energy
spectral weight is shifted to a finite wave-vector, $\pm \sigma$ (the
magnetization). The gap at $q=0$ is of order $h$ for transverse excitations. 
So there is a ``protected region" of zero transverse spectral  weight at
small $q$ and $\omega$ inside of which the longitudinal structure function
could perhaps be observed.

Finally, we would like to emphasize that the formulas (\ref{eq:identity1}) and (\ref{eq:identity2}) for the coupling constants of the irrelevant operators are also valid for nonintegrable models. This allows us to predict the width of the dynamical structure factor $S^{zz}(q,\omega)$ at small $q$ once the field dependence of $v$ and $K$ is determined from thermodynamic quantities. For example, the value of $\sigma_{inv}$, below which we expect to see nontrivial effects due to strong interactions, can be increased by adding a ferromagnetic next-nearest neighbour interaction. The isotropic $J_1-J_2$ model also contains a marginally irrelevant operator, whose amplitude can be tuned by the $J_2$ interaction. A ferromagnetic $J_2$ ($J_2<0$) would increase the constant $\sigma_0$ inside the logarithm in equation (\ref{divergingetas}). On more general grounds, the nonmonotonic behavior of the width $\delta\omega_q$ and the inversion of the lineshape should occur whenever the derivatives of the velocity and the Luttinger parameter with respect to magnetic field/chemical potential have opposite signs and the latter one is singular. In principle, this could also be observed in the dynamical structure factor of quantum wires, since $\partial K/\partial n$, where $n$ is the electron density, changes sign and diverges in the low-density limit (Wigner crystal regime) \cite{hausler}.  This suggests that the evolution of the lineshape as a function of density could be richer than what was proposed in \cite{pustilnik2}.

Open questions that we leave for future work include the behavior of
$S^{zz}(q,\omega)$ near the upper and lower threshold, which will require a
careful analysis of the interplay between more irrelevant operators in the
effective model, a field theory calculation of the width for zero field, the thermodynamic limit of the form factors calculated in the Bethe Ansatz and further effects of integrability on the lineshape.

\ack
We thank C. Broholm, R. Coldea, L. I. Glazman,  M. Pustilnik and D. A. Tennant for helpful discussions. This research was supported by CNPq (Brazil) through Grant No. 200612/2004-2 (R.G.P.), the DFG (J.S.), FOM (J.-S.C.), CNRS and the EUCLID network (J.M.M.), the NSF under No. DMR-060544 (S.R.W.), and NSERC (R.G.P., J.S., I.A.) and the CIfAR (I.A.).

\appendix 
\section{High-frequency tail for the zero field case}
In this appendix we derive the results (\ref{tail_zeta+}) and (\ref{betas_umklapp}). 

\subsection*{Tail from $\zeta_+$ interaction}
\begin{figure}
\begin{center} \includegraphics[%
  scale=0.9]{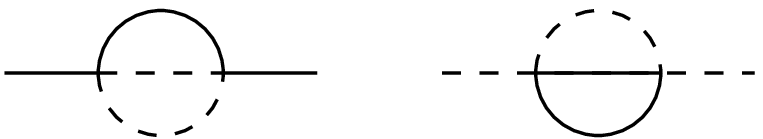}\end{center}

\caption{Diagrams at $O(\zeta_+^2)$ for the calculation of the tail.
\label{zeta+diagrams}}
\end{figure}

Since the $\zeta_+$ vertex in (\ref{deltaH_h=0}) has two $R$ and two $L$ legs, the correction to $\chi(q,i\omega)$ is separable into $\delta\chi=\delta\chi_{RR}+\delta\chi_{LL}$, where\be
\delta\chi_{RR}(q,i\omega)=\frac{K}{2\pi}\left[D^{(0)}_R(q,i\omega)\right]^2\Pi_{RLL}(q,i\omega).\label{XiRRzeta+}
\ee
$\Pi_{RLL}$ is the bubble with one right- and two left-moving bosons (first diagram of figure \ref{zeta+diagrams}) given by\be
\Pi_{RLL}(q,i\omega)=-2\pi^2\zeta_+^2\int_0^Ldx \, e^{-iqx}\int_0^\beta d\tau e^{i\omega\tau} D^{(0)}_R(x,\tau)\left[D^{(0)}_L(q,\tau)\right]^2.\label{PiRLL}
\ee
The expression for $\delta\chi_{LL}$ is obtained from (\ref{XiRRzeta+}) and (\ref{PiRLL}) by exchanging $R\leftrightarrow L$. After doing the Fourier transform and integrating over the internal frequencies, we find\be
\Pi_{RLL}(q,i\omega)=\frac{2\pi^2\zeta_+^2}{L^2}\sum_{k_1,k_2>0}\frac{k_1k_2(q+k_1+k_2)}{i\omega-vq-2v(k_1+k_2)},
\ee
where $k_{1,2}=2\pi n_{1,2}/L$, with $n_{1,2}$ integers. Taking the imaginary part of the retarded self-energy, we have \bea
-\textrm{Im }\Pi^{ret}_{RLL}(q,\omega)&=&\frac{2\pi^3\zeta_+^2}{vL^2}\left(\frac{2\pi}{L}\right)^2\nonumber\\& &\times\sum_{n_1,n_2>0}n_1n_2(n+n_1+n_2)\delta(\ell-n-2n_1-2n_2),\label{sumnis}
\eea
where we have used $q=q_n=2\pi n/L$ and $\omega=\omega_\ell=2\pi v\ell/L$. Notice that this implies that the energy levels in the tail are discrete and separated by $4\pi v/L$. We evaluate the sum on the righthand side of (\ref{sumnis}) as follows\bea
& &\sum_{n_1,n_2>0}n_1n_2(n+n_1+n_2)\, \delta(\ell-n-2n_1-2n_2)\nonumber\\
& &=\sum_{m=1}^\infty(n+m)\, \delta(\ell-n-2m)\sum_{n_1=0}^mn_1(m-n_1)\nonumber\\
& &=\sum_{m=1}^\infty\frac{m^3(n+m)}{6}\left(1-\frac{1}{m^2}\right)\, \delta(\ell-n-2m)\nonumber\\& &={1\over 6}\sum_{\ell}\left(\frac{\ell-n}{2}\right)^3\left(\frac{\ell+n}{2}\right)\left[1-\left(\frac{2}{\ell-n}\right)^2\right]\frac{2\pi v}{L} \delta(\omega-\omega_\ell),\label{resultsum}
\eea
with $\omega_\ell=2\pi v\ell/L$, $\ell=n+2,n+4,\dots$ . Substituting (\ref{resultsum}) in (\ref{sumnis}) and using (\ref{XiRRzeta+}), we find \bea
-2\textrm{Im}\chi^{ret}_{RR}(q,\omega)&=&\frac{K\zeta_+^2}{192v^2}\left(\frac{2\pi}{L}\right)^5\nonumber \\ & &\times\sum_{\ell} n^2\left(\ell^2-n^2\right) \left[1-\left(\frac{2}{\ell-n}\right)^2\right]\delta(\omega-\omega_\ell).
\eea
Likewise, we have
\bea
-2\textrm{Im}\chi^{ret}_{LL}(q,\omega)&=&\frac{K\zeta_+^2}{192v^2}\left(\frac{2\pi}{L}\right)^5\nonumber\\& &\times \sum_{\ell}n^2\left(\ell^2-n^2\right)\left[1-\left(\frac{2}{\ell+n}\right)^2\right] \delta(\omega-\omega_\ell).
\eea

Finally, the contribution of the $\zeta_+$ interaction to the high-frequency tail is\be
\delta S^{zz}_{\zeta_+}(q,\omega)=\frac{2\pi}{L}\sum_{\ell} F_{\zeta_+}^2(q_n,\omega_\ell)\delta(\omega-\omega_\ell),
\ee
with \be
F_{\zeta_+}^2(q_n,\omega_\ell)=\frac{K\zeta_+^2}{96v^2}\left(\frac{2\pi}{L}\right)^4 n^2\left(\ell^2-n^2\right)\left[1-\frac{2}{(\ell-n)^2}-\frac{2}{(\ell+n)^2}\right].
\ee
In the thermodynamic limit $L\to\infty$ (and $\ell\pm n\gg 1$), we obtain\be
\delta S^{zz}_{\zeta_+}(q,\omega)=\frac{K(\zeta_+/v)^2}{192v}q^2\left(\frac{\omega^2-v^2q^2}{v^2}\right)\theta(\omega-vq).
\ee

\subsection*{Tail from $\lambda_1$ interaction}
The perturbation theory in the Umklapp interaction for a finite system requires that we treat the zero mode operators. So consider the expressions for spin operators\begin{eqnarray}
S_{j}^{z} & \sim & \sqrt{\frac{K}{\pi}}\partial_{x}\phi+\left(-1\right)^{j}\textrm{const}\times\cos\left(\sqrt{4\pi K}\phi\right),\label{eq:Sz}\\
S_{j}^{-} & \sim & \textrm{const}\times e^{-i\sqrt{\pi/K}\theta}\left[\left(-1\right)^{j}+\cos\left(\sqrt{4\pi K}\phi\right)\right].\end{eqnarray}
Periodic boundary conditions for
the spin operators imply that we can regard $\phi$ and $\theta$
as compactified fields with radius $R=\left(1/4\pi K\right)^{1/2}$
and $\tilde{R}=\left(K/\pi\right)^{1/2}$, respectively. In general,
we can have \begin{eqnarray}
\phi\left(x+L\right) & = & \phi\left(x\right)+\mathcal{S}^{z}\sqrt{\frac{\pi}{K}}\\
\theta\left(x+L\right) & = & \theta\left(x\right)+m\sqrt{4\pi K},\end{eqnarray}
where $\mathcal{S}^{z}$ and $m$ are integers. 
In a finite system with periodic boundary conditions we use the mode expansion for the bosonic fields
\begin{eqnarray}
\phi\left(x,t\right) & = & \phi_{0}+\Pi_{0}\frac{vt}{L}+Q_0\frac{x}{L}\nonumber \\
 &  & +\sum_{n>0}\frac{1}{\sqrt{2q_nL}}\left[-a_{n}^{R}e^{-iq_n(vt-x)}+a_{n}^{L}e^{-iq_n(vt+x)}+h.c.\right],\label{eq:modeexpansion}
\end{eqnarray}
where $q_n=2\pi n/L$. 
The operators $\phi_{0}$ and $\Pi_{0}$
are associated with the zero mode and satisfy $\left[\phi_{0},\Pi_{0}\right]=i$.
The compactification of $\phi$ quantizes the eigenvalues of $Q_0$ to be $\mathcal{S}^z\sqrt{\pi/K}$. It follows from Eqs. (\ref{eq:Sz}) and (\ref{eq:modeexpansion})
that $\mathcal{S}^{z}$ corresponds to the total spin in the chain.
We shall be restricted to the subspace $\mathcal{S}^{z}=0$, to which
the ground state for even $L$ belongs. From $\partial_{t}\phi=v\partial_{x}\theta$,
we get\begin{eqnarray}
\theta\left(x,t\right) & = & \theta_{0}+\Pi_{0}\frac{x}{L}+Q_0\frac{vt}{L}\nonumber\\
 &  & +\sum_{n>0}\frac{1}{\sqrt{2q_nL}}\left[a_{n}^{R}e^{-iq_n(vt-x)}+a_{n}^{L}e^{-iq_n(vt+x)}+h.c.\right],\end{eqnarray}
with $[\theta_0,Q_0]=i$. The eigenvalues of $\Pi_0$ are then $m\sqrt{4\pi K}$, $m$ integer. Therefore, for
the Hamiltonian\begin{eqnarray}
H & = & \frac{v}{2}\int dx\,\left[\left(\partial_x \theta\right)^{2}+\left(\partial_{x}\phi\right)^{2}\right]\\
 & = & \frac{v}{2L}\left(\Pi_{0}^{2}+Q_0^2\right)+\frac{2\pi v}{L}\sum_{n>0}n\left[a_{n}^{R\dagger}a_{n}^{R}+a_{n}^{L\dagger}a_{n}^{L}\right],\end{eqnarray}
we obtain the spectrum ($\mathcal{S}^z=0$)\begin{equation}
E=\frac{2\pi v}{L}\left[m^{2}K+\sum_{n>0}n\left(m_{n}^{R}+m_{n}^{L}\right)\right],\end{equation}
where $m_{n}^{R,L}=0,1,2,\dots$. The corresponding wave function
is\begin{equation}
\left|\Psi\right\rangle =\exp\left[im\sqrt{4\pi K}\phi_{0}\right]\prod_{n>0}\left(a_{n}^{R\dagger}\right)^{m_{n}^{R}}\left(a_{n}^{L\dagger}\right)^{m_{n}^{L}}\left|0\right\rangle .\end{equation}
Since translation by one site takes $\phi\to\phi+\pi R$ and $\left|\Psi\right\rangle\to (-1)^m\left|\Psi\right\rangle$ \cite{eggert}, this symmetry implies that only intermediate states with even $m$ couple to the ground state via $S_q^z$.  

For the Umklapp interaction defined in (\ref{deltaH_h=0}), the $O(\lambda_1^2)$ correction to $\chi(q,i\omega)$ is\begin{eqnarray}
\delta\chi\left(q,i\omega\right) & = & -\frac{K}{8\pi}\left(\frac{\lambda_1}{2\pi}\right)^2\int_0^Ldx\, e^{-iqx}\int_0^\beta d\tau \, e^{i\omega\tau}\int d^{2}x_{1}\int d^{2}x_{2}\nonumber\\ & & \times \left\langle \partial_{x}\phi(x)e^{i4\sqrt{\pi K}\phi(1)}e^{-i4\sqrt{\pi K}\phi(2)}\partial_{x}\phi(0)\right\rangle +(1\leftrightarrow 2).\label{orderlambda2}\end{eqnarray}
Following \cite{Oshikawa} we can show that \begin{eqnarray}
& & \left\langle \phi\left(x\right)\phi\left(0\right)e^{i4\sqrt{\pi K}\phi\left(1\right)}e^{-i4\sqrt{\pi K}\phi\left(2\right)}\right\rangle_{\textrm{con}}\nonumber\\&= & 16\pi K\left[\left\langle \phi(x)\phi(1)\right\rangle \left\langle \phi(0)\phi(1)\right\rangle-\left\langle \phi(x)\phi(1)\right\rangle \left\langle \phi(0)\phi(2)\right\rangle+(1\leftrightarrow2)\right]\nonumber\\ & &\qquad\times  \left\langle e^{i4\sqrt{\pi K}\phi(1)}e^{-i4\sqrt{\pi K}\phi(2)}\right\rangle .\end{eqnarray}
As a result, $\delta\chi$ can be cast in the form\be
\delta\chi\left(q,i\omega\right)=2\left(\frac{\lambda_1 K}{2\pi}\right)^2\left[\frac{D^{(0)}(q,i\omega)}{q}\right]^2\left[\Pi(q,i\omega)-\Pi(0,0)\right],
\ee
where \be
\Pi(q,i\omega)=-\int_0^Ldx\, e^{-iqx}\int_0^\beta d\tau\, e^{i\omega\tau}  \left\langle e^{i4\sqrt{\pi K}\phi(x,\tau)}e^{-i4\sqrt{\pi K}\phi(0,0)}\right\rangle.
\ee

The correlation function $\Pi(x,\tau)= \left\langle e^{i4\sqrt{\pi K}\phi(x,\tau)}e^{-i4\sqrt{\pi K}\phi(0,0)}\right\rangle$ for a finite system has to be calculated using the mode expansion (\ref{eq:modeexpansion}) including the zero mode. Note that the operators in (\ref{orderlambda2}) couple the
ground state to states with $m=\pm2$, since we are calculating matrix
elements of the form\be
\left\langle \alpha\left|\partial_{x}\phi\, e^{i2\sqrt{4\pi K}\phi_{0}+\cdots}\right|0\right\rangle .\ee
Since $e^{i4\sqrt{\pi K}\phi(z,\bar{z})}$ is a primary field of holomorphic weight $(2K,2K)$, the correlation function in the infinite complex plane is given by\be
\Pi(z,\bar{z})=\left(\frac{1}{z}\right)^{4K}\left(\frac{1}{\bar{z}}\right)^{4K}.
\ee
We use the ``CFT normalization condition" of \cite{lukyanov}. The correlation function for a finite system is obtained using the conformal mapping $z=e^{2\pi\xi/L}$, $\bar{z}=e^{2\pi\bar{\xi}/L}$, where $\xi=v\tau+ix$ and $\bar{\xi}=v\tau-ix$, with $0<x<L$. The result is \be
\Pi(x,\tau)=\Pi(\xi,\bar{\xi})=\left[\frac{\pi/L}{\sin\pi (x-iv\tau)/L}\right]^{4K}\left[\frac{\pi/L}{\sin\pi (x+iv\tau)/L}\right]^{4K}.
\ee
In order to calculate $\textrm{Im }\Pi^{ret}$ at zero temperature, we switch back to real time with the prescription $iv\tau\to vt-i\alpha$, $\alpha\to0^+$. We then calculate\bea
\Pi(q,\omega)&\equiv&-i\int_0^Ldx\, e^{-iqx}\int_{-\infty}^{+\infty}dt \,e^{i\omega t}\left[\frac{\pi/L}{\sin\pi (x-vt+i\alpha)/L}\right]^{4K}\nonumber\\& &\times\left[\frac{\pi/L}{\sin\pi (x+vt-i\alpha)/L}\right]^{4K},
\eea
which has the property $\Pi(q,\omega)=2i\textrm{Im}\Pi^{ret}(q,\omega)$. We also use the fact that for a periodic function with discrete modes $q_{n}=2\pi n/L$\be
f\left(x\right)=\sum_{n}\frac{f_{n}}{L}e^{i2\pi nx/L}.\ee
We have \begin{eqnarray}
\int_{-\infty}^{+\infty}dx\, e^{-iqx}f\left(x\right) & = & \sum_{n}f_{n}\frac{1}{L}\int_{-\infty}^{+\infty}dx\, e^{-i(q-2\pi n/L)x}\nonumber\\
 & = & \sum_{n}f_{n}\delta\left(\frac{qL}{2\pi}-n\right)=f\left(q\right)\sum_{n}\delta\left(\frac{qL}{2\pi}-n\right).\end{eqnarray}
So we will consider the integral on the entire plane and eventually
cancel a sum over delta functions for the discrete momenta. Performing a change of variables\begin{eqnarray}
\tilde{\Pi}\left(q,\omega\right)&=&\Pi(q,\omega) \sum_{n}\delta\left(\frac{qL}{2\pi}-n\right)\nonumber\\& = & -\frac{i}{2v}\left(\frac{\pi}{L}\right)^{8K}\int_{-\infty}^{+\infty} dx_{+}e^{i\left(\omega-vq\right)x_{+}/2v}\left[\sin\frac{\pi\left(x_{+}-i\alpha\right)}{L}\right]^{-4K}\nonumber\\
 &  & \times\int_{-\infty}^{+\infty} dx_{-}e^{-i\left(vq+\omega\right)x_{-}/2v}\left[\sin\frac{\pi\left(x_{-}+i\alpha\right)}{L}\right]^{-4K},\end{eqnarray}
where $x_\pm\equiv x\pm vt$, we are left with integrals of the form\bea
I_{1}&=&\int_{-\infty}^{+\infty}du\, e^{ir u}\left[\sin\frac{\pi\left(u-i\alpha\right)}{L}\right]^{-4K}\nonumber \\&=&-\theta\left(r\right)\sum_{n=-\infty}^{+\infty}\int_{BC_{n}}dz\, e^{ir z}\textrm{Disc}\left[\sin\frac{\pi\left(z-i\alpha\right)}{L}\right]^{-4K},\eea
where $BC_{n}$ is the branch cut $z=nL+i\alpha+iy$, $0<y<\infty$ and $\textrm{Disc}\,f(z)\equiv f(z-0^-)-f(z+0^+)$ is the discontinuity of the function across the branch cut.
By shifting to $z^{\prime}=z-nL$, we get \be
I_{1}=-\theta\left(r\right)\sum_{n}e^{ir nL}e^{-in4\pi K}\int_{BC_{0}}dz\, e^{ir z}\textrm{Disc}\left[\sin\frac{\pi\left(z-i\alpha\right)}{L}\right]^{-4K}.\ee
We have\be
\textrm{Disc}\left[\sinh\frac{\pi\left(z-i\alpha\right)}{L}\right]^{-4K}=\left|\sinh\frac{\pi y}{L}\right|^{-4K}2i\sin4\pi K.\ee
Then\bea
I_{1}&=&-\theta\left(r\right)2i\sin4\pi K\, e^{i4\pi K}\left[\sum_{n}e^{in\left(r L-4\pi K\right)}\right]ie^{-r\alpha}\nonumber\\& &\times\int_{0}^{\infty}dy\, e^{-r y}\left|\sinh\frac{\pi y}{L}\right|^{-4K}.\eea
We use\be
\sum_{n}e^{in\left(r L-4\pi K\right)}=\sum_{m}\delta\left(\frac{r L-4\pi K}{2\pi}-m\right).\ee
and (from \cite{schulz})\be
\int_{0}^{\infty}ds\left[\sinh\left(\pi Ts\right)\right]^{-4K}e^{isz}=\frac{2^{4K-1}}{\pi T}B\left(2K-i\frac{z}{2\pi T},1-4K\right),\ee
where $B(x,y)=\Gamma(x)\Gamma(y)/\Gamma(x+y)$ is the Euler Beta function, and finally get\bea
I_{1}&=&\frac{2^{4K}L}{\pi}\,\sin(4\pi K)\,\theta\left(r\right)B\left(2K+\frac{r L}{2\pi},1-4K\right)\nonumber \\ & &\times\sum_{m}\delta\left(\frac{r L-4\pi K}{2\pi}-m\right).\eea
Likewise, the integral\be
I_{2}=\int_{-\infty}^{+\infty}du\, e^{-i\tilde{r}u}\left[\sin\frac{\pi\left(u+i\alpha\right)}{L}\right]^{-4K}\ee
is given by\bea
I_{2}&=&\frac{2^{4K}L}{\pi}\,\sin(4\pi K)\, \theta\left(\tilde{r}\right)B\left(2K+\frac{\tilde{r}L}{2\pi},1-4K\right)\nonumber\\ & &\times\sum_{m^{\prime}}\delta\left(\frac{\tilde{r}L-4\pi K}{2\pi}-m^{\prime}\right).\eea
In our case, $r=\left(\omega-vq\right)/2v$ and $\tilde{r}=\left(\omega+vq\right)/2v$.
The two delta functions can be recombined to replace the second condition by $q=q_n=2\pi n/L$. \bea
& &\sum_{m}\delta\left(\frac{r L-4\pi K}{2\pi}-m\right)\sum_{m^{\prime}}\delta\left(\frac{\tilde{r}L-4\pi K}{2\pi}-m^{\prime}\right)\nonumber\\& &=\sum_{m}\delta\left(\frac{\left(\omega-vq\right)L}{4\pi v}-m-2K\right)\sum_{n}\delta\left(\frac{qL}{2\pi}-n\right).\eea
We can then cancel the second delta function and write\begin{eqnarray}
\Pi\left(q,\omega\right) & = & -4i\left(\frac{2\pi}{L}\right)^{8K-2}\sin^{2}(4\pi K) B\left(2K+\frac{\left(\omega-vq\right)L}{4\pi v},1-4K\right)\nonumber\\
 &  & \times  B\left(2K+\frac{\left(\omega+vq\right)L}{4\pi v},1-4K\right)\frac{2\pi}{L}\sum_{\ell}\delta\left(\omega-\omega_{\ell}\right),\end{eqnarray}
where $\omega_\ell=2\pi v(\ell+4K)/L$, $\ell=n,n+2,\dots $. Finally, the contribution from the Umklapp operator to the high-frequency tail is\be
\delta S^{zz}_{\lambda_1}(q,\omega)=\frac{2\pi}{L}\sum_{\ell} F_{\lambda_1}^2(q_n,\omega_\ell)\delta(\omega-\omega_\ell),
\ee
with\bea
F_{\lambda_1}^2(q_n,\omega_\ell) & = & 2\left(\frac{2\lambda_1K}{\pi v}\right)^2\left(\frac{2\pi}{L}\right)^{8K-4}\sin^{2}(4\pi K)\frac{n^2}{\left(\ell^2-n^2\right)^2} \times \nonumber\\& &B\left(4K+\frac{\ell-n}{2},1-4K\right) B\left(4K+\frac{\ell+n}{2},1-4K\right).
\eea
In the limit $L\rightarrow\infty$, we can use $B(x,y)\sim\Gamma(y)x^{-y}$
for $x\to\infty$ and we find\be
\delta S^{zz}_{\lambda_1}(q,\omega)=\frac{2\lambda_1^2K^2}{\Gamma^2(4K)}(2v)^{3-8K}q^2\left(\omega^2-v^2q^2\right)^{4K-3}\theta(\omega-vq).
\ee

\subsection*{Infrared-divergent tail for $\zeta_3\neq 0$}
\begin{figure}
\begin{center} \includegraphics[%
  scale=0.9]{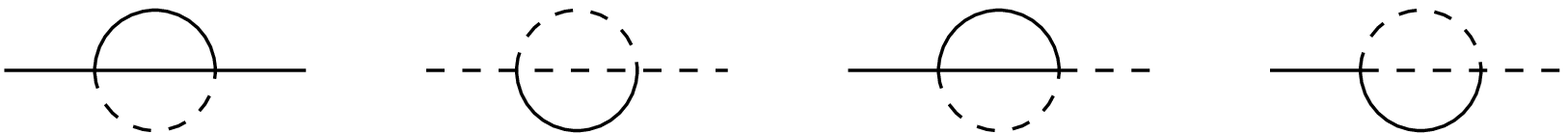}\end{center}

\caption{Diagrams for the high-frequency tail involving the $\zeta_3$ interaction.
\label{zeta3diagrams}}
\end{figure}

The tail of order $\zeta_+^2$ vanishes at $\omega\to vq$ because the $\omega$ dependence of $\Pi_{RLL}$ (equation \ref{resultsum}) cancels the factor of $(\omega-vq)^{-2}$ from the external legs of the corresponding diagram in $\delta\chi_{RR}$. It is easy to see that if the internal bubble had a different combination of $R$ and $L$ bosons ({\it e.g.}, if we replaced $\Pi_{RLL}$ by $\Pi_{RRL}$) this cancellation would not happen, leading to a divergence at $\omega\to vq$. The dangerous combinations of external legs and three-boson bubbles are excluded for the XXZ model, but are allowed for non-integrable models with $\zeta_3\neq 0$. The extra diagrams that contribute to the tail ($\delta\omega_q\ll \omega-vq\ll J$) to second order in the coupling constants are illustrated in figure \ref{zeta3diagrams}. The calculation is similar to the $O(\zeta_+^2)$ diagram. The result in the thermodynamic limit is \be
\delta S^{zz}_3(q,\omega)=\frac{3K}{128v}q^2\left[\frac{\zeta_3^2}{v^2}\, \frac{\omega^4+6v^2q^2\omega^2+v^4q^4}{v^2\left(\omega^2-v^2q^2\right)}+\frac{2\zeta_3\zeta_+}{v^2}\, \frac{\omega^2+v^2q^2}{v^2}\right].
\ee
Note the divergence as $\omega\to vq$ in the $O(\zeta_3^2)$ term. As in the finite field case, this divergence stems from the breakdown of perturbation theory in the band curvature terms at $\omega\sim vq$. We expect that this $\zeta_3$ contribution, which again is only finite for nonintegrable models, smoothes out the behavior of $S^{zz}(q,\omega)$ near the upper threshold $\omega_U(q)$, where the high-frequency tail joins the on-shell peak.

\bibliography{Lit_Sqw}

\end{document}